\documentclass[12pt,amsmath,amssymb,aps]{article}
\pdfoutput=1
\usepackage[T1]{fontenc}
\usepackage[left=1.9cm,right=1.9cm,top=2.2cm,bottom=2.2cm]{geometry}
\usepackage{hyperref}
\usepackage{cite}
\usepackage[utf8]{inputenc}
\usepackage[dvipsnames]{xcolor}
\usepackage{epsfig,latexsym}
\usepackage[symbol]{footmisc}
\usepackage[utf8]{inputenc}
\usepackage[dvipsnames]{xcolor}
\usepackage{epsfig,latexsym}
\usepackage{amsmath}
\usepackage{adjustbox}
\usepackage{tensor}
\usepackage{graphicx}
\usepackage{verbatim}
\usepackage{mathrsfs}
\usepackage{amssymb}
\usepackage{multirow}
\usepackage{color,colordvi}
\usepackage{appendix}
\usepackage{slashed}
\usepackage{array}
\usepackage[flushleft]{threeparttable}
\usepackage{cancel}
\usepackage{float}
\usepackage[font=footnotesize, justification=centering]{caption}
\usepackage{epsf}
\usepackage{tikz-feynhand}
\usetikzlibrary{shapes,shadows,arrows}
\usetikzlibrary{arrows}
\tikzset{arrow data/.style 2 args={%
 decoration={%
 markings,
 mark=at position #1 with \arrow{#2}},
 postaction=decorate}
 }%
\usetikzlibrary{svg.path}
\usetikzlibrary{patterns}
\usepackage{physics}
\usepackage{MnSymbol}
\usepackage{rotating}
\usepackage{multirow}
\usepackage[normalem]{ulem}
\usepackage{tcolorbox}
\usepackage{bbold}
\usepackage{cancel}
\usepackage{ytableau}
\usepackage[vcentermath]{youngtab}
\usepackage{tensor}
\usepackage{autobreak}
\usepackage{tikz}
\usepackage{pgfplots}
\usepackage[figurename=Diagram]{caption}
\pgfplotsset{width=3.3cm,compat=1.8,domain = min:max}
\usepgfplotslibrary{polar}
\usepackage{verbatim}
\usetikzlibrary{positioning, fit, calc}
\DeclareMathAlphabet{\mathpzc}{OT1}{pzc}{m}{it}

\csname @addtoreset\endcsname{equation}{section}

\newcolumntype{x}[1]{>{\centering\arraybackslash\hspace{0pt}}p{#1}}
\newcommand{\beq}{\begin{equation}}
\newcommand{\eeq}{\end{equation}}
\renewcommand{\[}{\left[}

\newcommand{\be}{\begin{eqnarray}\small}
\newcommand{\ee}{\end{eqnarray}}
\newcommand{\bea}{\begin{eqnarray}\small}
\newcommand{\eea}{\end{eqnarray}}
\newcommand{\bi}{\begin{itemize}}
\newcommand{\ei}{\end{itemize}}
\newcommand{\ben}{\begin{enumerate}}
\newcommand{\een}{\end{enumerate}}
\def\bes{\begin{equation*}}
\def\ees{\end{equation*}}
\def\bead{\begin{aligned}}
\def\eead{\end{aligned}}
\def\bmat{\left(\begin{matrix}}
\def\emat{\end{matrix}\right)}

\def\centerarc[#1](#2)(#3:#4:#5);%
 {
 \draw[#1]([shift=(#3:#5)]#2) arc (#3:#4:#5);
 }

\definecolor{dred}{HTML}{D95F02}
\colorlet{red}{white!15!dred}
\definecolor{darkgreen}{HTML}{1B9E77}
\definecolor{lightgray}{gray}{0.90}

\definecolor{Ecolor}{RGB}{106,157,235}
\definecolor{lightgray}{RGB}{220,220,220}

\definecolor{Gray}{gray}{0.95}

\DeclareOldFontCommand{\rm}{\normalfont\rmfamily}{\mathrm}

\usepackage{xparse}
\usepackage{tikz}
\usetikzlibrary{calc,matrix,patterns,shadings,backgrounds}
\pgfdeclarelayer{myback}
\pgfsetlayers{myback,background,main}

\tikzset{myfillcolor/.style = {draw,fill=#1}}%

\NewDocumentCommand{\highlight}{O{blue!40} m m}{%
\draw[myfillcolor=#1] (#2.north west)rectangle (#3.south east);
}

\NewDocumentCommand{\vshade}{O{blue!40} O{white} m m}{%
\draw[bottom color =#1,top color=#2] (#3.north west)rectangle (#4.south east);
}

\NewDocumentCommand{\oshade}{O{blue!40} O{white} m m}{%
\draw[right color =#1,left color=#2] (#3.north west)rectangle (#4.south east);
}

\NewDocumentCommand{\inshade}{O{blue!40} O{white} m m}{%
\draw[inner color =#1,outer color=#2] (#3.north west)rectangle (#4.south east);
}

\NewDocumentCommand{\fillpattern}{O{north west lines} O{blue!50} m m}{\draw[pattern=#1, pattern color=#2] (#3.north west)rectangle (#4.south east);
}

\NewDocumentCommand{\pt}{O{north west lines}}{\draw[pattern=north west lines, pattern color=red] (0,0) rectangle (0.3,0.3)}

\hypersetup{
 colorlinks=true,
 citecolor=black,
 linkcolor=blue,
 filecolor=magenta,
 urlcolor=blue,
}

\title{\textbf{\Large Identifying CP Basis Invariants in SMEFT}}
\date{ }
\author{Neda Darvishi$^a$\thanks{neda.darvishi@rhul.ac.uk}, Yining Wang$^{b,c}$\thanks{wangyining@itp.ac.cn}, Jiang-Hao Yu$^{b,c,d,e,f}$\thanks{jhyu@itp.ac.cn}}

\newcommand{\affiliations}{\centering
\noindent\textit{$^a$Department of Physics, Royal Holloway, University of London, Egham, Surrey, TW20 0EX, UK}
\\
\textit{$^b$CAS Key Laboratory of Theoretical Physics, Institute of Theoretical Physics, Chinese Academy of Sciences, Beijing 100190, China}
\\
\textit{$^c$School of Physical Sciences, University of Chinese Academy of Sciences, Beijing 100049, China}
\\
\textit{$^d$Center for High Energy Physics, Peking University, Beijing 100871, China}\\
\textit{$^e$School of Fundamental Physics and Mathematical Sciences, Hangzhou Institute for Advanced Study, UCAS, Hangzhou 310024, China}
\\
\textit{~~~~~~~~~~~~~~$^f$International Centre for Theoretical Physics Asia-Pacific, Beijing/Hangzhou, China}
}

\begin{document} 
\maketitle

{\centering\small\affiliations}


\begin{abstract}
\noindent 
Building on our automated framework that uses ring diagrams for classifying CP basis invariants~\cite{Darvishi:2023ckq}, this paper broadens the application of the methodology with more extensive examples and a wider scope of theoretical frameworks. Here, we showcase its versatility through detailed analyses of specific operators  in the Standard Model Effective Field Theory (SMEFT), such as a four-fermion operator at dimension-6 and a Yukawa operator extended up to dimension-$2n$ terms while maintaining a dimension-6 core, as well as in SMEFT with sterile neutrinos ($\nu$SMEFT) up to dimension-7. By integrating the ring-diagram technique with the Cayley-Hamilton theorem, we have developed a system that not only simplifies the process of identifying basic and joint invariants but also enables the automatic differentiation between CP-even and CP-odd invariants from the lowest orders.
Additionally, this work presents a comparison of our results with those derived using the traditional Hilbert-Poincar\'e series and its Plethystic logarithm. While these conventional approaches primarily yield the numerical count of invariants, our framework provides a complete structure of invariants, thereby surpassing the limitations of these traditional~methods.
\end{abstract} 

\newpage
\section{Introduction}\label{sec:intro}

Numerous experiments and observations have hinted at the presence of new physics beyond the Standard Model (BSM), including phenomena like neutrino masses, matter-antimatter asymmetry, and dark matter. Various theories of new physics (NP) have been proposed to account for these observations by extending the Standard Model (SM)'s field content and/or symmetries. As is known, the CP violation in the SM is not enough to explain the matter-antimatter asymmetry, and thus an additional source of the CP violation is needed by extending the SM. The CP violation is closely related to the flavor structure, in which arguably it is necessary to choose a certain basis and parametrization to formulate the Lagrangian in the theory. However, physical observables cannot depend on arbitrary choices of basis and can only depend on basis invariant quantities, such as rephasing invariants. In the context of group theory, basis invariants are objects which do not transform under the action of the group of basis changes~\cite{Jarlskog:1985ht,Jarlskog:1985cw,Dunietz:1985uy,Wu:1985ea,Bernabeu:1986fc,Jenkins:2009dy}. Thus, the classification of invariants is essential for analyzing theories that explain flavor and CP violation.

One of the most well-known examples of a basis invariant is the Jarlskog invariant~\cite{Jarlskog:1985ht,Jarlskog:1985cw}, which signals the presence of CP violation in the SM. The corresponding CP-odd invariant, ${J}^-$, is given by
\begin{align}
{J}^-\,=\,3 {\rm Im}\, \mathrm{Det} \Big[ Y_u{Y_u}^{\dagger}, Y_d{Y_d}^{\dagger} \Big],
\end{align}
where $Y_u$ and $Y_d$ are the Yukawa matrices for up-type and down-type quarks, respectively. This invariant comprehensively captures the effects of flavor transformations, while remaining independent of phase rotations. Furthermore, it has been extended to describe CP violation in quark and lepton mixing, as well as in BSM extensions. To generalize the description of CP violation in a basis-independent way, weak-basis invariants were proposed~\cite{Bernabeu:1986fc,Branco:1987mj}. These invariants are used to obtain rephasing-invariant CP-violating quantities in various NP models, such as the two-Higgs-doublet model~\cite{Mendez:1991gp,Lavoura:1994fv,Botella:1994cs,Gunion:2005ja,Trautner:2018ipq}, vector-like quark extensions~\cite{Aguilar-Saavedra:1997sbo,Branco:2011iw,Branco:1986my,Botella:1985gb,Gronau:1986xb,Darvishi:2016gvm,Darvishi:2016tni}, supergravity and supersymmetry models~\cite{Branco:1986kf,Lebedev:2002wq,Botella:2004ks}, and even in axion models~\cite{DiLuzio:2020oah,Valenti:2021rdu,Darvishi:2022wnd}.

In recent years, given the considerable energy gap between the NP scale and the electroweak scale, it is more popular to utilize effective field theory to describe NP. The standard model effective field theory (SMEFT) ~\cite{Weinberg:1979sa,Buchmuller:1985jz,Grzadkowski:2010es,Lehman:2014jma,Henning:2015alf,Li:2020gnx,Murphy:2020rsh,Li:2020xlh,Liao:2020jmn,Liao:2016hru,Harlander:2023psl} provides the most general parametrization on various NP effects using the higher dimensional operators, including a systematic organization of the CP violation effects. Since there are many CP-violating operators at higher dimensions, how to count and obtain these basis invariants becomes a challenge. The Hilbert-Poincar\'e series (HS) and its Plethystic logarithm (PL)~\cite{Pouliot_1999,Benvenuti:2006qr,Feng:2007ur,Jenkins:2009dy,Hanany:2014dia,Lehman:2015via,Lehman:2015coa,Henning:2015daa,Henning:2015alf} help us to investigate the basis invariant~\cite{Jenkins:2009dy,Hanany:2010vu,Wang:2021wdq,Yu:2021cco,Bonnefoy:2021tbt,Yu:2022ttm}, beside the weak basis invariant construction. This aids in finding the number of invariants, but cannot provide the explicit structure of invariants as well as distinguish between CP-even and CP-odd invariants. Particularly, when the CP properties of blocks are non-trivial.

While the ungraded Hilbert series is useful for counting invariants, it generally does not provide explicit structural information or distinguish between CP-even and CP-odd invariants, especially when the CP properties of building blocks are nontrivial. We acknowledge, however, that the graded Hilbert series can encode additional information, such as the spurion content of each invariant and, in some contexts, even their CP properties. This has been explored, for example, in~\cite{Henning:2017fpj,Grinstein:2023njq}. Nevertheless, the use of the graded Hilbert series becomes increasingly complex and computationally demanding in practice, particularly in theories involving high-rank tensors or many spurions. Our motivation, therefore, is to develop an explicit and systematic method that can classify invariants and their CP properties more transparently, while complementing the Hilbert series approach~\cite{Hanany:2010vu}.

In cases where the building blocks have trivial CP properties, distinguishing CP-odd invariants is relatively straightforward, for instance, CP-odd invariants can often be identified by the presence of an odd number of CP-odd blocks. However, for more general cases, particularly at higher orders, identifying the full set of independent invariants typically involves using the Hilbert series and its Plethystic Logarithm, along with constructing a linear ansatz to uncover syzygies. This makes the formulation and classification of basis invariants a compelling and still incomplete problem, motivating the development of more systematic approaches.

In \cite{Darvishi:2023ckq}, we introduced a systematic and generic procedure to explicitly classify these CP basis invariants. This approach uses fundamental blocks which can conveniently be constructed based on the recently developed so-called Ring-diagrams. These blocks are created as orthogonal trivial singlets, leading to the minimal possible invariants by design. This method is accompanied by the Cayley-Hamilton theorem and ring sets (petals) enabling us to identify basic and joint invariants as well as defining their CP properties. In this formalism, with the aid of the Cayley-Hamilton theorem, a set of rules over petals is applied. Therefore, with petals all higher-order invariants that are not independent quantities are distinguished. 
Particularly, our approach has the following advantages:
\begin{itemize}
\item Facilitates the recovery of the Cayley-Hamilton theorem using petals.
\item Offers a structured method for invariant classification across various theories.
\item Automatically distinguishes CP-odd from CP-even basis invariants.
\item Allows for the construction of CP-odd invariants from the lowest possible orders.
\item Can be applied to high-rank tensors and complex structures.
\end{itemize} 
Additionally, the minimal set of leading order CP-odd invariants is correlated to the number of phases that are expected for the sources of CP violation. 

In this paper, we explicitly apply the application of Ring-diagram in various examples in the SMEFT framework. Furthermore, to illustrate CP violation arising from the interference between operators of different dimensions, we derive the invariants in the SMEFT extended with a sterile neutrino ($\nu$SMEFT)~\cite{delAguila:2008ir,Aparici:2009fh,Bhattacharya:2015vja,Liao:2016qyd,Li:2021tsq}, including operators of dimensions 5, 6, and 7, as well as in a representative Type-I seesaw model.

The outline of the paper is as follows. In Section~\ref{sec:SM} we introduce our formalism to construct SM invariants. Afterwards, we generalise this formalism to identify invariants beyond the SM scenarios. We also show a simple method to distinguish the CP properties of invariants. In Section~\ref{sec:SMEFT-dim6}, we apply our method for the classification of invariants in the framework of SMEFT with dim-6 core and showcase two examples namely, (i) with the Yukawa operator (ii) with the Four-fermion operator. In Section~ \ref{sec:SSMEFT}, we showcase our explicit and automated construction of invariants in the low energy $\nu$SMEFT including three operators namely (i) with only one dim-5 operator, (ii) two dim-5 and dim-6 operators and finally (iii) with two dim-5 and dim-7 operators at the tree-level matching. We also include an example based on the Type-I seesaw model with a ${\rm U}(3) \times {\rm U}(3)$ flavor symmetry, to demonstrate the scalability of our method in more complex symmetry settings. Additionally, in all sections we also provide the traditional HS and PL of the ring to verify the number of invariants.
Finally, Section~\ref{sec:con} presents a summary of our findings and concluding remarks.

\section{Introduction of Ring-diagrams}
\label{sec:SM}

In this section, we elaborate on a methodology to categorize basis invariants within the SM and beyond, focusing particularly on flavor matrices. We commence with the SM Yukawa Lagrangian involving up-type and down-type quarks:
\bea
\mathcal{L} \supset Y_u \, \overline Q_{{\sf L}\,i} \widetilde{H} u_{{\sf R}\,j}+Y_d \, \overline Q_{{\sf L}\,i} ~{H} d_{{\sf R}\,j}+\text{h.c.},
\label{SML}
\eea
where the flavor matrices, from which masses and mixing angles are derived through diagonalization, are fundamental. These matrices, however, depend on the chosen basis, as quark and lepton fields in the Lagrangian can undergo unitary transformations. Under chiral flavor transformations, the Yukawa couplings transform as follows:
\begin{eqnarray}\small
Y_u & \to & {\rm U}(3)_Q \; Y_u \; {{\rm U}(3)_u}^{\dagger}, \nonumber \\
Y_d & \to & {\rm U}(3)_Q \; Y_d \; {{\rm U}(3)_d}^{\dagger},
\label{Q-trans}
\end{eqnarray}
with ${\rm U}(3)_Q$ and ${\rm U}(3)_{u,d}$ representing the unitary transformations on the quark doublet $Q_{{\sf L}}$ and singlet fields $u_{{\sf R}},~u_{{\sf R}}$, respectively, ensuring the Lagrangian's invariance. Direct comparisons of predictions of the mass matrices with experimental values are not straightforward due to their basis-dependency. Observable quantities must remain invariant under basis changes as described by~\eqref{Q-trans}. This necessitates expressing observables in a basis-independent manner using invariant quantities. The obvious invariants emerge from combinations $U \equiv Y_uY^\dagger_{u}$ and $D \equiv Y_d Y^\dagger_{d}$, transforming as:
\begin{eqnarray}\small
U & \to & {\rm U}(3)_Q \; U \; {{\rm U}(3)_Q}^{\dagger}, \nonumber \\
D & \to & {\rm U}(3)_Q \; D \; {{\rm U}(3)_Q}^{\dagger},
\label{UD-trans}
\end{eqnarray}
leading to traces of $U$ and $D$. By listing all possibilities from fundamental blocks $U$ and $D$, basic quark invariants can be derived. Higher order invariants are constrained by the Cayley–Hamilton theorem, which allows any $n$th power of an $n \times n$ matrix $A$ to be expressed in terms of powers less than $n$. For a $3 \times 3$ matrix $A$, this is exemplified by~\cite{Jenkins:2009dy}:
\begin{align}\small
A^3=&A^2\Tr(A)-\frac{1}{2}A\left[\Tr(A)^2-\Tr(A^2)\right]+\frac{1}{6}\big[\Tr(A)^3
\nonumber \\
&-3\Tr(A^2)\Tr(A)+2\Tr(A^3)\big]\mathbb{1}_{3\times 3}\,
	\label{eq:CHA3}	
\end{align}
with $\Tr(A^3)$ being non-reducible and the highest order for a single $3 \times 3$ matrix $A$ as
\begin{align}\small
	\Tr(A^4)=\frac{1}{6}\Tr(A)^4-\Tr(A^2)\Tr(A)^2+\frac{4}{3}\Tr(A^3)\Tr(A)+\frac{1}{2}\Tr(A^2)^2.
	\label{eq:CHA4}
\end{align}
Taking the above constraints into account, SM invariants $J_{nm}$, where $n$ and $m$ denote the orders of $U$ and $D$ respectively, include:
\begin{subequations}
\label{eq:SMinv}
\small
\begin{align}
 J_{10} &= \Tr(U), \label{subeq:1} \\
 J_{01} &= \Tr(D), \label{subeq:2} \\
 J_{20} &= \Tr(U^2), \label{subeq:3} \\
 J_{02} &= \Tr(D^2), \label{subeq:4} \\
 J_{30} &= \Tr(U^3), \label{subeq:5} \\
 J_{03} &= \Tr(D^3), \label{subeq:6} \\
 J_{11} &= \Tr(UD) = \Tr(DU), \label{subeq:7} \\
 J_{21} &= \Tr(U^2D) = \Tr(DU^2), \label{subeq:8} \\
 J_{12} &= \Tr(UD^2) = \Tr(D^2U), \label{subeq:9} \\
 J_{22} &= \Tr(U^2D^2). \label{subeq:10}
\end{align}
\end{subequations}

A CP-odd invariant and its CP-even counterpart are also identifiable:
\begin{align}\small
J_{33}&=\Tr\left(U^2 D^2 U D \right),
\\
{J_{33}}'&=\Tr\left(D^2 U^2 D U\right),
\end{align}
where ${\rm CP} \, (J_{33})\to {J_{33}}'$, so by rearranging the above invariants one can deduce the following CP-odd and CP-even invariants 
\begin{align}\small
J_{CP-odd}&= (J_{33})-{(J_{33})}',
\label{J33m}
 \\
J_{CP-even}&= (J_{33})+{(J_{33})}'.
\label{J33p}
\end{align}
Hence, the CP violation can be parameterised as
\begin{eqnarray}\small
J^-&\equiv & {\rm Im}\, \mathrm{Tr} \left[ U, D \right]^3=3 \, {\rm Im}\, \mathrm{Det} \left[ U, D \right].
\label{J4full}
\end{eqnarray}
Concluding that the CP-violating invariants $J^-$ can be written by 
$$J^-={\rm Im} (\text{independent quantities})\neq 0.$$
This the invariant appearing in the parametrization of the Cabibbo--Kobayashi--Maskawa (CKM) Matrix relevant to CP violation in the renormalizable theory of weak interaction~\cite{Cabibbo:1963yz,Kobayashi:1973fv,Wolfenstein:1983yz,Chau:1984fp,Greenberg:1985mr}.
Note that Eq.~\eqref{J33p} can be rewritten in terms of the lower primary invariants defined in Eq.~\eqref{eq:SMinv}, so disregarded as a basic invariant.

Now, we use an alternative procedure that systemically can classify invariants in the SM and beyond, using a Ring-diagram that is accompanied by the Cayley--Hamilton theorem and equivalently by ring-set (petals) enabling us to identify basic invariants. Particularly, we show our approach automatically distinguishes the CP properties of invariants. Additionally, the structure of interrelations between basis invariants (syzygies) can be expressed through the chain procedure providing an explicit cancellation between terms where only basic invariants remain in them.
$${\rm Syzygies} = f({\rm Basic\;Invs}) +\overbrace{\sum_{oreder-n} {\rm Joint\;Invs}\times \Tr(A^n,B^n,C^n,D^n)^m}^{f(\rm Basic\;Inv)}\;.$$ 
Considering the outlined structure, it becomes apparent that when classifying invariants, beginning with the lowest order and progressively extending to higher orders, there comes a juncture where the summation of various permutations of a generating set can be articulated in terms of fundamental invariants. These invariants, which are algebraically dependent, are referred to as ``Joint" invariants. The visual representations of the sequence of steps that we follow to obtain and distinguish invariants are summarized in Flowcharts ~\ref{F1} and ~\ref{F2}.
The visual representations of the sequence of steps that we follow to obtain and distinguish invariants are summarized in Flowcharts ~\ref{F1} and ~\ref{F2}.

{\renewcommand{\figurename}{Flowchart}
\begin{figure}[t]
{\centering\includegraphics[width=0.9\textwidth]{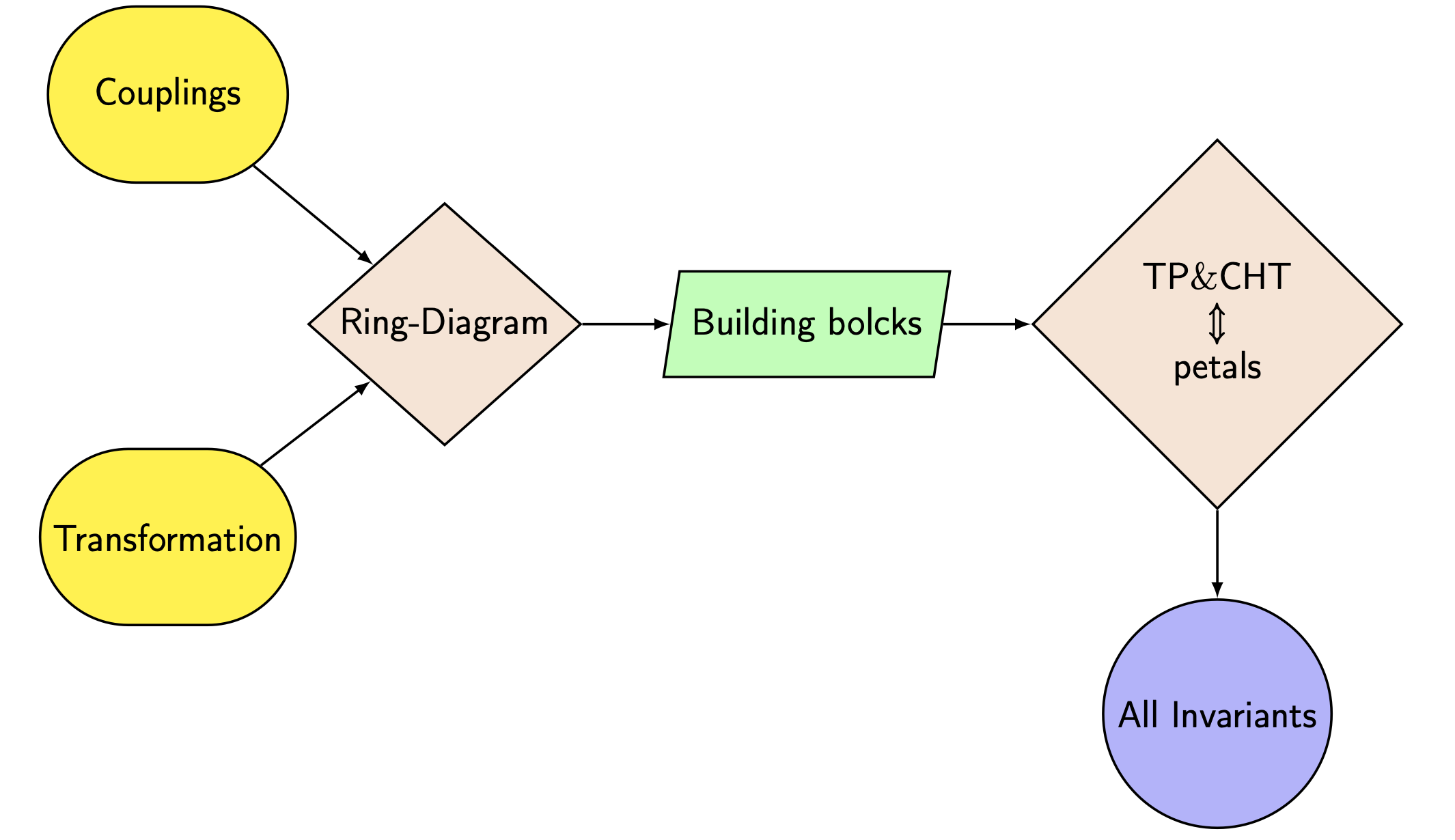}}
\caption{The steps for derivation of invariants are visualised. In the fifth block, the acronyms TP and CHT are used for tensor products and the Cayley-Hamilton Theorem.}
\label{F1}\end{figure}}

\begin{table}[t]
	\centering
	\begin{tabular}{c|c|c|c}
		& ${\rm U}(3)_Q$ & ${\rm U}(3)_u$ & ${\rm U}(3)_d$ \\\hline
		$Y_u$ &$\mathbf{3}$ & $ \mathbf{\overline{3}}$ & $\mathbf{1}$ \\[0.1cm]
		$Y_d$ &$\mathbf{3}$ & $ \mathbf{1}$ & $\mathbf{\overline{3}}$ \\[0.1cm]
	\end{tabular}
	\caption{\it Flavor transformation of the Yukawa matrices.}
	\label{tab:tab1}
\end{table} 
To initiate, we distribute three rings corresponding to the flavor transformations ${\rm U}(3)_Q$, ${{\rm U}(3)_u}$, and ${{\rm U}(3)_d}$ for the Yukawa matrices as displayed in Table~\ref{tab:tab1}. In this configuration, elements $y_u$ and $y_u^\dagger$ are placed in the ${{\rm U}(3)_u}$-ring, and $y_d$ and $y_d^\dagger$ in the ${{\rm U}(3)_d}$-ring, with the ${\rm U}(3)_Q$-ring linking the two. This structure and its simplified version are depicted in Diagram ~\ref{D1}.
\setcounter{figure}{0}
\renewcommand{\figurename}{Diagram}
\begin{figure}[t]
\begin{center}
\includegraphics[width=0.4\textwidth]{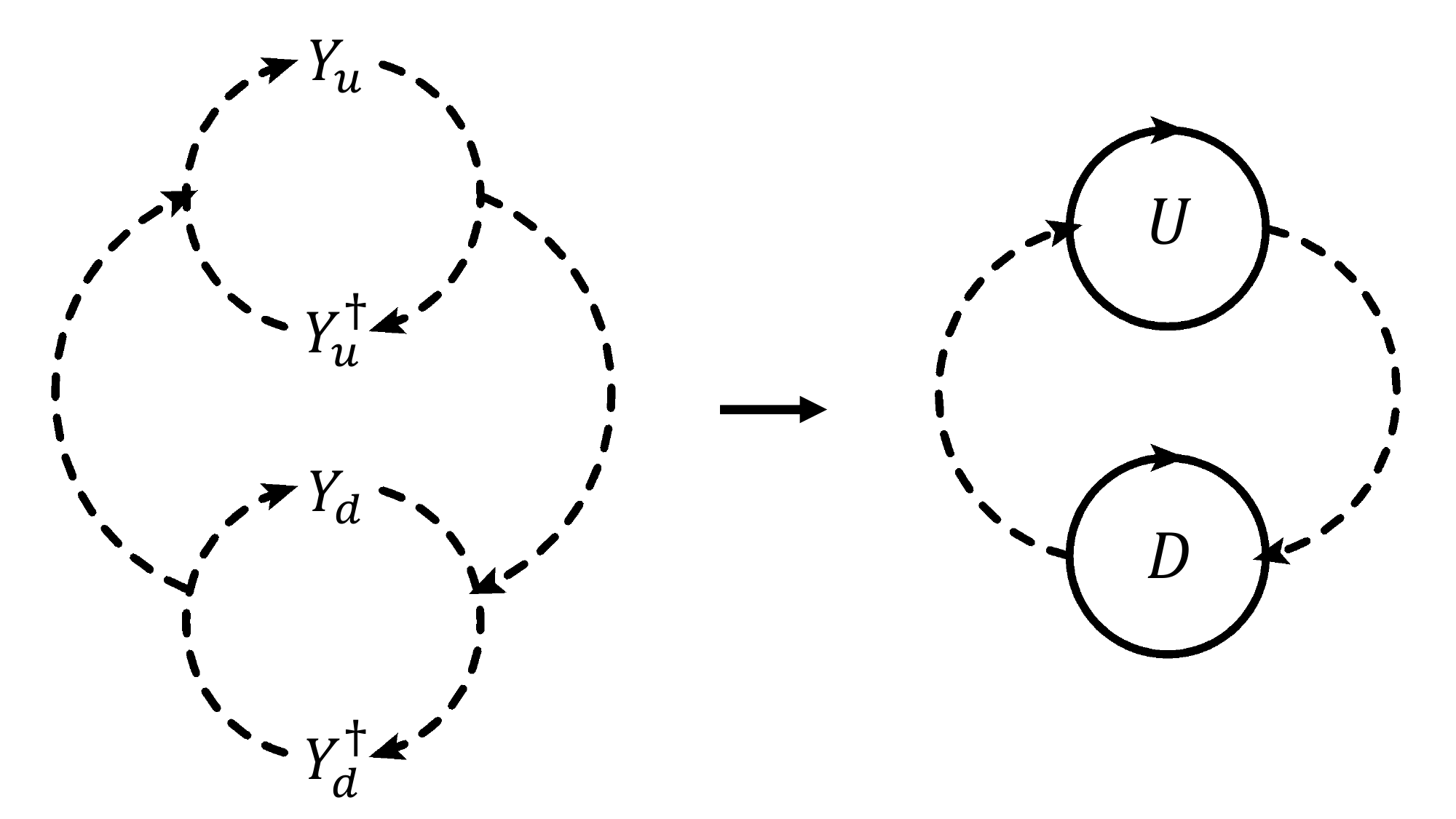}
\end{center}
\caption{Ring-diagram for the SM.}
\label{D1}
\end{figure}
We then define three building blocks relevant to these rings, allowing only a single passage through them, as follows:
\ytableausetup{smalltableaux}
\bea
&M_u=\text{diag}(U,\,0) \equiv \hspace{0.2cm}{\raisebox{-0.7em}{\begin{tikzpicture}
\def \radius {0.4cm}
 \node[draw, circle,style={thick},-latex] {$U$};
 \draw[-latex,->,thick] ({180}:\radius);
\end{tikzpicture}}}\,,
\quad M_d=\text{diag}(0,\,D)\equiv \hspace{0.2cm}{\raisebox{-0.7em}{\begin{tikzpicture}
\def \radius {0.4cm}
 \node[draw, circle,style={thick},-latex] {$D$};
 \draw[-latex,->,thick] ({180}:\radius);
\end{tikzpicture}}}\,
\nonumber \\
& M_{ud}^\pm \equiv \begin{pmatrix}
 U D & 0\\
 0 & \pm \,DU \\
\end{pmatrix}~ \equiv \raisebox{-0.5em}{ 
\begin{tikzpicture}[transform shape,line width=0.9pt]
\node (A) at (0,0) {$U$}; 
\node (B) at (1.2,0) {$D$}; 
\path[draw] 
(A) edge[bend right=-40,black,dashed,->] (B)
(B) edge[bend right=-40,black,-latex,dashed,->] (A);
\end{tikzpicture}} \,,
 \label{blocks-SM}
\eea
where traces over rings form so-called petals, represented as follows:
\bea
\ytableausetup{boxsize=1.6em}\ytableausetup{smalltableaux}
\Tr {\raisebox{-0.7em}{\begin{tikzpicture}
\def \radius {0.4cm}
 \node[draw, circle,style={thick},-latex] {$U$};
 \draw[-latex,->,thick] ({180}:\radius);
\end{tikzpicture}}~} & :=& \begin{tikzpicture}
 \begin{polaraxis}[grid=none, axis lines=none]
 	\addplot+[mark=none,domain=0:90,samples=50,color=red,style={ultra thick}] 
		{sin(2*x)}; 
 \end{polaraxis}
 \end{tikzpicture} 
= \Tr (U), \nonumber\\
 \Tr{\raisebox{-0.7em}{\begin{tikzpicture}
\def \radius {0.4cm}
 \node[draw, circle,style={thick},-latex] {$D$};
 \draw[-latex,->,thick] ({180}:\radius);
\end{tikzpicture}}~}& :=& \hspace{0.0em}{ \begin{tikzpicture}
 \begin{polaraxis}[grid=none, axis lines=none]
 	\addplot+[mark=none,domain=0:90,samples=50,color=blue,style={ultra thick}] 
		{sin(2*x)}; 
 \end{polaraxis}
 \end{tikzpicture}}= \Tr(D), 
 \nonumber\\
\Tr {\hspace{-0.2em}{\raisebox{-0.5em}{ 
\begin{tikzpicture}[transform shape,line width=0.9pt]
\node (A) at (0,0) {$U$}; 
\node (B) at (1.2,0) {$D$}; 
\path[draw] 
(A) edge[bend right=-40,black,dashed,->] (B)
(B) edge[bend right=-40,black,-latex,dashed,->] (A);
\end{tikzpicture}}} }^+ &:=& \hspace{0.0em}{\raisebox{-1.5
em}{\begin{tikzpicture}
 \begin{polaraxis}[grid=none, axis lines=none]
 	\addplot+[mark=none,domain=0:90,samples=50,color=red,style={ultra thick}] 
		{sin(2*x)}; 
 \end{polaraxis}
 \begin{polaraxis}[grid=none, axis lines=none]
 	\addplot+[mark=none,domain=0:90,samples=50,color=blue,style={ultra thick}] 
		{-sin(2*x)}; 
 \end{polaraxis}
 \end{tikzpicture}}}
= \Tr (U D), \nonumber \\
 \Tr {\hspace{-0.2em}{\raisebox{-0.5em}{ 
\begin{tikzpicture}[transform shape,line width=0.9pt]
\node (A) at (0,0) {$U$}; 
\node (B) at (1.2,0) {$D$}; 
\path[draw] 
(A) edge[bend right=-40,black,dashed,->] (B)
(B) edge[bend right=-40,black,-latex,dashed,->] (A);
\end{tikzpicture}}}}^-& :=& \hspace{-0.2em}{\raisebox{-1.5
em}{\begin{tikzpicture}
 \begin{polaraxis}[grid=none, axis lines=none]
 	\addplot+[mark=none,domain=0:90,samples=50,color=red,style={ultra thick}] 
		{sin(2*x)}; 
 \end{polaraxis}
 \begin{polaraxis}[grid=none, axis lines=none]
 	\addplot+[mark=none,domain=0:90,samples=50,color=blue,style={ultra thick}] 
		{-sin(2*x)}; 
 \end{polaraxis}
 \end{tikzpicture}}}
= 0.
 \label{read-ring1}
\eea
As discussed earlier, the components of $U$ and $D$ are $y_u$, $y_u^\dagger$, and $y_d$, $y_d^\dagger$, which are $3 \times 3$ matrices following U(3)$_u$ and U(3)$_d$ transformations, respectively. The Yukawa couplings $y_u$ and $y_d$ are indeed $3 \times 3$ matrices; however, once the blocks $U$, $D$, and $UD$ are constructed directly from the symmetries and transformations, as illustrated in Diagram \ref{D1}, we deal with $2 \times 2$ blocks. These blocks significantly simplify the representation while maintaining the integrity of the physical information. Now, using the fundamental nature of these blocks and rings, the higher-order invariants can be organized using the following tensor product: \begin{equation} \label{x123SM} J_{{u}^{x_1} {d}^{x_2} {ud}^{x_n}} = \Tr \left({M_u}^{x_1} \otimes {M_d}^{x_2} \otimes {M_{ud}}^{x_n} \right), \end{equation} or equivalently, using the concept of ``petal combinations" representation. The nature of the products/diagram formalism provides the framework for combining these blocks into more complex structures. Given still the $3 \times 3$ nature of $U$ and $D$ as matrices, this approach permits up to three maximum passes over a ring, or combinations thereof, as long as the count of each petal-type does not surpass three. To formulate higher-order invariants with this petal methodology, up to two of the three petals of the same colour can be repeated. Consequently, the third petal of the same colour must be sequenced after a petal of a different type. Furthermore, simplification of a petal set is feasible by substituting two adjacent petals with a double-layer petal. This procedure is extendable to repetitions of multiple petals up to two times, e.g.
\bea
 \hspace{-0.1cm}{\raisebox{-1.1em}{\includegraphics[width=0.2\textwidth]{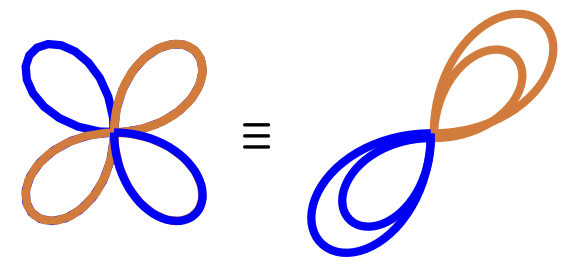}}}
 \nonumber \eea
After this process of simplification, these arrangements can be translated into relational expressions by selecting an arbitrary starting point within the rings and then reading the entire cycle clockwise $\pm$ anti-clockwise i.e. 
\bea J^\pm \equiv (\circlearrowright +\searrow\swarrow \nwarrow \cdots) \pm (\circlearrowleft+\nearrow\nwarrow\swarrow \cdots). \label{arrows}
 \eea 
This process repeats until identical terms are produced. The clockwise arrow ($\circlearrowright$) indicates reading through the terms in a forward direction around the cycle, while the anti-clockwise arrow ($\circlearrowleft$) indicates reading through the terms in reverse. The arrows ($\searrow\swarrow\cdots$) represent diagonal paths through the cycle. For example, in the case of a four-petal diagram with three distinct components:
\bea  \hspace{-0.1cm}{\raisebox{-1.1em}{\includegraphics[width=0.09\textwidth]{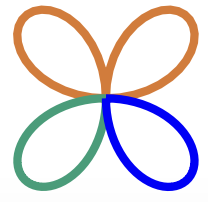}}}
 \nonumber
 \eea
By picking a starting point (for instance, the top left), $J^+$ and $J^-$ expressions are constructed as follows:
\bea
J^+ \equiv R\, R\, B\, G \,(\circlearrowright) + R\, B\, R\, G \,(\searrow\swarrow) +  G\, B \,R\, R\, (\circlearrowleft) + G\, R\, B\, R \,(\nearrow\nwarrow),
\nonumber \\
J^- \equiv R\, R\, B\, G \,(\circlearrowright) + R\, B\, R\, G \,(\searrow\swarrow) -  G\, B \,R\, R\, (\circlearrowleft) - G\, R\, B\, R \,(\nearrow\nwarrow),
\nonumber
\eea
for $J^-$, the diagonal terms cancel each other out, while in $J^+$, one of the repeated diagonal terms is dropped to avoid redundancy. This results in the simplified expressions:
\bea \hspace{-0.1cm}{\raisebox{-1.1em}{\includegraphics[width=0.09\textwidth]{R2BG.png}}}
\equiv &(R\, R\, B\, G + R\, R\, G\, B + R\, B\,R\, G) \quad\text{and}\quad (R\, R\, B\, G - R\, R\, G\, B).
 \eea

In the next section, we will elucidate the link between these operational rules and the principles underlying the Cayley-Hamilton~theorem. Utilising the fundamental blocks~\eqref{read-ring1} the SM invariants can be recovered as:
{\allowdisplaybreaks
\bea
&\left.
 \begin{array}{rl}
J_{u^1}=\Tr( M_u) & := \begin{tikzpicture}
 \begin{polaraxis}[grid=none, axis lines=none]
 \addplot+[mark=none,domain=0:90,samples=50,color=red,style={ultra thick}] 
 {sin(2*x)}; 
 \end{polaraxis}
 \end{tikzpicture} 
= \Tr (U) \\
J_{d^1}=\Tr(M_d) & := \hspace{0.0em}{ \begin{tikzpicture}
 \begin{polaraxis}[grid=none, axis lines=none]
 \addplot+[mark=none,domain=0:90,samples=50,color=blue,style={ultra thick}] 
 {sin(2*x)}; 
 \end{polaraxis}
 \end{tikzpicture}}= \Tr (D)
 \end{array}
\right\} & \to \text{order-1}
 \nonumber \\
 &
 \nonumber \\
&\left. \begin{array}{rl}
J_{ud^1}=\Tr (M_{ud}) & := \hspace{0.0em}{\raisebox{-1.5
em}{\begin{tikzpicture}
 \begin{polaraxis}[grid=none, axis lines=none]
 	\addplot+[mark=none,domain=0:90,samples=50,color=red,style={ultra thick}] 
		{sin(2*x)}; 
 \end{polaraxis}
 \begin{polaraxis}[grid=none, axis lines=none]
 	\addplot+[mark=none,domain=0:90,samples=50,color=blue,style={ultra thick}] 
		{-sin(2*x)}; 
 \end{polaraxis}
 \end{tikzpicture}}}
= \Tr (U D) \nonumber \\
J_{u^2}=\Tr(M_u^{ 2}) & := \hspace{0.0em}{\raisebox{-1.5
em}{\begin{tikzpicture}
 \begin{polaraxis}[grid=none, axis lines=none]
 	\addplot+[mark=none,domain=0:90,samples=50,color=red,style={ultra thick}] 
		{sin(2*x)}; 
 \end{polaraxis}
 \begin{polaraxis}[grid=none, axis lines=none]
 	\addplot+[mark=none,domain=0:90,samples=50,color=red,style={ultra thick}] 
		{-sin(2*x)}; 
 \end{polaraxis}
 \end{tikzpicture}}}= \Tr (U^2) \nonumber \\
J_{d^2} =\Tr(M_d^{2})& := \hspace{0.0em}{\raisebox{-1.5
em}{\begin{tikzpicture}
 \begin{polaraxis}[grid=none, axis lines=none]
 	\addplot+[mark=none,domain=0:90,samples=50,color=blue,style={ultra thick}] 
		{sin(2*x)}; 
 \end{polaraxis}
 \begin{polaraxis}[grid=none, axis lines=none]
 	\addplot+[mark=none,domain=0:90,samples=50,color=blue,style={ultra thick}] 
		{-sin(2*x)}; 
 \end{polaraxis}
 \end{tikzpicture}}}=\Tr (D^2)
 \end{array}
\right\} & \to \text{order-2}
 \nonumber \\
 &
 \nonumber \\
&\left. \begin{array}{rl} 
J_{u^1{ud}^1}=\Tr(M_u \otimes M_{ud} ) & := \hspace{0cm}{\raisebox{-1.1em}{ 
\begin{tikzpicture}
 \begin{polaraxis}[grid=none, axis lines=none]
 	\addplot+[mark=none,domain=0:60,samples=50,color=blue,style={ultra thick}] 
		{-sin(3*x)}; 
 \end{polaraxis}
 \begin{polaraxis}[grid=none, axis lines=none]
 	\addplot+[mark=none,domain=61:180,samples=50,color=red,style={ultra thick}] 
		{-sin(3*x)}; 
		\end{polaraxis}
 \end{tikzpicture}}}= \Tr (U^2 D) 
\nonumber \\
J_{{ud}^1d^1}=\Tr(M_d \otimes M_{ud}) & := \hspace{0cm}{\raisebox{-1.1em}{ 
\begin{tikzpicture}
 \begin{polaraxis}[grid=none, axis lines=none]
 	\addplot+[mark=none,domain=60:120,samples=50,color=red,style={ultra thick}] 
		{-sin(3*x)}; 
 \end{polaraxis}
 \begin{polaraxis}[grid=none, axis lines=none]
 	\addplot+[mark=none,domain=0:60,samples=50,color=blue,style={ultra thick}] 
		{-sin(3*x)}; 
		\end{polaraxis}
		 \begin{polaraxis}[grid=none, axis lines=none]
 	\addplot+[mark=none,domain=120:180,samples=50,color=blue,style={ultra thick}] 
		{-sin(3*x)}; 
		\end{polaraxis}
 \end{tikzpicture}}}= \Tr( D^2U) \nonumber \\
J_{u^3}=\Tr(M_u^{ 3}) & := \hspace{0cm}{\raisebox{-1.1em}{ 
\begin{tikzpicture}
 \begin{polaraxis}[grid=none, axis lines=none]
 	\addplot+[mark=none,domain=0:180,samples=50,color=red,style={ultra thick}] 
		{-sin(3*x)}; 
 \end{polaraxis}
 \end{tikzpicture}}}= \Tr (U^3)
 \nonumber \\
J_{d^3} =\Tr(M_d^{3})& := \hspace{0cm}{\raisebox{-1.1em}{ 
\begin{tikzpicture}
 \begin{polaraxis}[grid=none, axis lines=none]
 	\addplot+[mark=none,domain=0:180,samples=50,color=blue,style={ultra thick}] 
		{-sin(3*x)}; 
 \end{polaraxis}
 \end{tikzpicture}}}= \Tr (D^3)
 \end{array}
\right\} & \to \text{order-3}
 \nonumber \\
 &
 \nonumber \\
&\left. \begin{array}{rl}
J_{ud^2}=\Tr(M_{ud}^{ 2}) & := \hspace{0.0cm}{\raisebox{-1.4em}{ \begin{tikzpicture}
\begin{polaraxis}[grid=none, axis lines=none]
\addplot+[mark=none,domain=0:180,samples=50,color=red,style={ultra thick}] 
		{+sin(2*x)}; 
 \end{polaraxis}
 \begin{polaraxis}[grid=none, axis lines=none]
\addplot+[mark=none,domain=180:360,samples=50,color=blue,style={ultra thick}] 
		{+sin(2*x)}; 
 \end{polaraxis}
 \end{tikzpicture}}}= \Tr ((U D)^2) \end{array}
\right\} & \to \text{order-4}
\label{Inv-SM}
\eea
With $x_1 = x_2 = 0$ and $x_3 = 3$ in Eq. \eqref{x123SM}, an order-6 CP-odd invariant related to \eqref{J33m} can be deduced from 
\bea
\small
{J^-}_{ud^3} = &\Tr(M_{ud}^{ 3}):= \hspace{-0.1cm}{\raisebox{-1.1em}{\includegraphics[width=0.09\textwidth]{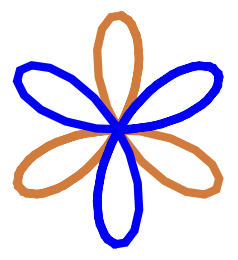}}} = \Big[\Tr (U^2 D^2 U D) - \Tr (D^2 U^2 D U)\Big], \label{Inv-CPO-SM}
\eea
by traversing the entire cycle of the hexafoil with two ingredients both clockwise and counter-clockwise, as 
\bea
 &&R\, B\, R\, B \,R\, B \,(\circlearrowright) + R\, B\, B \,R\, R\, B\,(\downarrow\swarrow\nwarrow) - B \,R\, B \,R\, B \,R\, \,(\circlearrowleft) - B \,R\, R\, B\,B \,R\,(\searrow\nearrow\uparrow)\nonumber\\
 && \equiv  R\, B\, B \,R\, R\, B\,(\downarrow\swarrow\nwarrow) - B \,R\, R\, B\,B \,R\,(\searrow\nearrow\uparrow),
\nonumber
\eea
which we derive the non-reducible CP-odd invariant  ${J^-}_{ud^3}$. The CP violation is parametrised as $ \mathrm{Im} \,{J^-}_{ud^3}$. Note that ${J^+}_{ud^3}$ can be delineated in terms of basic invariants, a relationship that will be detailed in the subsequent section in Eq.~\eqref{eq:CP-evenA3B3}. This occurs because, in the CP-even construction, petals of the same color can always sit adjacent to each other, forming a $3+3$ structure, which makes the expression reducible. 

Furthermore one can compare the number of the generating set of invariants using HS by the assignment of $z_1 + z_2 + z_3$ and $z_1^{-1} + z_2^{-1} + z_3^{-1}$ to represent the ${\bf 3}$ and ${\bf 3^*}$, respectively, which are pertinent to the flavor transformation of the Yukawa matrices as shown in Table~\ref{tab:tab1}. Thus, the Character functions of the building block $U$ and $D$, are
\begin{eqnarray}\small
\chi_{U,D}(z_1,z_2,z_3)&=&\left(z_1 +z_2 +z_3 \right)\left(z_1^{-1}+z_2^{-1}+z_3^{-1}\right).
\end{eqnarray}
Here, $z_1,z_2,z_3$ are coordinates on the maximum torus of ${\rm U}(3)$. Thus, we can find the plethystic exponential (PE), as
\begin{eqnarray}\small
{\rm PE}\left(z_1,z_2,z_3 \right.&;&\left.U D\right)
={\rm PE}\left[\chi_{U} (z_1,z_2,z_3 )U\right] {\rm PE}\left[\chi_{D} (z_1,z_2,z_3 )D\right].
\end{eqnarray}
Therefore, the multi-graded HS for the quark sector are
\begin{eqnarray}\small
{\cal H}(U,D) &=& ({1+U^3D^3})^{-1} \times \Big\{ (1-U)(1-U^2)(1-U^3)(1-D)(1-D^2) \nonumber \\
&&(1-D^3) (1-UD)(1-U^2D)(1-UD^2)(1-U^2D^2) \Big\}.
\label{H-SM}
\end{eqnarray}
Here, if $U=D=q$ the one-variable HS takes on the following form 
\begin{eqnarray}\small
{\cal H}_{\rm SM}(q,q)&=&\frac{1+q^{6}}{(1-q)^2(1-q^2)^3(1-q^3)^4(1-q^4)}.
\label{H-SM-q}
\end{eqnarray}
Given the function ${\cal H}\left(q\right)$, the PL is defined by
\begin{eqnarray}\small
\label{eq:PL def}
{\rm PL}\left[{\cal H}\left(q\right))\right]\equiv
\sum_{k=1}^{\infty} \frac{\mu(k)}{k}\,{\rm ln}\left[{\cal H}\left(q^k\right)\right]\;,
\end{eqnarray}
with $\mu(k)$ being the M{\"o}bius function,
\begin{equation}
\mu(k)\equiv
\begin{cases}
0 & \text{$k$ contains any repeated prime factors} \\
1 & k=1 \\
(-1)^n & \text{$k$ is a product of $n$ distinct primes}
\end{cases}\;.
\end{equation}
Hence, PL related to Eq.~\eqref{H-SM-q} reads
\begin{eqnarray}\small
{\rm PL}\left[{\cal H}(q)\right]=2 q + 3 q^2 + 4 q^3 + q^4 + q^6 - q^{12}\;,
\end{eqnarray}
indicating the total number of $(2 \,+\, 3 \,+\, 4 \,+\,1 \,+\, 1)$ invariants that are in agreement with the sets of invariants given in Eqs.~\eqref{Inv-SM} and ~\eqref{Inv-CPO-SM}.

{\renewcommand{\figurename}{Flowchart}
\begin{figure}[t]
{\centering\includegraphics[width=0.9\textwidth]{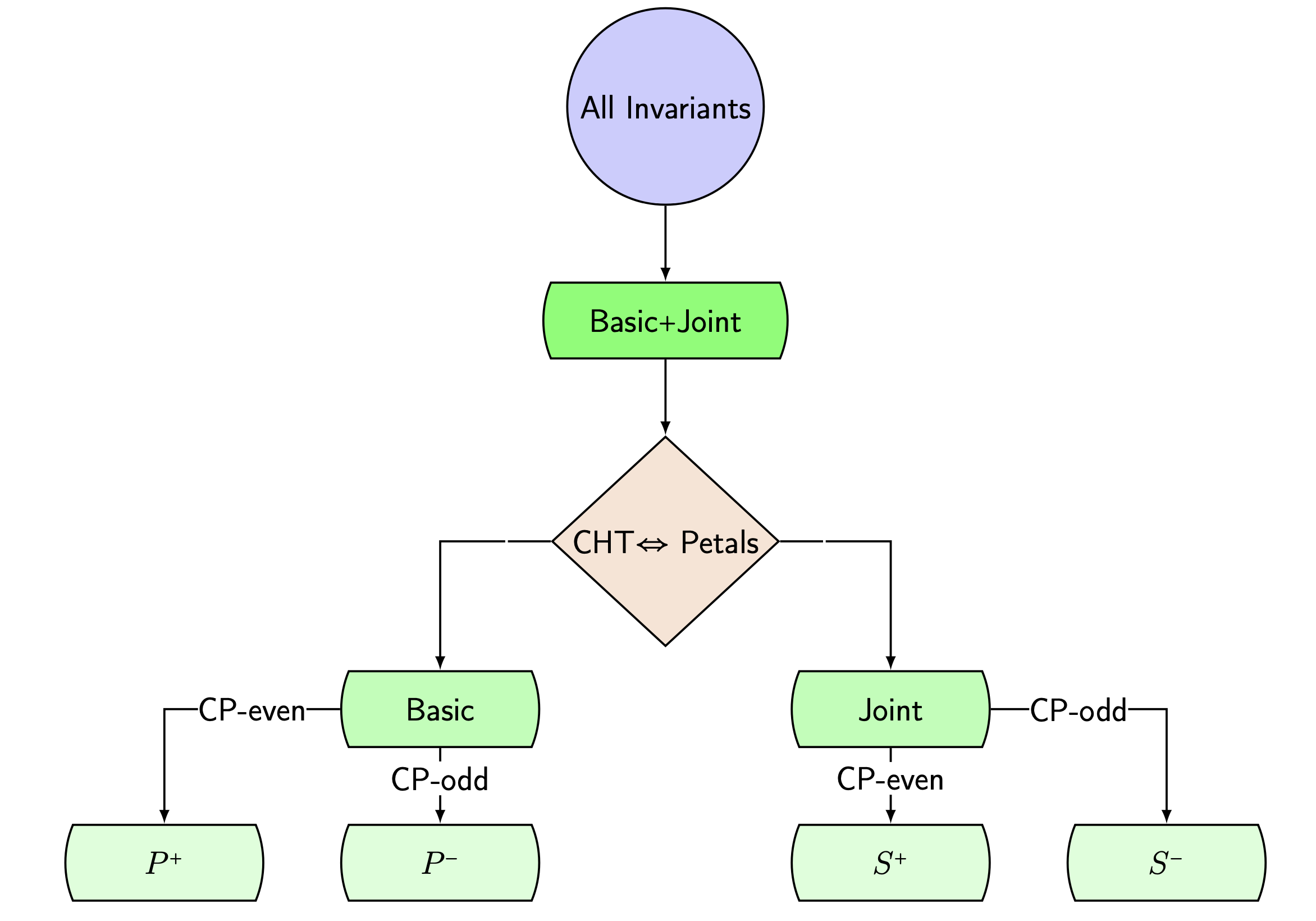}}
\caption{The steps for distinguishing invariants are visualised.}
\label{F2}
\end{figure}}

This formalism can be extended to identify possible invariants in a generic scenario, as well as distinguishing the CP-violating invariants. Let us start with a generic example with $n=m+k$ main elements $p_1, \cdots, p_m$ and $q_1, \cdots,q_k$, where $p$s are rotating with $U_1$, $U_2$ and $q$s with $U_1$, $U_3$. In analogy with $U_{1,2,3}$, it is convenient to draw three rings where the first ring is labelled with $U_2$ related to rotations of $p$s with $U_2$, the second one is $U_3$-ring pertinent to rotations of $q$s with $U_3$, whereas the third ring connects all rings relevant to mutual $U_1$. This structure is shown in the following Ring-diagram~\ref{R1}, where the number of these rings can be increased due to the presence of more rotations with $U_i$.
\setcounter{figure}{1}
\begin{figure}
\centering
\includegraphics[width=0.2\textwidth]{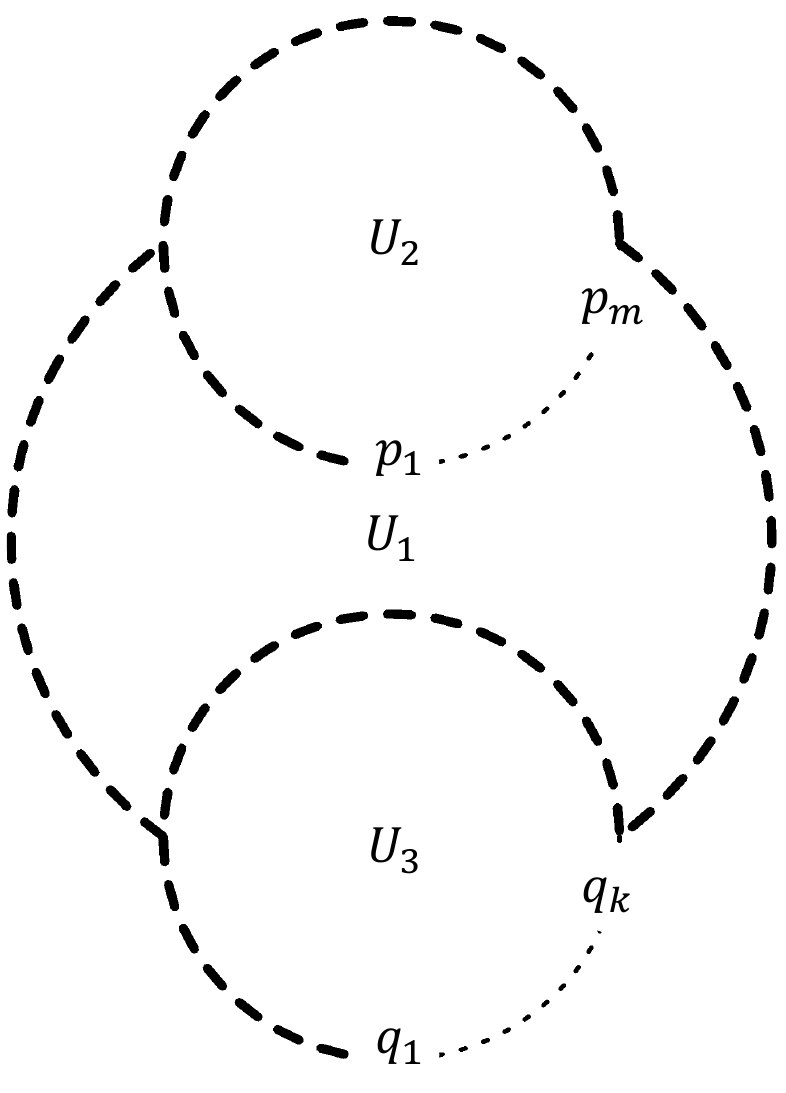}
\caption{The basic structure of a Ring-diagram.}
\label{R1}
\end{figure}
The ring setup is designed around two types of inputs for crafting invariants. First, we have the first-order blocks, derived from main elements within a $U_i$-ring that are effectively cancelling $U_i$. These blocks are placed directly in their respective ring and are marked with a small circle ``$\circlearrowright$'', linked by dashed lines, for example, ``$\begin{tikzpicture}[transform shape,line width=1pt]
\node (A) at (2,2) {};
\node (B) at (3,2) {};
\path[draw]
(A) edge[bend right=-30,black,dashed,->] (B)
(B) edge[bend right=-30,black,-latex,dashed,->] (A);
\end{tikzpicture}$''.
The second type involves elements that need combinations from other rings to cancel out $U_i$ rotations. These elements enter unchanged, connecting to others or blocks with one-way dashed lines, like ``$\begin{tikzpicture}[transform shape,line width=1pt]
\node (A) at (2,2) {};
\node (B) at (3,2) {};
\path[draw]
(A) edge[bend right=-30,black,dashed,->] (B);
\end{tikzpicture}$''. A dashed line loop indicates a building block, allowing only one pass through these lines.
In this case, a cycle of dashed lines forms a building block, with only one turn passing through the connection lines i.e. $$\begin{tikzpicture}[transform shape,line width=0.8pt]
\node (A) at (1,1) { x}; 
\node (B) at (3,1) { x}; 
\path[draw] 
(A) edge[bend right=-30,black,dashed,->] (B)
(B) edge[bend right=-30,black,dashed,->] (A);
\end{tikzpicture},\,
\begin{tikzpicture}[transform shape,line width=0.8pt]
\node (A) at (1,1) {x }; 
\node (B) at (2,2) {x }; 
\node (C) at (3,1) { x}; 
\path[draw] 
(A) edge[bend right=-30,black,dashed,->] (B) 
(B) edge[bend right=-30,black,dashed,->] (C)
(C) edge[bend right=-30,black,dashed,->] (A);
\end{tikzpicture}, \,
\begin{tikzpicture}[transform shape,line width=0.8pt]
\node (A) at (1,1) {x }; 
\node (B) at (2,2) { x}; 
\node (C) at (3,1) {x };
\node (D) at (2,0) { x}; 
\node (E) at (1, 1) {x }; 
\path[draw] 
(A) edge[bend right=-30,black,dashed,->] (B) 
(B) edge[bend right=-30,black,dashed,->] (C)
(C) edge[bend right=-30,black,dashed,->] (D) 
(D) edge[bend right=-30,black,dashed,->] (E);
\end{tikzpicture}, \, \cdots.$$ 
The mutual $U_1$ ring's design limits pass through connection lines based on the number of blocks in $U_{2,3}$ rings. This approach streamlines block combination creation.

Therefore, in a ring diagram one can identify first-order blocks and building blocks from dashed cycles as $m_1$ and $m_2$, respectively. These blocks serve as the foundation for creating tensors and higher-order invariants. First-order blocks can be aligned into sets of $m_1$ orthogonal vectors. For example, in a scenario with four blocks ($A$, $B$, $C$, and $D$) subject to three rotations, as shown Ring-diagram~\ref{r2}, the foundational block $M_a$ is defined as a diagonal matrix with $A$ in the first position and zeros elsewhere, representing a direct translation from the ring input. Similarly, $M_b$, $M_c$, and $M_d$ are constructed to hold $B$, $C$, and $D$ in their respective primary positions within diagonal matrices, ensuring each element's unique pathway through the system. 
\begin{figure}[H]
\begin{center}
\includegraphics[width=0.23\textwidth]{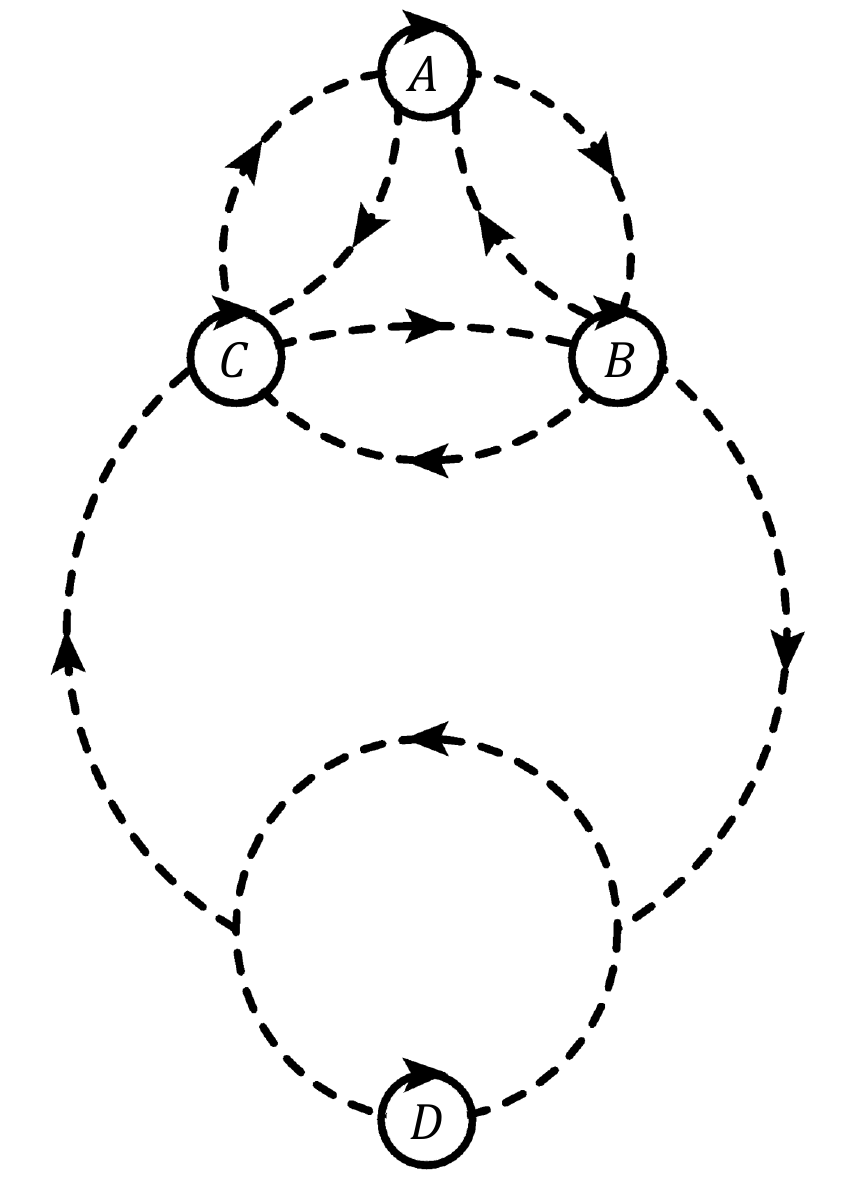}
\end{center}
\caption[Diagram]{Ring-diagram for three rotations and four blocks A,B,C and D}
\label{r2}
\end{figure} 

The exploration of higher-order elements through dashed cycles introduces a method for constructing advanced mathematical structures in a compact and intuitive manner. Here, we define tensor products symbolized by dashed connections between different elements. These connections represent the interactions between blocks e.g. 
\begin{align}\small
M_{ab}^{\pm} \equiv {\sigma_0}\otimes \begin{pmatrix}
\raisebox{-0.9em\begin{tikzpicture}[transform shape,line width=0.6pt]
\node (A) at (0,0) {A }; 
\node (B) at (0.9,0) {B}; 
\path[draw] 
(A) edge[bend right=-40,black,dashed,-> ] (B);
\end{tikzpicture}}
& \,0 \\
0 \,& \pm\raisebox{-0.9em}{\begin{tikzpicture}[transform shape,line width=0.6pt]
\node (A) at (0,0) {A}; 
\node (B) at (0.9,0) {B}; 
\path[draw] 
(B) edge[bend right=-40,black, dashed,->] (A);
\end{tikzpicture}}
\end{pmatrix},
\label{Mab}
\end{align}
with $\sigma_0=\bf{1}_2$. In detail, the second-order blocks according to 
dashed cycles become $M_{ab}$, $M_{bc}$, $M_{ca}$, $M_{ad}$, $M_{bd}$, and $M_{cd}$ expanding upon the foundational constructs.These matrices form the basis for further operations, allowing for the combination and manipulation of elements to explore a wide range of invariant possibilities within the given framework.

In Subsections~\ref{sec:dim-5},~\ref{sec:dim-6} and~\ref{sec:dim-7}, we provide various examples with more elements in the cycle of building blocks. Additionally, due to the properties of blocks, we can directly represent the building blocks in terms of the trace of rings (petals). The trace of one cycle of first-order blocks tagged by ``$\circlearrowright $'' in Ring-diagram~\ref{r2} can be translated into a single petal representation. 
For higher-order cycles depending on their $n$ connection lines $n$ petals are required e.g. $n=2$ for ``$\begin{tikzpicture}[transform shape,line width=1pt]
\node (A) at (2,2) { }; 
\node (B) at (3,2) { }; 
\path[draw] 
(A) edge[bend right=-30,black,dashed,->] (B)
(B) edge[bend right=-30,black,-latex,dashed,->] (A);
\end{tikzpicture}$''. This structure from the ring diagram can be summarized as follows:
{\allowdisplaybreaks
\ytableausetup{boxsize=1.6em}\begin{eqnarray}\small
\hspace{0cm}{\raisebox{-0.7em}{\begin{tikzpicture}
\def \radius {0.4cm}
 \node[draw, circle,style={thick},-latex] {A};
 \draw[-latex,->,thick] ({180}:\radius);
\end{tikzpicture}}}~:=&~
\raisebox{-0.2em}{\begin{tikzpicture}
 \begin{polaraxis}[grid=none, axis lines=none]
 	\addplot+[mark=none,domain=0:90,samples=50,color=black,style={ultra thick}] 
		{sin(2*x)}; 
 \end{polaraxis}
 \end{tikzpicture}}\;,\qquad \qquad
\raisebox{-0.9em}{ \begin{tikzpicture}[transform shape,line width=1pt]
\node (A) at (0,0) {A}; 
\node (B) at (1.5,0) {B}; 
\path[draw] 
(A) edge[bend right=-40,black,dashed,->] (B)
(B) edge[bend right=-40,black,-latex,dashed,->] (A);
\end{tikzpicture}} 
 :=&
\raisebox{-1em}{\begin{tikzpicture}
 \begin{polaraxis}[grid=none, axis lines=none]
 	\addplot+[mark=none,domain=0:90,samples=50,color=black,style={ultra thick}] 
		{sin(2*x)}; 
 \end{polaraxis}
 \begin{polaraxis}[grid=none, axis lines=none]
 	\addplot+[mark=none,domain=0:90,samples=50,color=black,style={ultra thick},style=dashed] 
		{-sin(2*x)}; 
 \end{polaraxis}
 \end{tikzpicture}},
 \nonumber \\ 
\raisebox{-0.8em}{\begin{tikzpicture}[transform shape,line width=1pt]
\node (A) at (1,1) {C }; 
\node (B) at (2,2) {A}; 
\node (C) at (3,1) {D}; 
\path[draw] 
(A) edge[bend right=-30,black,dashed,->] (B) 
(B) edge[bend right=-30,black,dashed,->] (C)
(C) edge[bend right=-30,black,dashed,->] (A);
\end{tikzpicture}} :=&
\raisebox{-1em}{ \begin{tikzpicture}
 \begin{polaraxis}[grid=none, axis lines=none]
\addplot+[mark=none,domain=0:60,samples=50,color=black,style={ultra thick},style=dotted] 
		{sin(-3*x)}; 
 \end{polaraxis}
 \begin{polaraxis}[grid=none, axis lines=none]
\addplot+[mark=none,domain=60:120,samples=50,color=black,style={ultra thick}] 
		{sin(-3*x)}; 
 \end{polaraxis}
 \begin{polaraxis}[grid=none, axis lines=none]
\addplot+[mark=none,domain=120:180,samples=50,color=black,style={ultra thick},style=dashdotted] 
		{sin(-3*x)}; 
 \end{polaraxis}
 \end{tikzpicture}}, \qquad \qquad
\raisebox{-2.8em}{ \begin{tikzpicture}[transform shape,line width=1pt]
\node (A) at (1,1) {G}; 
\node (B) at (2,2) {A}; 
\node (C) at (3,1) {E};
\node (D) at (2,0) {F}; 
\node (E) at ( 1, 1) {}; 
\path[draw] 
(A) edge[bend right=-30,black,dashed,->] (B) 
(B) edge[bend right=-30,black,dashed,->] (C)
(C) edge[bend right=-30,black,dashed,->] (D) 
(D) edge[bend right=-30,black,dashed,->] (E);
\end{tikzpicture}} :=&
\raisebox{-1.3em}{ \begin{tikzpicture}
 \begin{polaraxis}[grid=none, axis lines=none]
\addplot+[mark=none,domain=0:90,samples=50,color=black,style={ultra thick}] {sin(2*x)}; 
 \end{polaraxis}
 \begin{polaraxis}[grid=none, axis lines=none]
\addplot+[mark=none,domain=90:180,samples=50,color=black,style={ultra thick},style=densely dashed] {sin(2*x)}; 
 \end{polaraxis}
 \begin{polaraxis}[grid=none, axis lines=none]
\addplot+[mark=none,domain=180:270,samples=50,color=black,style={ultra thick},style=densely dashdotdotted] {sin(2*x)}; 
 \end{polaraxis}
 \begin{polaraxis}[grid=none, axis lines=none]
\addplot+[mark=none,domain=270:360,samples=50,color=black,style={ultra thick},style=loosely dashed] {sin(2*x)}; 
 \end{polaraxis}
 \end{tikzpicture}}, \, \cdots\,.
 \label{read-ring}
\end{eqnarray}}
Furthermore, the petals associated with the blocks are distinguished by different colors/patterns, reflecting their unique paths within the ring diagram. This visual differentiation is crucial for understanding the directional flow and relationships between elements in the construction of higher-order invariants. By design, the ring diagram's structure inherently prevents the redundancy of repetitive elements, ensuring that only new combinations contribute to the formation of invariants. This approach not only illustrates the interconnection among various invariants but also facilitates the identification of CP properties within higher-order constructs, even when the CP characteristics of the original blocks remain unspecified. 

The methodology allows for the expansion of building blocks to higher-orders, either through direct extension in the diagrammatic representation or via the multiplication of existing blocks. The generalized invariant form is given by:
\begin{equation}
J_{{n_a}^{x_1},\,{n_b}^{x_2},\,\cdots,\,{n_{ab}}^{x_n}} = \Tr \left({M_{a}}^{x_1} \otimes {M_b}^{x_2} \otimes \cdots \otimes {M_{ab}}^{ x_n} \right).
\label{J-general}
\end{equation}
To develop invariants through diagrams as illustrated in \eqref{read-ring}, one could initially select a random starting point within rings allocated to a basic unit and proceed to trace an entire cycle clockwise $\pm$ anti-clockwise $(\circlearrowright +\searrow\swarrow \nwarrow \cdots) \,\pm\, (\circlearrowleft+\searrow\nearrow\nwarrow \cdots)$ and repeated until identical terms produced.

Additionally, we apply the limit obtained from the Cayley-Hamilton theorem shown in Eqs. \eqref{eq:CHA3} and \eqref{eq:CHA4} over rings, enabling us to construct higher-order invariants based on the ring diagram (this can be used as an alternative approach in the construction of invariants using tensors products). 
Hence, in diagrams each cycle is allowed to repeat itself only up to three times when a single block is involved i.e.
{\ytableausetup {boxsize=1.7em}\begin{eqnarray}\small
J_{{a}^3}=\Tr(A^3) := 
\hspace{-0.7cm}{\raisebox{-1.6em}{ \begin{tikzpicture}
\hspace{0.2cm}{\raisebox{0.25em}{ \begin{polaraxis}[width=4.5cm,grid=none, axis lines=none]
 	\addplot+[mark=none,domain=0:90,samples=50,color=black,style={ultra thick}] 
		{ sin(2*x)};
 \end{polaraxis}}
\hspace{0.65cm}{\raisebox{1.8em}{ \begin{polaraxis}[grid=none, axis lines=none]
 	\addplot+[mark=none,domain=0:90,samples=50,color=black,style={ultra thick}] 
		{ sin(2*x)};
 \end{polaraxis}}}}
\hspace{-0.4cm}{\raisebox{0.9em}{ \begin{polaraxis}[width=4cm,grid=none, axis lines=none]
 	\addplot+[mark=none,domain=0:90,samples=50,color=black,style={ultra thick}] 
		{ sin(2*x)};
 \end{polaraxis}}}
 \end{tikzpicture}}}\equiv \hspace{0.1cm}{\raisebox{-0.3cm}{ \begin{tikzpicture}
\begin{polaraxis}[grid=none, axis lines=none]
 	\addplot+[mark=none,domain=0:270,samples=50,color=black,style={ultra thick}] 
		{sin(-3*x)}; 
 \end{polaraxis}
 \end{tikzpicture}}}\,.
\end{eqnarray}}

Otherwise, in the presence of other $3 \times 3$ matrices (other petals that can distinguish with a different sign or colour), the invariants can be reduced to the lower powers following the relation below
{\small\begin{align}\small
J_{{a}^2{ab}^1}&=\Tr(A^3B)={1\over 6} \Big[ \Tr(A^3)\Tr(B)
-3 \Tr(A^2)\Tr(A) \Tr(B)+6 \Tr(A) \Tr(A^2B)
\nonumber \\&
+2\Tr(A)^3 \Tr(B)+3 \Tr(A B) \big(\Tr(A^2)-\Tr(A)^2\big)\Big],
\label{eq:A3C}
\end{align}}
indicating the next invariant out of two fundamental blocks 
\ytableausetup {boxsize=1.5em}\bea
J_{{a}^1{ab}^1}=\Tr(A^2B) := 
\hspace{-0.1cm}{\raisebox{-1.1em}{ 
\begin{tikzpicture}
 \begin{polaraxis}[grid=none, axis lines=none]
 	\addplot+[mark=none,domain=0:60,samples=50,color=black,style={ultra thick},style=dashed] 
		{-sin(3*x)}; 
 \end{polaraxis}
 \begin{polaraxis}[grid=none, axis lines=none]
 	\addplot+[mark=none,domain=61:180,samples=50,color=black,style={ultra thick}] 
		{-sin(3*x)}; 
		\end{polaraxis}
 \end{tikzpicture}}}\,,
\eea
and in the same fashion for $J_{{a b}^1{b}^1}=\Tr(A B^2)$. Among the two matrices, one non-reducible CP-conserving invariant emerges from
{\small\begin{align}
J_{{a b}^2}&= \Tr ((AB)^2)+\Tr(A^2 B^2)\ =\ \Tr(A B)^2 + 
2\Tr(B) \Tr(A^2 B) 
\nonumber \\
& + {1\over 2}\big[ \Tr(A)^2 (\Tr(B)^2 - \Tr(B^2)) + \Tr(A^2) (\Tr(B^2)-\Tr(B)^2 ) \big]\nonumber \\
&-2\Tr(A) \Tr(B) \Tr(A B)+2\Tr(A) \Tr(A B^2).
\end{align}}
Equivalently, this may be as
\ytableausetup {boxsize=1.5em}\bea
J_{{ab}^2}=\Tr((A B)^2) := 
\hspace{-0.6cm}{\raisebox{-3.2em}{ 
\begin{tikzpicture}
\hspace{0.7cm}{\raisebox{1.8em}{ \begin{polaraxis}[grid=none, axis lines=none]
 	\addplot+[mark=none,domain=0:90,samples=50,color=black,style={ultra thick}] 
		{ sin(2*x)};
 \end{polaraxis}}}
 \hspace{-0.3cm}{\raisebox{1.1em}{ \begin{polaraxis}[width=3.9cm,grid=none, axis lines=none]
 	\addplot+[mark=none,domain=0:90,samples=50,color=black,style={ultra thick}] 
		{ sin(2*x)};
 \end{polaraxis}}}
 \hspace{-0.42cm}{\raisebox{0em}{ 
\hspace{0.7cm}{\raisebox{1.8em}{ \begin{polaraxis}[grid=none, axis lines=none]
 	\addplot+[mark=none,domain=0:90,samples=50,color=black,style={ultra thick},style=dashed] 
		{ -sin(2*x)};
 \end{polaraxis}}}
 \hspace{-0.3cm}{\raisebox{1.1em}{ \begin{polaraxis}[width=3.9cm,grid=none, axis lines=none]
 	\addplot+[mark=none,domain=0:90,samples=50,color=black,style={ultra thick},style=dashed] 
		{-sin(2*x)};
 \end{polaraxis}}}}}
 \end{tikzpicture}}}~~\equiv \hspace{-0.1cm}{\raisebox{-1.3em}{ \begin{tikzpicture}
\begin{polaraxis}[grid=none, axis lines=none]
\addplot+[mark=none,domain=0:270,samples=50,color=black,style={ultra thick},style=dashed] 
		{-sin(2*x)}; 
 \end{polaraxis}
 \begin{polaraxis}[grid=none, axis lines=none]
\addplot+[mark=none,domain=0:270,samples=50,color=black,style={ultra thick},style=dashed] 
		{+sin(2*x)}; 
 \end{polaraxis}
 \begin{polaraxis}[grid=none, axis lines=none]
\addplot+[mark=none,domain=0:180,samples=50,color=black,style={ultra thick}] 
		{sin(2*x)}; 
 \end{polaraxis}
 \end{tikzpicture}}}\,.
 \eea
The highest invariant consisting of two matrices is a joint invariant, where the summation between different permutations produces basic invariants or lower-order joint invariants. 

In the case of two matrices, one may have
{\small\begin{align}\small
J_{{a b}^3}:=& \Tr ((A B)^3) \,+\,{3\over 2}~ \big[\Tr(A^2 B^2 A B)+\Tr(B^2 A^2 B A) \big]
\nonumber\\ 
&= {1\over 2}\Big[\Tr(B^2 A^2 B A) + \Tr(A^2 B^2 A B)\Big]
+{1\over 12} \bigg[\big(\Tr(A^2-\Tr(A)^2)\big) \Big[9 \Tr(B^2) \Tr(A B)
 \nonumber\\ &-2 \Tr(B^3)\Tr(A) \Big]+ 3 \Tr(B)^2\,\Big[4 \Tr(A) \Tr(A^2 B)-\Tr(A B)\Big(\Tr(A)^2 + 3 \Tr(A^2)\Big)\Big] 
\nonumber\\ &
+ 12\Big[ \Tr(A B^2) \Tr(A^2 B) 
+ \Tr(A B)^3 + \Tr(A B) \Big(3 \Tr(A) \Tr(A B^2) 
- 2 \Tr(A^2 B^2)\Big)\Big] 
\nonumber\\ &
+ \Tr(B) \Big[\Tr(B^2) \Big( 9 \Tr(A) \Tr(A^2)- 7 \Tr(A)^3- 2 \Tr(A^3)\Big) + 12 \Big(\Tr(A)^2 \Tr(A B^2)
\nonumber\\ &
 - \Tr(A)~ \big(3 \Tr(A B)^2 + \Tr(A^2 B^2)\big)+ 3 \Tr(A B) \Tr(A^2 B)\Big)\Big]\,+\, \Tr(B)^3~\Big(5 \Tr(A)^3
\nonumber\\ & 
- 7 \Tr(A) \Tr(A^2)+ 2 \Tr(A^3)\Big)\bigg]
\nonumber\\ &
\equiv \; {1\over 2}\; \big[\Tr(A^2 B^2 A B)+\Tr(B^2 A^2 B A) \big]+f(\rm Basic~Invs).
\end{align} }
Here, the summation $\big[\Tr(B^2 A^2 B A) \;+\; \Tr(A^2 B^2 A B)\big]$ can be explicitly written in terms of basic invariants, as
{\small\begin{align}
\small 
& \Tr(A^2 B^2 A B)+\Tr(B^2 A^2 B A) \,=\, {1\over 3} \Big[\Tr(B)^3 \Tr(A)\Big(\Tr(A)^2 - \Tr(A^2)\Big) 
\nonumber\\ &
+ \Tr(B^3)~\Big(\Tr(A) \Tr(A^2) - \Tr(A^3)\Big) 
- 3 \Tr(B)^2 \Tr(A) \Big(\Tr(A) \Tr(A B) 
\nonumber\\ &-\Tr(A^2 B)\Big) 
+ \Tr(B)\Big(\Tr(B^2) (\Tr(A^3)-\Tr(A)^3 ) 
- 3 \Tr(A) \Tr(A^2 B^2)
\nonumber\\ &+ 3 \Tr(A)^2 \Tr(A B^2)\Big) 
+ 3~ \Big(\Tr(A B^2) \Tr(A^2 B) + \Tr(A B) \Tr(A^2 B^2)\Big)\Big]
\nonumber \\ &
= f(\rm Basic~Invs).
\label{eq:CP-evenA3B3}
\end{align}}
Nevertheless, the CP-odd combinations of the above terms become non-reducible that can be represented in terms of building blocks as
 \bea
{J^-}_{{a b}^3}:= \big[\Tr(A^2 B^2 A B)\,-\,\Tr(B^2 A^2 B A)\big]
 \equiv \hspace{0.1cm}{\raisebox{-1.5em}{\begin{tikzpicture}
 \begin{polaraxis}[grid=none, axis lines=none]
 	\addplot+[mark=none,domain=0:180,samples=50,color=black,style={ultra thick}] 
		{-sin(3*x)}; 
 \end{polaraxis}
 \begin{polaraxis}[grid=none, axis lines=none]
 	\addplot+[mark=none,domain=0:180,samples=50,color=black,style={ultra thick},style=dashed] 
		{+sin(3*x)}; 
		\end{polaraxis}
 \end{tikzpicture}}}\,,
 \label{eq:CP1}
 \nonumber 
\eea
where the presence of non-zero imaginary values in the aforementioned quantity signals the existence of a non-vanishing CP-violating phase.

Furthermore, this procedure for the construction of higher-order invariants can be extended by including additional $3\times 3$ matrices based on Eq.~\eqref{eq:CHA4}. In this context, the lowest invariants involving an additional $3\times 3$ matrix $C$ become
 \bea
{J^\pm}_{a^1{b c}^1}:= \big[\Tr(A B C)\,\pm\,\Tr(A C B)\big]
\equiv 
\hspace{0.2cm}{\raisebox{-1em}{\begin{tikzpicture}
 \begin{polaraxis}[grid=none, axis lines=none]
 	\addplot+[mark=none,domain=61:120,samples=50,color=black,style={ultra thick}] 
		{-sin(3*x)}; 
 \end{polaraxis}
 \begin{polaraxis}[grid=none, axis lines=none]
 	\addplot+[mark=none,domain=0:60,samples=50,color=black,style={ultra thick},style=dashed] 
		{-sin(3*x)}; 
		\end{polaraxis}
 \begin{polaraxis}[grid=none, axis lines=none]
 	\addplot+[mark=none,domain=120:180,samples=50,color=black,style={ultra thick},style=dotted] 
		{-sin(3*x)}; 
\end{polaraxis}
 \end{tikzpicture}}}\;,
 \label{eq:ABC}
 \nonumber 
 \eea
where $J^+$ and $J^-$ indicate CP-even and CP-odd invariants, respectively. The occurrence of non-zero values in the CP-odd invariant introduces a CP-violating phase that does not vanish. Note that these relations are not reducible and can be regarded as basic invariants. This is contrary to the case of $2\times 2$ matrices where these could be written in terms of lower invariants.

The next invariants out of three matrices $ A, B, C$ are joint invariants of order four in the form, as
\begin{align}\small
J_{{ab}^1{ac}^1}&+{J^+}_{a^2{bc}^1}=\Tr(A^2 B C)+\Tr(A^2 C B) +\Tr(A B A C)
\nonumber\\
&={1\over2}\Big[ \Tr(A)^2 \Tr(B) \Tr(C)- \Tr(BC) (\Tr(A)^2- \Tr(A^2))\Big]+ \Tr(C){\Tr(A^2B)}
\nonumber\\
& - \Tr(A) \Big[\Tr(AB) \Tr(C)\,+\,\Tr(AC)\Tr(B) \,-\,\Tr(ABC+ACB) \Big]
\nonumber\\
&+ \Tr(B) \Big[ \Tr(A^2C)\,-\,\Tr(A^2) \Tr(C)\Big]+ \Tr(AB)\Tr(BC),
\label{ABAC}
\end{align}
realizing the combination of three terms $\Tr(A^2 B C)+\Tr(A^2 C B) +\Tr(A B A C)$ can be expressed in terms of non-reducible invariants (basic invariants). On the other hand, there is a non-reducible CP-odd invariant in the following form
\begin{align}\small
{J^-}_{a^2{bc}^1}=\Tr(A^2 B C)\,-\,\Tr(A^2 C B).
\end{align}
Therefore, the two CP-odd and CP-even invariants in order four out of three different matrices in the form of petals are
\bea
{J^\pm}_{a^2{b c}^1}:= \big[\Tr(A^2 B C)\,\pm\,\Tr(A^2 C B)\big]
\equiv 
\hspace{+0.2cm}{\raisebox{-1em}{\begin{tikzpicture}
 \begin{polaraxis}[grid=none, axis lines=none]
 	\addplot+[mark=none,domain=90:270,samples=50,color=black,style={ultra thick}] 
		{-sin(2*x)}; 
 \end{polaraxis}
 \begin{polaraxis}[grid=none, axis lines=none]
 	\addplot+[mark=none,domain=0:90,samples=50,color=black,style={ultra thick},style=dashed] 
		{-sin(2*x)}; 
		\end{polaraxis}
 \begin{polaraxis}[grid=none, axis lines=none]
 	\addplot+[mark=none,domain=270:360,samples=50,color=black,style={ultra thick},style=dotted] 
		{-sin(2*x)}; 
\end{polaraxis}
 \end{tikzpicture}}}\;.
 \label{eq:CP2}
 \eea
 
Similarly, it is possible to derive higher-order combinations of three matrices and their permutations i.e. $A^2 B^2 C$, $A^3 B C$, etc. For example, at the order five invariant out of three different matrices reads
\begin{align}\small
{J^+}_{a^3{cb}^1}+&{J^+}_{a^1{ac}^1{ab}^1}=\Tr(A^3 C B)+ \Tr(A^3 B C) + 2 \Tr(A^2 B A C)+2 \Tr(A^2 C A B)
\nonumber \\
=&- 3 \Tr(A)\big[\Tr(A^2 B C)+ \Tr(A^2 C B)\big]+{1\over 6}\bigg[4\big(\Tr(A)^3-\Tr(A^3)\big) \Tr(B) \Tr(C)
\nonumber \\
+&6 \Tr(A)\Big[ (\Tr(A^2) - \Tr(A)^2) \Tr(B C) + 2 \Tr(A^2 B) \Tr(C)+2 \Tr(B) \Tr(A^2 C)
\nonumber \\
+ & 4 \Tr(A B) \Tr(A C)\Big] 
- 12 \Tr(A)^2 \Big[ \Tr(C) \Tr(A B)+ \Tr(B) \Tr(A C) \Big]+ 3 \big(5\Tr(A)^2
\nonumber \\
- &\Tr(A^2)\big) \big[\Tr(A B C)+\Tr(A C B)\big] - 12 \Big( \Tr(A^2 C) \Tr(A B)+ \Tr(A^2 B) \Tr(A C)\Big)\bigg]
 \nonumber \\
= & {\rm Basic~Invs}.
\label{eq:A3BC}
\end{align}
However, from the above combination along with the following CP-odd relation
\begin{align}\small
{J^-}_{a^3{bc}^1}=&\Tr(A^3 C B)-\Tr(A^3 B C) =2 \Tr(A) \Big[\Tr(A^2 C B)- \Tr(A^2 B C)\Big] 
\nonumber \\
&+{1\over 2} \Big[\Tr(A^2) - \Tr(A)^2\Big] \Big[\Tr(A C B) - \Tr(A B C)\Big],
\label{eq:A3BC}
\end{align}
while $\Tr(A^2 B A C)- \Tr(A^2 C A B)$ is non-reducible, so one can account for only one CP-odd invariant. From Eq.~\eqref{eq:A3BC}, the CP-odd $\Tr(A^3 C B)-\Tr(A^3 B C)$ invariants can be written in terms of lower CP-odd invariants (the basic invariants $\Tr(A C B) - \Tr(A B C)$ and the CP-odd $\Tr(A^2 C B)- \Tr(A^2 B C)$), thus this contribution is not considered as a distinct invariant. In terms of rings and building blocks, we have
\bea
{J^-}_{a^3{bc}^1}:= \big[\Tr(A^2 B A C)\,-\,\Tr(A^2 C A B)\big]
 \equiv 
\hspace{0.2cm}{\raisebox{-1.4em}{\begin{tikzpicture}
 \begin{polaraxis}[grid=none, axis lines=none]
 	\addplot+[mark=none,domain=216:324,samples=50,color=black,style={ultra thick}] 
		{-sin(5*x)}; 
 \end{polaraxis}
 \begin{polaraxis}[grid=none, axis lines=none]
 	\addplot+[mark=none,domain=0:72,samples=50,color=black,style={ultra thick},style=dotted] 
		{-sin(5*x)}; 
		\end{polaraxis}
 \begin{polaraxis}[grid=none, axis lines=none]
 	\addplot+[mark=none,domain=270:360,samples=50,color=black,style={ultra thick},style=dashed] 
		{-sin(5*x)}; 
\end{polaraxis}
 \end{tikzpicture}}}\,.
 \nonumber 
 \eea

Furthermore, with the inclusion of an additional $3\times 3$ matrix $D$, by altering $A\to A+B+C+D$ in Eq.~\eqref{eq:CHA4}, the six permutations of $ABCD$ can be summed and simplified into fundamental invariants as follows:
\begin{align}\small
&\, {J_{{ab}^1{cd}^1}} + {J_{{bc}^1{ad}^1}}+ {J_{{ca }^1{bd}^1}}\,=\, 
\nonumber \\ 
& \Tr(A B C D)+ \Tr(A B D C)+ \Tr(A C D B)+ \Tr(A C B D)
\nonumber \\ 
&+ \Tr(A D C B)+ \Tr(A D B C)\,=\,\Tr(A) \Tr(B) \Tr(C) \Tr(D) 
\nonumber \\ 
&-\Tr(A) \Tr(B) \Tr(C D)+ \Tr(B)\,\big[ \Tr(A C D) + \Tr(A D C)\big]\nonumber \\ 
&-\Tr(A) \Tr(D) \Tr(B C)+ \Tr(D)\, \big[\Tr(A B C)+ \Tr(A C B)\big]
\nonumber \\ 
&-\Tr(B) \Tr(C) \Tr(A D)+ \Tr(C) \,\big[\Tr(A B D)+ \Tr(A D B)]
\nonumber \\ 
&-\Tr(C) \Tr(D) \Tr(A B)+\Tr(A)\,\big[\Tr(B C D)+ \Tr(B D C)\big]
\nonumber \\ 
&+\Tr(A D) \Tr(B C)+\Tr(A C) \Tr(B D)+ \Tr(A B) \Tr(C D)
\nonumber \\ 
&-\Tr(B) \Tr(D) \Tr(A C)-\Tr(A)\Tr(C) \Tr(B D).
\label{ABCD}
\end{align}
Hence, five of these permutations may be regarded as CP-even joint invariants. It is important to note that the CP-odd relation is simplifiable to CP-odd invariants following the form presented in Eq.~\eqref{eq:ABC}.

This chain procedure is trivially extended up to higher-orders, where only basic invariants and the summation of the joint invariant are present.
Although, the summation of the lowest joint invariant can be expressed in terms of basic invariants.
Thus syzygies can be obtained through the chain procedure providing an explicit cancellation between terms where only basic invariants remain in them
$$
{\rm Syzygies} = {f(\rm Basic\;Invs)} +\overbrace{\sum_{oreder-n} {\rm Joint\;Invs}\times \Tr(A^n,B^n,C^n,D^n)^m}^{f(\rm Basic\;Invs)}\;.$$ 

This explicit construction of invariants can be always compared with the numbers obtained from the HS and PL.
 
Furthermore, from Eq.\eqref{eq:CP1}, Eq.\eqref{eq:CP2} and Eq.\eqref{eq:ABC}, one can realize in addition to the obvious CP-violating invariants by introducing a complex matrix ${\rm Im}\Tr(A-{\rm h.c.})$ two types of invariants can be built, namely
\begin{align}\small
	&{J^-}_{{a b}^3}={\rm Im} \big[\Tr(A^2 B^2 A B)\,-\,\Tr(B^2 A^2 B A)\big],
	\\
	&{J^-}_{a^x bc^y}={\rm Im}\Tr(A^x[B^y,\,C^y]).
	\label{eq:CPs}
\end{align}
Although from Eq. \eqref{eq:A3BC} it can be seen that Eq.~\eqref{eq:CPs} is stopped at power three as they are re-written in terms of lower order CP-odd invariants.

Ultimately, this technique provides a new and powerful systematic way for the construction of basis invariants.
In the following sections, we apply this procedure over various models including SM, SMEFT and $\nu$SMEFT with various operators up to dim-7.

\section{Invariants in SMEFT}\label{sec:SMEFT-dim6}
There have been many attempts to describe the low-energy dynamics of quantum field theories with the aid of Effective Field Theories (EFTs). This requires `integrating out' heavy degrees of freedom which cannot be generated on-shell in the relevant experimental setting. This suggests maintaining only light degrees of freedom in the theory as dynamical fields~\cite{egts,eftlectures}. In the SMEFT, the dynamical light fields are the observed SM particles and the integrated heavy degrees of freedom stand for hypothetical new particles~\cite{Buchmuller,Grzadkowski}. In this approach any new particles are assumed to be very heavy to produce directly, providing a model-independent way to constrain many beyond the SM signals~\cite{Dedes:2017zog,EFTBSM}. 

 In SMEFT, the SM Lagrangian, $\mathcal{L_\mathrm{SM}}$, is extended through a series of higher-order dim$-n$ ${\rm SU(3)}_c\otimes{\rm SU(2)}_{\sf L}\otimes{\rm U(1)}_Y$ gauge-invariant operators, ${\rm O}_i^{({\rm dim}=n)}$, describing higher-order interactions as follows
 \begin{equation}
 \mathcal{L}_{\rm SMEFT}=\mathcal{L}_{\rm SM}+\sum_{n>4}\sum_i\frac{C_i^{({\rm dim}=n)}}{\Lambda^{n-4}}{\rm O}_i^{({\rm dim}=n)},
 \end{equation}
where $\Lambda$ is some higher mass scale and $C_i$s stand for the corresponding dimensionless coupling constants, i.e.~the so-called Wilson coefficients. 

In this section, we apply our systematic construction of invariants in the framework of SMEFT and showcase two examples: the Yukawa operator up to dim-$2n$ and the four-fermion operator at dim-6. The invariants in the generic dim-$2n$ case are considered to demonstrate the full classification of invariants without the truncation required at dim-6. Note that this way of construction can be applied to all types of operators up to four involving matrices, the petals rule constructed out of Cayley--Hamilton theorem always follows the same structure.

The discussion related to the construction of invariants in SMEFT with odd (and even) powers of $\Lambda$ is delegated to the context of $\nu$SMEFT in the next section.

\subsection{SMEFT with One Yukawa Operator at Dim$-2n$}
\label{sec:SMEFT-dim6-1}
The SM Lagrangian combined with dim$-2n$ ($n>2$) operator has the following form
\beq
\mathcal{L} =\mathcal{L}_{\rm SM}+\sum_i\frac{C_i^{(2n)}}{\Lambda^{2n-4}}{\rm O}_i \,
\eeq
where $C_i^{(2n)}$ is Wilson coefficient. 

The Yukawa couplings in the SMEFT Lagrangian for the dim-4 and one dim-$2n$ operator at the order $1/\Lambda^{2n-4}$ reads
\bea
\mathcal{L} & \supset Y_u \overline Q_{{\sf L},i} \widetilde{H}\, u_{{\sf R},j}\,+\, \frac{C_{HQu}^{(2n)}}{\Lambda^{2n-4}}\abs{H}^{2n-4} \overline Q_{{\sf L},i} \widetilde{H}\, u_{{\sf R},j}\,+\,Y_d \, \overline Q_{{\sf L},i} ~H \, d_{{\sf R},j}+{\rm h.c.}
 \nonumber \\
 & = \begin{pmatrix} \overline Q_{{\sf L},i} & 0\\ 0 & \overline Q_{{\sf L},i} \end{pmatrix} \begin{pmatrix}Y_u \,+\, C^{(2n)} & 0 \\ 0 & Y_d \end{pmatrix}
\begin{pmatrix} \widetilde{H} u_{{\sf R},j}& 0\\0 & H d_{{\sf R},j}\end{pmatrix}+{\rm h.c.}\,,
\eea
with the dimensionless coupling constants $C_{HQu}^{(2n)}$ and $C^{(2n)} \equiv \frac{\abs{H}^{2n-2}}{\Lambda^{2n-4}} C_{HQu}^{(2n)}$, this construction consistently maintains a dim-6 core while extending to higher orders via the Higgs and energy scale factors.

From the above relation appears that $Y_u$ and $C^{(2n)}$ transforming both with ${\rm U}(3)_Q$ and ${\rm U}(3)_u$.
Accordingly, the ring diagram related to three transformations ${\rm U}(3)_Q,\,{{\rm U}(3)_u}$ and ${{\rm U}(3)_d}$ is given in Diagram \ref{RD-dim6-1}. In this diagram, in the top and the bottom rings ${{\rm U}(3)_u}$ and ${{\rm U}(3)_d}$ are cancelled out and only those blocks remaining at dim-$2n$ are retained i.e. $C_u = C^{(2n)}Y_u^{\dagger}$ and $C_u^\dagger = Y_u {C^{(2n)}}^{\dagger}$. These can be re-expressed in the following forms
\begin{equation}
C\,=\,{C_u+C_u^\dagger \over 2}, \quad T\,=\,{C_u-C_u^\dagger \over 2}.
\end{equation}
\begin{figure}[H]
\begin{center}
\includegraphics[width=0.23\textwidth]{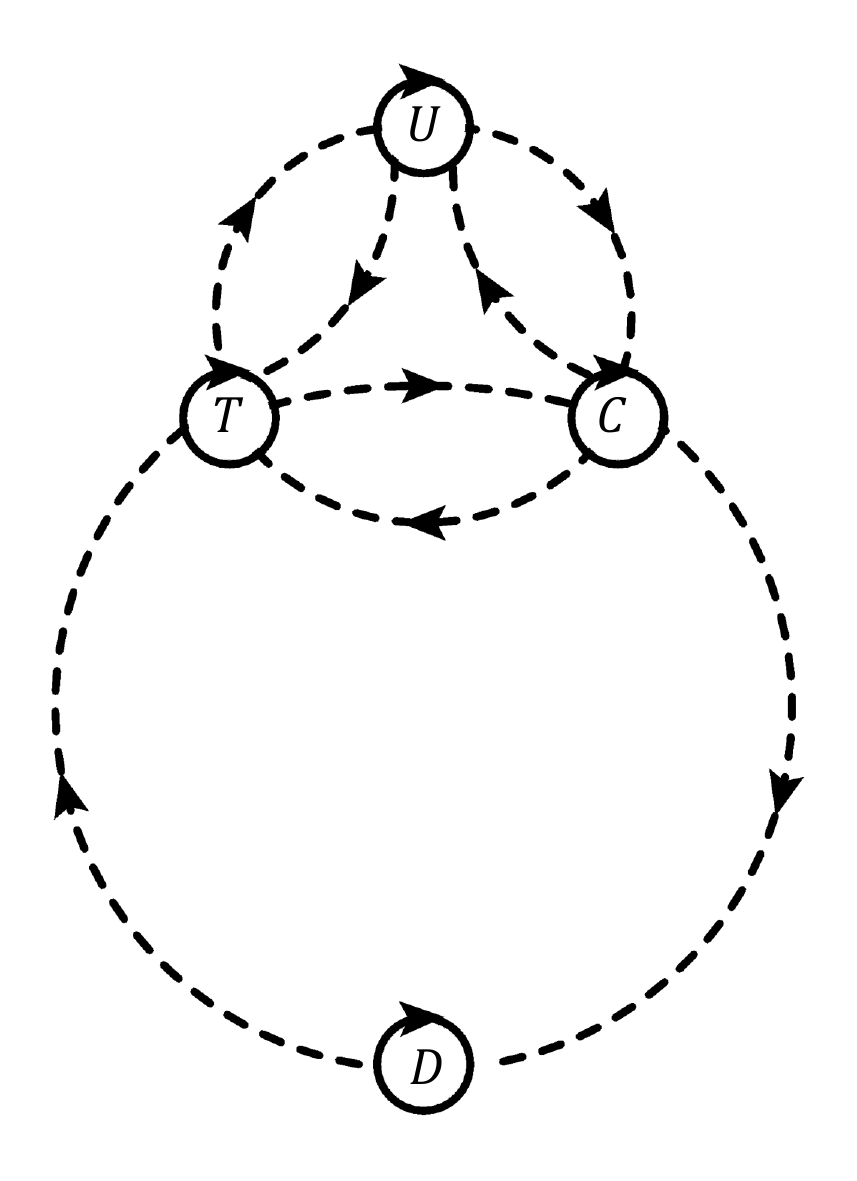}
\end{center}
\caption[Diagram]{The Ring-diagram for dim-$2n$ SMEFT with Yukawa operator.}
\label{RD-dim6-1}
\end{figure}
Therefore, as appears in Diagram~\ref{RD-dim6-1} in addition to the SM blocks $M_u,\,M_d$ and $M_{ud}$ presented in Eq.~\eqref{blocks-SM} the building blocks read
\begin{eqnarray}\small
&{M_c^+} \equiv \text{diag} \begin{pmatrix} 0 &C&0&0 \end{pmatrix}\equiv \hspace{0.2cm}{\raisebox{-0.7em}{\begin{tikzpicture}
\def \radius {0.4cm}
 \node[draw, circle,style={thick},-latex] {$C$};
 \draw[-latex,->,thick] ({180}:\radius);
\end{tikzpicture}}}\,,
\qquad\qquad
&{M_t^-} \equiv \text{diag} \begin{pmatrix} 0 & 0 & T&0 \end{pmatrix}\equiv \hspace{0.2cm}{\raisebox{-0.7em}{\begin{tikzpicture}
\def \radius {0.4cm}
 \node[draw, circle,style={thick},-latex] {$T$};
 \draw[-latex,->,thick] ({180}:\radius);
\end{tikzpicture}}}\,,
\nonumber \\
&{M_{uc}^\pm} \equiv \sigma_0 \otimes \begin{pmatrix}
 U \,C & 0 \\
0 & \pm C \, U \\
\end{pmatrix}~ \equiv \raisebox{-0.5em}{ 
\begin{tikzpicture}[transform shape,line width=0.9pt]
\node (A) at (0,0) {$U$}; 
\node (B) at (1.2,0) {$C$}; 
\path[draw] 
(A) edge[bend right=-40,black,dashed,->] (B)
(B) edge[bend right=-40,black,-latex,dashed,->] (A);
\end{tikzpicture}} \,,
\qquad
&{M^\pm_{ct}} \equiv \sigma_0 \otimes \begin{pmatrix}
C\,T& 0\\
0 &\, \mp T\,C \\
\end{pmatrix}~ \equiv \raisebox{-0.5em}{ 
\begin{tikzpicture}[transform shape,line width=0.9pt]
\node (A) at (0,0) {$C$}; 
\node (B) at (1.2,0) {$T$}; 
\path[draw] 
(A) edge[bend right=-40,black,dashed,->] (B)
(B) edge[bend right=-40,black,-latex,dashed,->] (A);
\end{tikzpicture}} \,,
\nonumber \\
&{M_{tu}^\pm} \equiv \sigma_0\otimes \begin{pmatrix}
T\, U &\, 0 \\
0& \mp U \, T \\
\end{pmatrix}~ \equiv \raisebox{-0.5em}{ 
\begin{tikzpicture}[transform shape,line width=0.9pt]
\node (A) at (0,0) {$T$}; 
\node (B) at (1.2,0) {$U$}; 
\path[draw] 
(A) edge[bend right=-40,black,dashed,->] (B)
(B) edge[bend right=-40,black,-latex,dashed,->] (A);
\end{tikzpicture}} \,,
\qquad
&{M_{cd}^\pm} \equiv {\sigma_0} \otimes \begin{pmatrix}
 C\, D&0\\
0&\pm D\,{C} \\
\end{pmatrix}~ \equiv \raisebox{-0.5em}{ 
\begin{tikzpicture}[transform shape,line width=0.9pt]
\node (A) at (0,0) {$C$}; 
\node (B) at (1.2,0) {$D$}; 
\path[draw] 
(A) edge[bend right=-40,black,dashed,->] (B)
(B) edge[bend right=-40,black,-latex,dashed,->] (A);
\end{tikzpicture}} \,,
\nonumber \\ 
&{M_{td}^\pm}\equiv {\sigma_0}\otimes \begin{pmatrix}
T\,D&0 \\
0& \mp D\, T \\
\end{pmatrix}~ \equiv \raisebox{-0.5em}{ 
\begin{tikzpicture}[transform shape,line width=0.9pt]
\node (A) at (0,0) {$T$}; 
\node (B) at (1.2,0) {$D$}; 
\path[draw] 
(A) edge[bend right=-40,black,dashed,->] (B)
(B) edge[bend right=-40,black,-latex,dashed,->] (A);
\end{tikzpicture}}\,.
\qquad &\,
 \label{blocks-dim-6}
\end{eqnarray}
Among the blocks mentioned above, $M_{ct}$ includes a dim-8 term, which is not relevant in a dim-6 operator. However, we retain this block to enable a comparison of the deduced number of invariants with those obtained from the HS method. Nonetheless, these blocks are still included in the dim-8 operator.

In this framework, in addition to SM invariants in \eqref{Inv-SM}, by the aid of fundamental blocks~\eqref{blocks-dim-6} the basic invariants can be organized in the following form
\begin{itemize}
\item Order-1:
\begin{center}\includegraphics[width=0.9\textwidth]{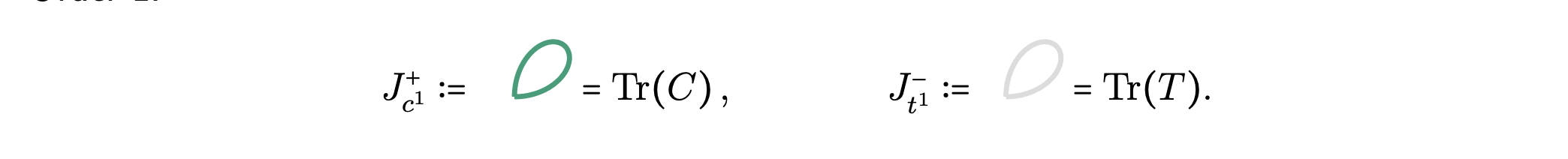}\end{center}
\item Order-2:
\begin{center}\includegraphics[width=0.9\textwidth]{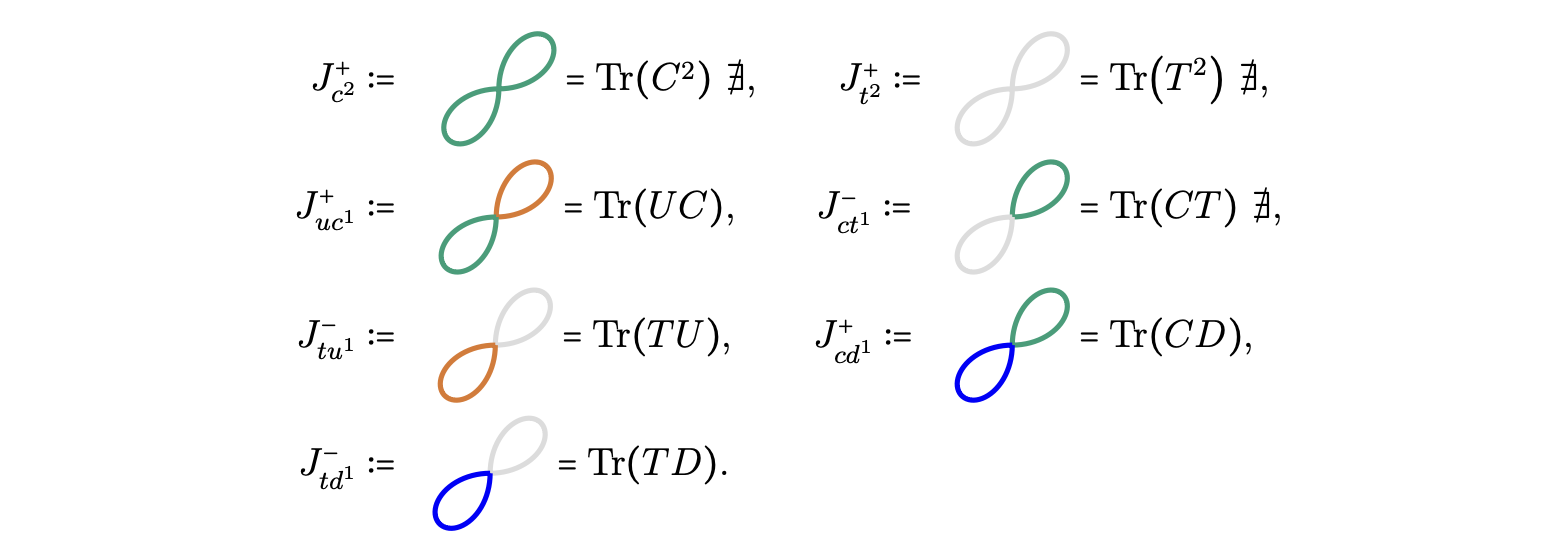}\end{center}
\item Order-3:
\begin{center}\includegraphics[width=0.9\textwidth]{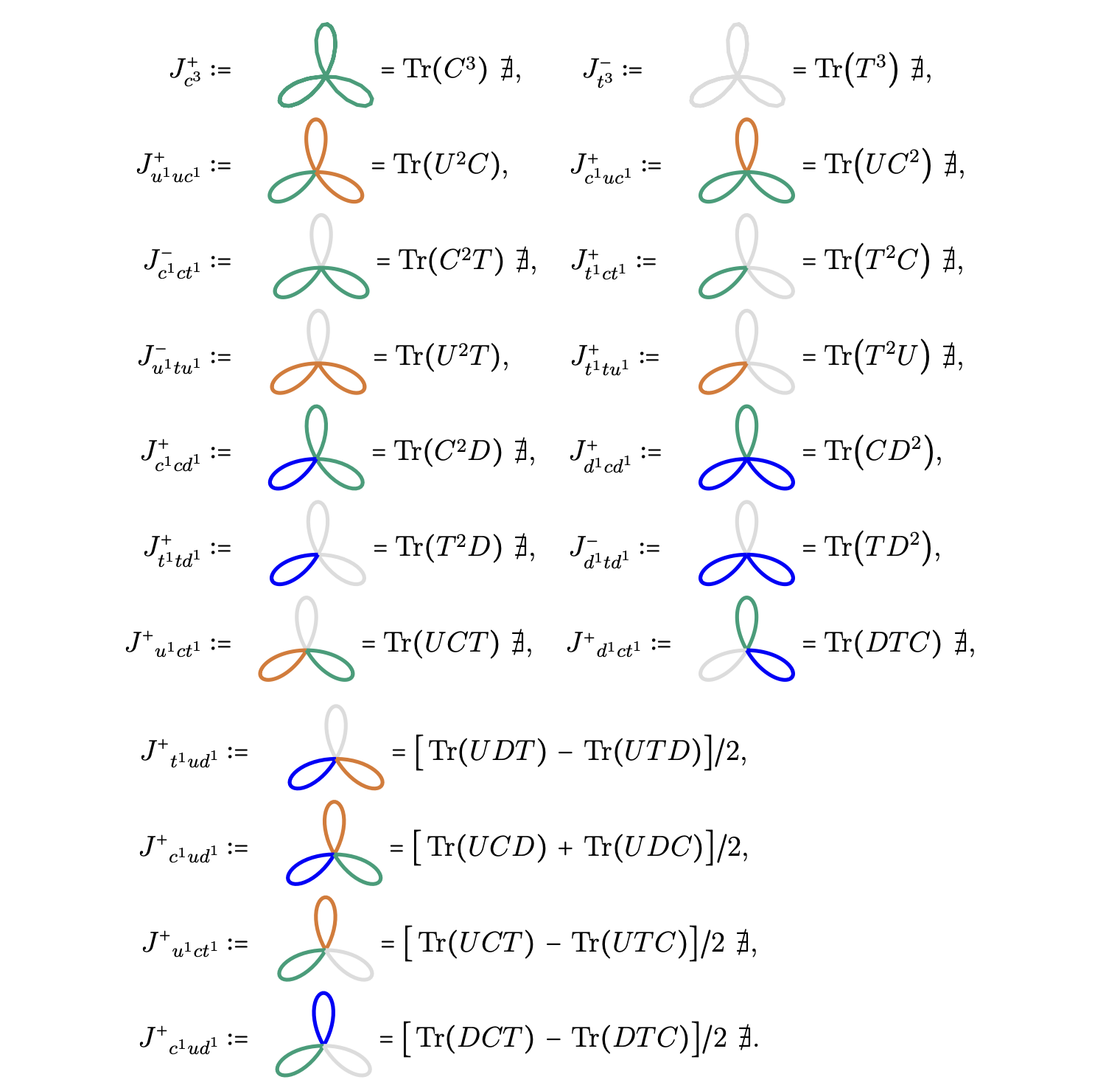}\end{center}
\item Order-4:
\begin{center}\includegraphics[width=0.9\textwidth]{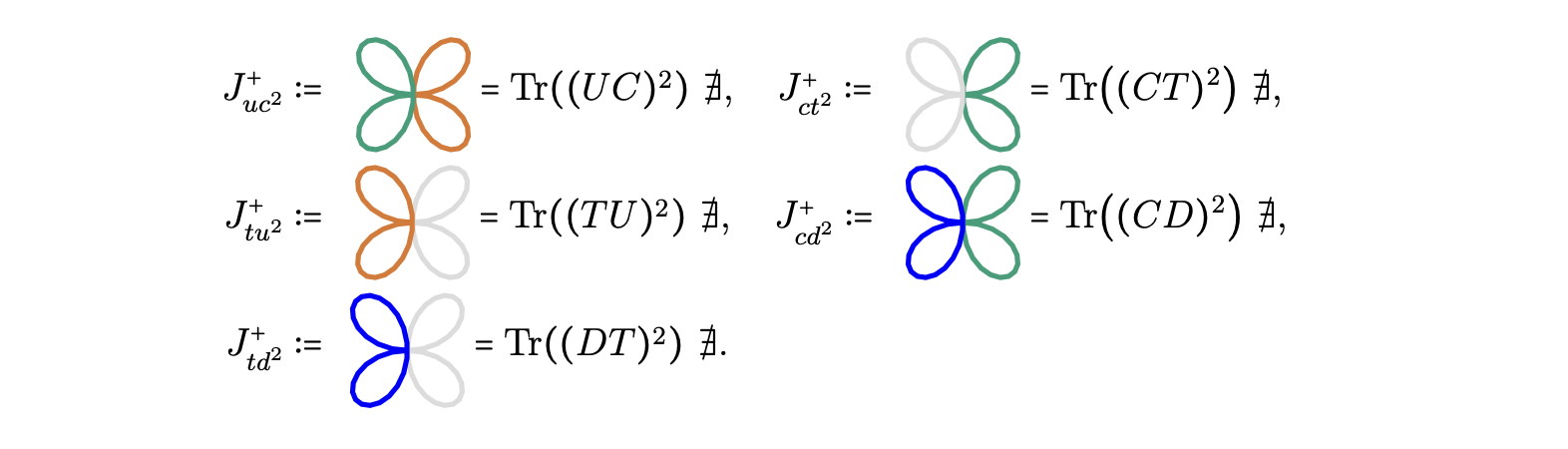}\end{center}
\end{itemize}
In the case of dim-6, in the list above, the invariants with dimensions higher than 6 are marked with ``$\nexists$'' and can be excluded. However, these invariants are considered in the dim-8 operator.

Obviously, due to the presents of the anti-hermitian block $T$, invariants containing odd powers of $T$ and $M_{ab}$ are CP-odd.

Additionally, by assigning $z_1 +z_2 +z_3 $ and $z_1^{-1}+z_2^{-1}+z_3^{-1}$ for ${\bf 3}$ and ${\bf 3^*}$, we can derive the Character functions of $U,D,C$ and $T$,
\begin{eqnarray}\small
\chi_{U,D,C,T} (z_1,z_2,z_3 )&=&\left(z_1 +z_2 +z_3 \right)\left(z_1^{-1}+z_2^{-1}+z_3^{-1}\right),
\end{eqnarray}
where each quantity $z_1,\,z_2,\,z_3$ are the coordinates on the maximum torus of $U(3)$. Thus we find
\begin{eqnarray}\small
{\rm PE}\left(z_1,z_2,z_3 \right.&;&\left.q_u,q_d,q_c, q_t \right)
={\rm PE}\left[\chi_{U} (z_1,z_2,z_3 )q_u\right] {\rm PE}\left[\chi_{D} (z_1,z_2,z_3 )q_d\right]
\nonumber\\
&& \times {\rm PE}\left[\chi_{C} (z_1,z_2,z_3 )q_c\right] {\rm PE}\left[\chi_{T} (z_1,z_2,z_3 )q_t\right]
\nonumber\\
&=&{\rm exp}\left(\sum_{k=1}^{\infty}\frac{\chi_{U} (z_1^{k},z_2^{k},z_3^{k})q_u^{k}}{k} \right){\rm exp}\left(\sum_{k=1}^{\infty}\frac{\chi_{D} (z_1^{k},z_2^{k},z_3^{k})q_d^{k}}{k} \right)\nonumber\\
&&\times {\rm exp}\left(\sum_{k=1}^{\infty}\frac{\chi_{C} (z_1^{k},z_2^{k},z_3^{k})q_c^{k}}{k} \right){\rm exp}\left(\sum_{k=1}^{\infty}\frac{\chi_{T} (z_1^{k},z_2^{k},z_3^{k})q_t^{k}}{k} \right)\nonumber\\
&=&\Big[ \left(1-q_u\right)^3\left(1-q_uz_2 z_1^{-1}\right)\left(1-q_uz_1 z_2^{-1}\right)\left(1-q_uz_3 z_1^{-1}\right)\left(1-q_uz_1 z_3^{-1}\right) \nonumber\\
&& \left(1-q_uz_2 z_3^{-1}\right)\left(1-q_u z_3 z_2^{-1}\right)
\left(1-q_d \right)^3\left(1-q_dz_2 z_1^{-1}\right)\left(1-q_dz_1 z_2^{-1}\right)
\nonumber\\
&&\left(1-q_dz_3 z_1^{-1}\right)\left(1-q_dz_1 z_3^{-1}\right)\left(1-q_dz_2 z_3^{-1}\right)\left(1-q_dz_3 z_2^{-1}\right) 
\nonumber\\
&& \left(1-q_c\right)^3\left(1-q_c z_2 z_1^{-1}\right)\left(1-q_cz_1 z_2^{-1}\right)\left(1-q_cz_3 z_1^{-1}\right)\left(1-q_cz_1 z_3^{-1}\right) \nonumber\\
&& \left(1-q_cz_2 z_3^{-1}\right)\left(1-q_cz_3 z_2^{-1}\right)
\left(1-q_t\right)^3\left(1-q_t z_2 z_1^{-1}\right)\left(1-q_tz_1 z_2^{-1}\right) \nonumber\\
&& \left(1-q_tz_3 z_1^{-1}\right)\left(1-q_tz_1 z_3^{-1}\right)\left(1-q_tz_2 z_3^{-1}\right)\left(1-q_t z_3 z_2^{-1}\right)
\Big]^{-1}\;.
\end{eqnarray}

Substituting the ${\rm PE}$ and the Haar measure of the $U(3)$ group, one can get
\begin{eqnarray}\small
{\cal H}\left(q\right)
&=& \int \left[{\rm d}\mu \right]_{{\rm U}(3)}{\rm PE}\left(z_1,z_2,z_3 ;q\right)\nonumber\\
&=& \frac{1}{6\left(2\pi {\rm i}\right)^3}\oint_{\left|z_1 \right|=1}\frac{{\rm d}z_1 }{z_1 }\oint_{\left|z_2 \right|=1}\frac{{\rm d}z_2 }{z_2 }\oint_{\left|z_3 \right|=1}\frac{{\rm d}z_3 }{z_3 }\nonumber\\
&& \times \left[-\frac{\left(z_2 -z_1 \right)^2\left(z_3 -z_1 \right)^2\left(z_3 -z_2 \right)^2}{z_1^2z_2^2z_3^2} \right]\times {\rm PE}\left(z_1,z_2,z_3,q\right)\;.
\end{eqnarray}
Applying the residue theorem to the contour integrals, one obtains
\begin{eqnarray}\small
{\cal H}\left(q\right)=\frac{{\cal N}\left(q\right)}{{\cal D}\left(q\right)}\;,
\label{Hdim6-1}
\end{eqnarray}
with
\begin{align}\small
{\cal N}=&
1 + 12 q^3 + 33 q^4 + 56 q^5 + 138 q^6 + 316 q^7 + 613 q^8 + 
 1052 q^9 + 1696 q^{10}+ 2540 q^{11}
 \nonumber\\
 & + 3566 q^{12} + 4612 q^{13} + 
 5538 q^{14} + 6116 q^{15} + 6318 q^{16} + 6116 q^{17} + 5538 q^{18}
 \nonumber\\
 & + 
 4612 q^{19} + 3566 q^{20} + 2540 q^{21} + 1696 q^{22} + 1052 q^{23} + 
 613 q^{24} + 316 q^{25} + 138 q^{26} 
 \nonumber\\
 &+ 56 q^{27} + 33 q^{28} + 12 q^{29} + q^{32},
 \end{align}
and
\begin{align}\small
{{\cal D}\left(q\right)}=(1-q)^4(1-q^2)^{10}(1-q^3)^{12}(1-q^4)^2\,.
\end{align}
The denominator of HS with $q_c,q_t, q_u, q_d$ can be written down explicitly as
\begin{align}\small
{{\cal D}\left(q_c q_t q_u q_d\right)}& \times \big\{ (1 + q_c q_d) (1 + q_c q_t) (1 + 
 q_d q_u) (1 + q_tq_u) (1 + q_d q_t) (1 + q_c q_u)\big\}
	 \nonumber\\
=&((1 - q_c) (1 - q_d) (1 - q_t) (1 - q_u) (1 - q_u^2)(1 - q_d^2)(1 - q_c^2) (1 - q_t^2)
	 \nonumber\\
	&(1 - q_c q_d) (1 - q_c q_t) (1 - 
 q_d q_u) (1 - q_tq_u) (1 - q_d q_t) (1 - q_c q_u) 
 \nonumber\\
	& (1 - q_t^3)(1 - q_u^3) (1 -q_c^3) (1 -q_d^3)(1 - q_d^2 q_u) (1 - q_c q_tq_u) (1 - q_d q_tq_u) 
 \nonumber\\
	& (1 - q_t^2 q_u) (1 - q_c^2 q_t) (1 - 
 q_c q_d q_t) (1 - q_d^2 q_t) (1 - q_c q_t^2) (1 - 
 q_d q_t^2) 
 \nonumber\\
	& (1 - q_c^2 q_d) (1 - q_c q_d^2) (1 - q_c^2 q_u) (1 - q_c q_d q_u) (1 - q_c q_u^2) (1 - q_d q_u^2) 
 \nonumber\\
	&(1 -q_tq_u^2))(1 - q_tq_u^2)) (1 - q_c^2 q_d^2) (1 - q_c^2 q_t^2) (1 - q_d^2 q_u^2) (1 - q_t^2 q_u^2) \nonumber\\
	& (1 - q_d^2 q_t^2) (1 - q_c^2 q_u^2).
\end{align}
The invariant shown in above relation matches with the list of basic invariants given at order-1,2,3,4.

 Now to step forward, the following non-reducible CP-odd invariants in addition to the SM CP-odd invariant in Eq.~\eqref{Inv-CPO-SM} up to dim-$2n$ can be constructed
 {\allowdisplaybreaks \ytableausetup{smalltableaux}
\bea
{J^-}_{t^1{ud}^1} &:=& \hspace{-0.1cm}{\raisebox{-1.1em}{\includegraphics[width=0.09\textwidth]{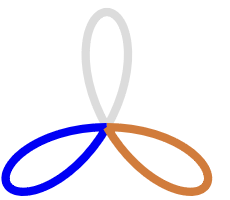}}}=\big[\Tr(U D T)+\Tr(U T D)\big]/2, 
 \label{Inv-dim6-CPO1} \\
 J^-_{u^1{cd}^1} &:=& \hspace{-0.1cm}{\raisebox{-1.1em}{\includegraphics[width=0.09\textwidth]{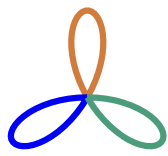}}}=\big[\Tr(U C D )-\Tr(U D C)\big]/2, 
 \label{Inv-dim6-CPO2} \\
 J^-_{u^2{td}^1} &:=& \hspace{-0.1cm}{\raisebox{-1.1em}{\includegraphics[width=0.09\textwidth]{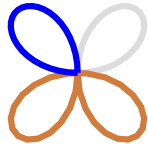}}}=\big[\Tr(U^2 D T)\,+\,\Tr(U^2 T D)\big]/2, 
\\ 
J^-_{u^2{cd}^1} &:=& \hspace{-0.1cm}{\raisebox{-1.1em}{\includegraphics[width=0.09\textwidth]{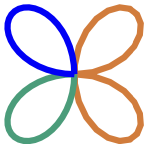}}}=\big[\Tr(U^2 C D )\,-\,\Tr(U^2 D C)\big]/2,
\\ 
J^-_{d^2{uc}^1} &:=& \hspace{-0.1cm}{\raisebox{-1.1em}{\includegraphics[width=0.09\textwidth]{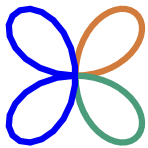}}}=\big[\Tr(D^2 U C )\,-\,\Tr(D^2 C U)\big]/2,
 \\ 
J^-_{d^2{tu}^1} &:=& \hspace{-0.1cm}{\raisebox{-1.1em}{\includegraphics[width=0.09\textwidth]{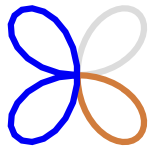}}}=\big[\Tr(D^2 T U )\,+\,\Tr(D^2 U T)\big]/2, 
 \\
{J^-}_{c^1{ud}^2} &:=& \hspace{-0.1cm}{\raisebox{-1.1em}{\includegraphics[width=0.09\textwidth]{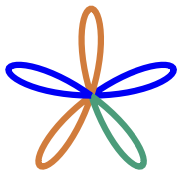}}}=\big[\Tr(U^2 D^2 C)-\Tr(U^2 C D^2 )\big]/2, 
 \\ 
{J^-}_{t^1{ud}^2} &:=&\hspace{-0.1cm}{\raisebox{-1.1em}{\includegraphics[width=0.09\textwidth]{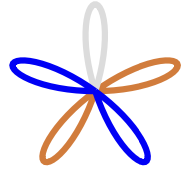}}}=\big[\Tr(T U^2 D^2 )+\Tr(T D^2 U^2 )\big]/2\,.
 \label{Inv-dim6-CPO}
\eea }
Accordingly, the total nine CP-odd invariants with the non-zero ${\rm Im}~J\neq 0$ can be accounted for the CP violation in this framework~\cite{Bonnefoy:2021tbt}.

Note that taking into account higher-orders of $C$ and $T$, the number of invariants is matched with the total number of $q^n$ invariants in PL of HS with $q=q_c=q_t= q_u= q_d$, as
\begin{eqnarray}\small
{\rm PL}[{\cal H}(q)]=4 q + 10 q^2 + 24 q^3 + 35 q^4 + 56 q^5 + 60 q^6-\mathcal{O}(q^7)\;.
\label{PL-dim-6}
\end{eqnarray}
For example, at order three there are in total 22 invariants (including four invariants from the SM lists~\eqref{Inv-SM},~eighteen from the dim-6 list at order-3 and two CP-odd invariants from Eqs.~\eqref{Inv-dim6-CPO1} and \eqref{Inv-dim6-CPO2}. Therefore, the aggregate count reaches 24 =(4+18+2) $q^3$ invariants that are matched with the given number in \eqref{PL-dim-6} at this order. 

This way of construction allows us to build invariants explicitly other than HS which only finds the numbers of invariants without any insight into their structures.

\subsection{The Dim-6 SMEFT with one Four-Fermions Operator}\label{sec:SMEFT-dim6-2}
Now, we apply our systematic construction over another dim-6 operator with one Four-Fermions operator involving quarks and leptons, with the following Lagrangian
\begin{eqnarray}\small
\mathcal{L} =\mathcal{L}_{\rm SM} +\frac{C_{ue,ijkl}}{\Lambda^{2}}(\bar u_i \gamma^\mu u_j)(\bar e_k \gamma_\mu e_l)
\end{eqnarray}
where $C^{6}\equiv {1\over \Lambda^{2}} C_{ue,ijkl}$. This example requires five rings to cancel out the chiral flavor transformations in the following form
\begin{eqnarray}\small\label{mumd}
Y_u&\to & {\rm U}(3)_Q \ Y_u \ {{\rm U}(3)_u}^{\dagger}, \nonumber \\
Y_d &\to & {\rm U}(3)_Q \ Y_d \ {{\rm U}(3)_d}^{\dagger},
\nonumber \\
Y_e &\to & {\rm U}(3)_L \ Y_d \ {{\rm U}(3)_e}^{\dagger},
\label{qtrans}
\end{eqnarray}
maintaining the Lagrangian invariant. Therefore, corresponding to ${\rm U}(3)_Q,\,{{\rm U}(3)_u},\,{{\rm U}(3)_d}$ and ${{\rm U}(3)_L},\,{{\rm U}(3)_e}$, it is more convenient to draw two separate ring diagrams with a common hermitian block $C \equiv C^{6} Y_{u} Y_{u}^\dagger Y_{e} Y_{e}^\dagger$ that participates in both diagrams. This structure is shown in Ring-diagram~\ref{RD-Dim6-2}.
\begin{figure}[H]
\begin{center}
\includegraphics[width=0.38\textwidth]{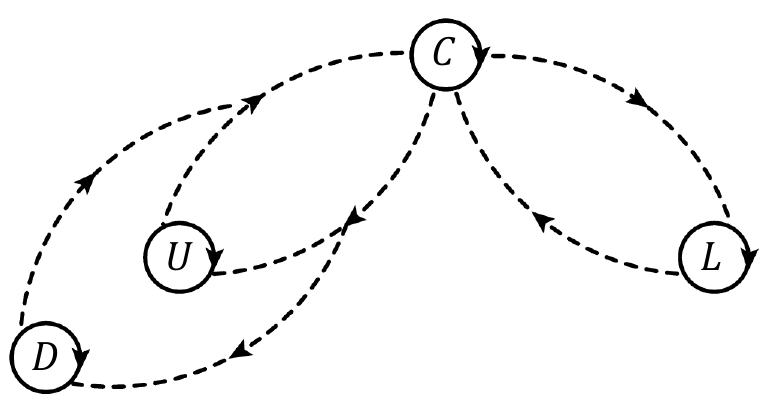}
\end{center}
\caption{The dim-6 SMEFT with Four-Fermions operator.}
\label{RD-Dim6-2}
\end{figure}

Therefore, based on Diagram \ref{RD-Dim6-2}
in addition to the orthogonal blocks $M_u,\,M_d$ and $M_{ud}$ presented in Eq.~\eqref{blocks-SM} the building blocks are
\begin{eqnarray}\small
&M_c \equiv \text{diag} \begin{pmatrix} 0 &C&0&0 \end{pmatrix}\equiv \hspace{0.2cm}{\raisebox{-0.7em}{\begin{tikzpicture}
\def \radius {0.4cm}
 \node[draw, circle,style={thick},-latex] {$C$};
 \draw[-latex,->,thick] ({180}:\radius);
\end{tikzpicture}}}\,,
\qquad\qquad
&M_{L} \equiv \text{diag} \begin{pmatrix} 0 & 0 &0&L \end{pmatrix}\equiv \hspace{0.2cm}{\raisebox{-0.7em}{\begin{tikzpicture}
\def \radius {0.4cm}
 \node[draw, circle,style={thick},-latex] {$L$};
 \draw[-latex,->,thick] ({180}:\radius);
\end{tikzpicture}}}\,,
\nonumber \\
&{M^\pm_{uc}} \equiv \sigma_0 \otimes \begin{pmatrix}
 U \,C & 0 \\
0 & \pm C \, U \\
\end{pmatrix}~ \equiv \raisebox{-0.5em}{ 
\begin{tikzpicture}[transform shape,line width=0.9pt]
\node (A) at (0,0) {$U$}; 
\node (B) at (1.2,0) {$C$}; 
\path[draw] 
(A) edge[bend right=-40,black,dashed,->] (B)
(B) edge[bend right=-40,black,-latex,dashed,->] (A);
\end{tikzpicture}}\,,
\qquad
&{M^\pm_{cl}} \equiv \sigma_0 \otimes \begin{pmatrix}
C\,L& 0\\
0 &\, \pm L\,C \\
\end{pmatrix}~ \equiv \raisebox{-0.5em}{ 
\begin{tikzpicture}[transform shape,line width=0.9pt]
\node (A) at (0,0) {$C$}; 
\node (B) at (1.2,0) {$L$}; 
\path[draw] 
(A) edge[bend right=-40,black,dashed,->] (B)
(B) edge[bend right=-40,black,-latex,dashed,->] (A);
\end{tikzpicture}} \,,
\nonumber \\
&{M^\pm_{cd}} \equiv {\sigma_0} \otimes \begin{pmatrix}
 C\, D&0\\
0&\pm D\,{C} \\
\end{pmatrix}~ \equiv \raisebox{-0.5em}{ 
\begin{tikzpicture}[transform shape,line width=0.9pt]
\node (A) at (0,0) {$C$}; 
\node (B) at (1.2,0) {$D$}; 
\path[draw] 
(A) edge[bend right=-40,black,dashed,->] (B)
(B) edge[bend right=-40,black,-latex,dashed,->] (A);
\end{tikzpicture}} \,.
\qquad &\,
 \label{blocks-dim-6-2}
\end{eqnarray}
Therefore, in addition to invariants in \eqref{Inv-SM}, using the building blocks~\eqref{blocks-dim-6-2} the basic invariants are:
\begin{itemize}
\item Order-1:
\begin{center}\includegraphics[width=0.9\textwidth]{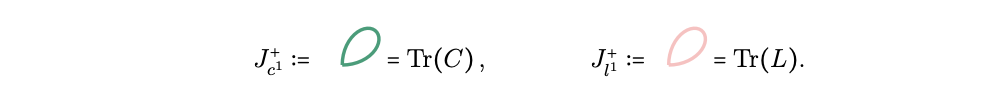}\end{center}
\item Order-2:
\begin{center}\includegraphics[width=0.9\textwidth]{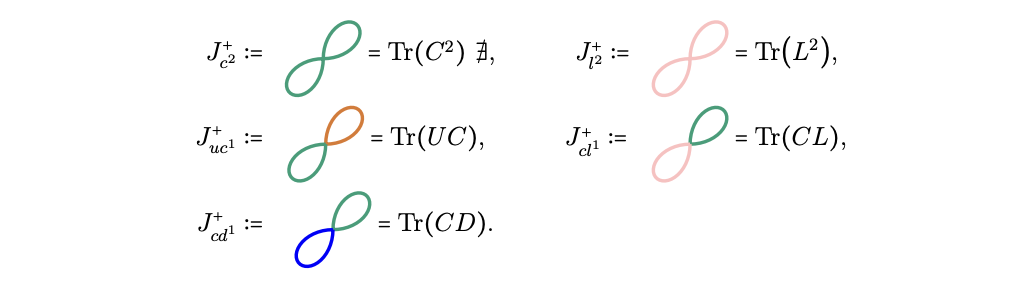}\end{center}
 \label{order-2-dim6-2}
\item Order-3:
\begin{center}\includegraphics[width=0.9\textwidth]{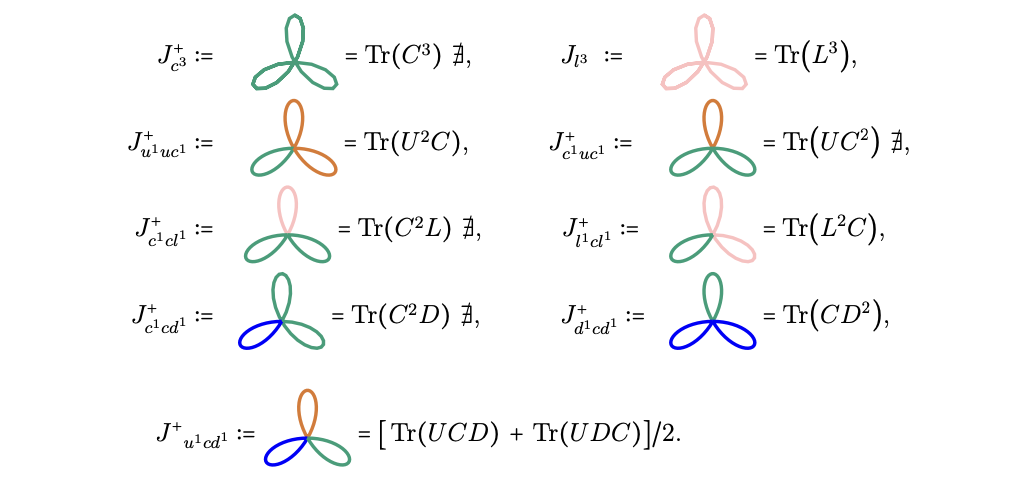} \label{order-3-dim6-2}\end{center}
\item Order-4:
\begin{center}\includegraphics[width=0.9\textwidth]{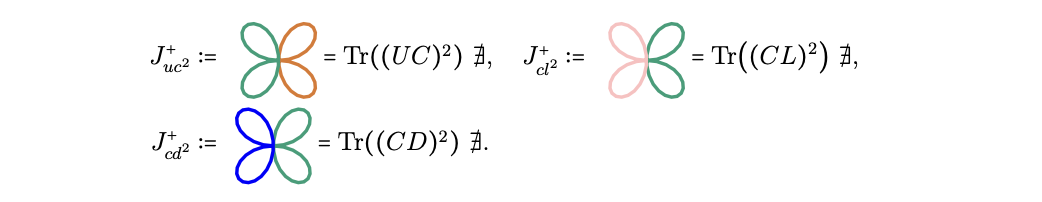}\end{center}
\label{order-4-dim6-2}
\end{itemize}
In a similar fashion, higher-orders of $C$ are distinct with ``$\nexists$'' and so they can be removed in the case of dim-6 operator. Note that some higher-order invariants can not form distinct invariants e.g.
\begin{center}\includegraphics[width=0.9\textwidth]{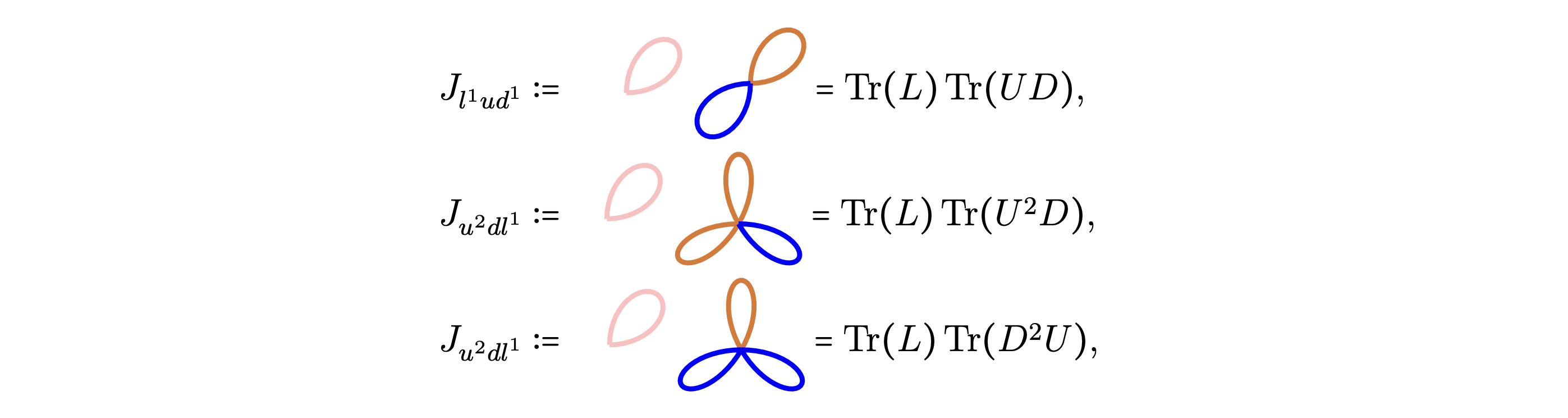}\end{center}
and similarly for higher-order invariants of this form. Additionally, up to dim-6 the following non-reducible CP-odd invariants can be constructed:
 {\allowdisplaybreaks \ytableausetup{smalltableaux}
\bea
{J^-}_{u^1{cd}^1} &:=& \hspace{-0.1cm}{\raisebox{-1.1em}{\includegraphics[width=0.09\textwidth]{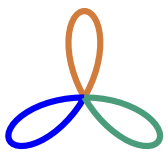}}}=\big[\Tr(U C D )-\Tr(U D C)\big]/2, 
 \label{Inv-dim6-2-CPO1} \\
 {J^-}_{u^1{cl}^1} &:=&  \hspace{-0.1cm}{\raisebox{-1.1em}{\includegraphics[width=0.09\textwidth]{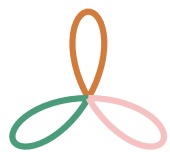}}}=\big[\Tr(U C L )-\Tr(U L C)\big]/2, 
 \label{Inv-dim6-2-CPO2} \\
 {J^-}_{d^1{cl}^1} &:=&  \hspace{-0.1cm}{\raisebox{-1.1em}{\includegraphics[width=0.09\textwidth]{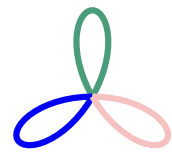}}}=\big[\Tr(D C L )-\Tr(D L C)\big]/2, 
 \label{Inv-dim6-2-CPO3} \\
 {J^-}_{l^2{cu}^1} &:=& \hspace{-0.1cm}{\raisebox{-1.1em}{\includegraphics[width=0.09\textwidth]{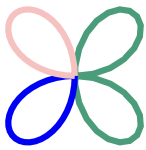}}}=\big[\Tr(L^2 U C )-\Tr(L^2 C U)\big]/2,
 \label{Inv-dim6-2-CPO5} \\
 {J^-}_{l^2{cd}^1} &:=& \hspace{-0.1cm}{\raisebox{-1.1em}{\includegraphics[width=0.09\textwidth]{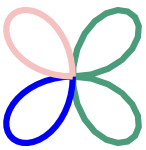}}}=\big[\Tr(L^2 C D )-\Tr(L^2 D C)\big]/2,
 \label{Inv-dim6-2-CPO3} \\
{J^-}_{u^2{cd}^1} &:=&  \hspace{-0.1cm}{\raisebox{-1.1em}{\includegraphics[width=0.09\textwidth]{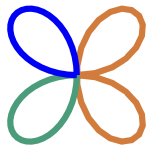}}}= \big[\Tr(U^2 C D )\,-\,\Tr(U^2 D C)\big]/2, 
\\ 
{J^-}_{d^2{uc}^1} &:=&  \hspace{-0.1cm}{\raisebox{-1.1em}{\includegraphics[width=0.09\textwidth]{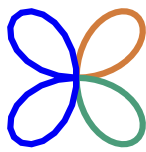}}}=\big[\Tr(D^2 U C )\,-\,\Tr(D^2 C U)\big]/2, 
\\ 
{J^-}_{c^1{ud}^2} &:=& \ \hspace{-0.1cm}{\raisebox{-1.1em}{\includegraphics[width=0.09\textwidth]{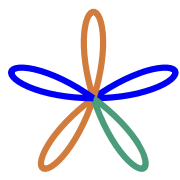}}}=\big[\Tr(U^2 D^2 C)\,-\,\Tr(U^2 C D^2 )\big]/2,
 \\
 {J^-}_{d^1{ud}^1{cd}^1} &:=&  \hspace{-0.1cm}{\raisebox{-1.1em}{\includegraphics[width=0.09\textwidth]{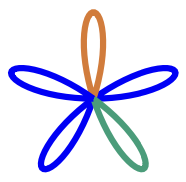}}}=\big[\Tr(D^2 U D C )\,-\,\Tr( D^2 C D U)\big]/2. 
 \label{Inv-dim6-CPO}
\eea }
This procedure can be straightforwardly applied in the other forms of four-fermion interactions as the petals rule always follows the same structure.
\section{Invariants in the Framework of $\nu$SMEFT} \label{sec:SSMEFT}
In this section, we consider the framework of $\nu$SMEFT that includes three right-handed neutrinos $N_{\sf R}$, which are singlets under the SM gauge group.

The Lagrangian of the type-I seesaw ~\cite{Minkowski:1977sc,Yanagida:1979as} that is the extension of SM with the Dirac neutrino Yukawa coupling $Y_\nu$ and the Majorana mass $M_{\sf R} $ of right-hand neutrinos may be written as
\begin{eqnarray}\small
	\label{eq:full lagrangian}
	{\cal L}_{\rm SS} ={\cal L}_{\rm SM} +\overline N_{\sf R} {\rm i}\slashed{\partial}N_{\sf R} -\left[Y_\nu \overline L_{\sf L} \widetilde{H} N_{\sf R} +{1\over 2}\overline N_{\sf R}^{\sf C} M_{\sf R} N_{\sf R} +{\rm h.c.}\right]\;,
\end{eqnarray}
where ${L }_{\sf L} \equiv (\nu_{\sf L},\; l_{\sf L})^{\sf T}$ and $N_{\sf R}^{\sf
 C}\;\equiv\; \mathcal{C} \overline N_{\sf R}^{\sf T}$ with $\mathcal{C}\equiv {\rm i}\gamma^2\gamma^0$ being the charge-conjugation operator. In the type-I seesaw model and its low-energy EFT with up to the dim-6 Weinberg operator, the basic flavor invariants have been investigated based on HS and PL~\cite{Yu:2021cco,Yu:2022ttm}.
 
Here, we explore the flavor invariants in $\nu$SMEFT up to dim-seven using our new systematic construction of invariants and compare them with HS and PL. In detail, we showcase this technique for the $\nu$SMEFT including three distinct combination of operators i.e. only one dim-5, two dim-5 and dim-6 operators and finally two dim-5 and dim-7 operators at the tree-level matching.

\subsection{One Dim-5 Operator}
\label{sec:dim-5}
In the $\nu$SMEFT, the leading flavor symmetry-breaking Lagrangian to the order of dim-5 can be given by
\begin{eqnarray}\small
	\label{eq:EFT-lag5}
\mathcal{L}= {\mathcal L}_{\rm SM} -\left(\frac{C_5 }{2\Lambda} \overline L_{\sf L }\widetilde{H}\widetilde{H}^{\sf T} L_{\sf L}^{\sf C} \,+\, {\rm h.c.}\right),
\end{eqnarray}
with the following Wilson coefficient
\begin{eqnarray}\small
	\label{eq:Neutino-wilson}
	C_5 \;=\; - Y_\nu {\Lambda\over M_{\sf R}} Y_\nu^{\sf T}\;.
\end{eqnarray}
The general unitary transformation in the flavor space in the leptonic sector is given as
\begin{eqnarray}\small
	\label{eq:field}
 L_{\sf L} &\to L_{\sf L}^{\prime} \;=\; {\rm U}(3)_{L} \; L_{\sf L}, \nonumber \\
	l_{\sf R} &\to l_{\sf R}^{\prime} \;=\; {\rm U}(3)_{l} \; l_{\sf R}, \nonumber \\
 N_{\sf R} &\to N_{\sf R}^{\prime} \,=\; {\rm U}(3)_{\sf R} \; N_{\sf R}.
\end{eqnarray}
Hence, these induce the transformation of Yukawa matrices, the Majorana masses and the Wilson coefficient as shown in Table~\ref{tab:tab2}.
\begin{table}[H]
\begin{center}
	\begin{tabular}{l|c|c|c}
		& \; ${\rm U}(3)_{L}$ \; & \; ${\rm U}(3)_e$ \; &\; ${\rm U}(3)_{\sf R}$ \; \\\hline
		$Y_e$ &$\mathbf{3}$ & $\mathbf{\overline{3}}$ & $\mathbf{1}$ \\[0.1cm]
		$Y_\nu$ &$\mathbf{3}$ & $ \mathbf{1}$ & $\mathbf{\overline{3}}$ \\[0.1cm]
		$ M_{\sf R}$ &$\mathbf{1}$ & $\mathbf{1}$ & $\mathbf{3}^*\times\mathbf{\overline{3}}$ \\[0.1cm]
		$C_5$ &$\mathbf{3}\times\mathbf{3}^{\sf T}$ & $\mathbf{1}$ & $\mathbf{1}$ \\[0.1cm]
	\end{tabular}
	\end{center}
\caption{\it Flavor transformation under ${\rm U}(3)_{L}$, ${\rm U}(3)_e$ and ${\rm U}(3)_{\sf R}$ of the Yukawa matrices $Y_{e,\nu}$, the Majorana masses $M_{\sf R}$ and the Wilson coefficient $C_5$.}
	\label{tab:tab2}
\end{table}
Henceforth, the Lagrangian in Eq.~\eqref{eq:EFT-lag5} remains unchanged under the above transformation. 
Therefore, the building blocks for the construction of flavor invariants are $\left\{L \equiv Y_e Y_e^\dagger, C_5 \right\}$ that taking place in Ring-diagram \ref{RD}.
\begin{figure}[H]
\begin{center}
\includegraphics[width=0.28\textwidth]{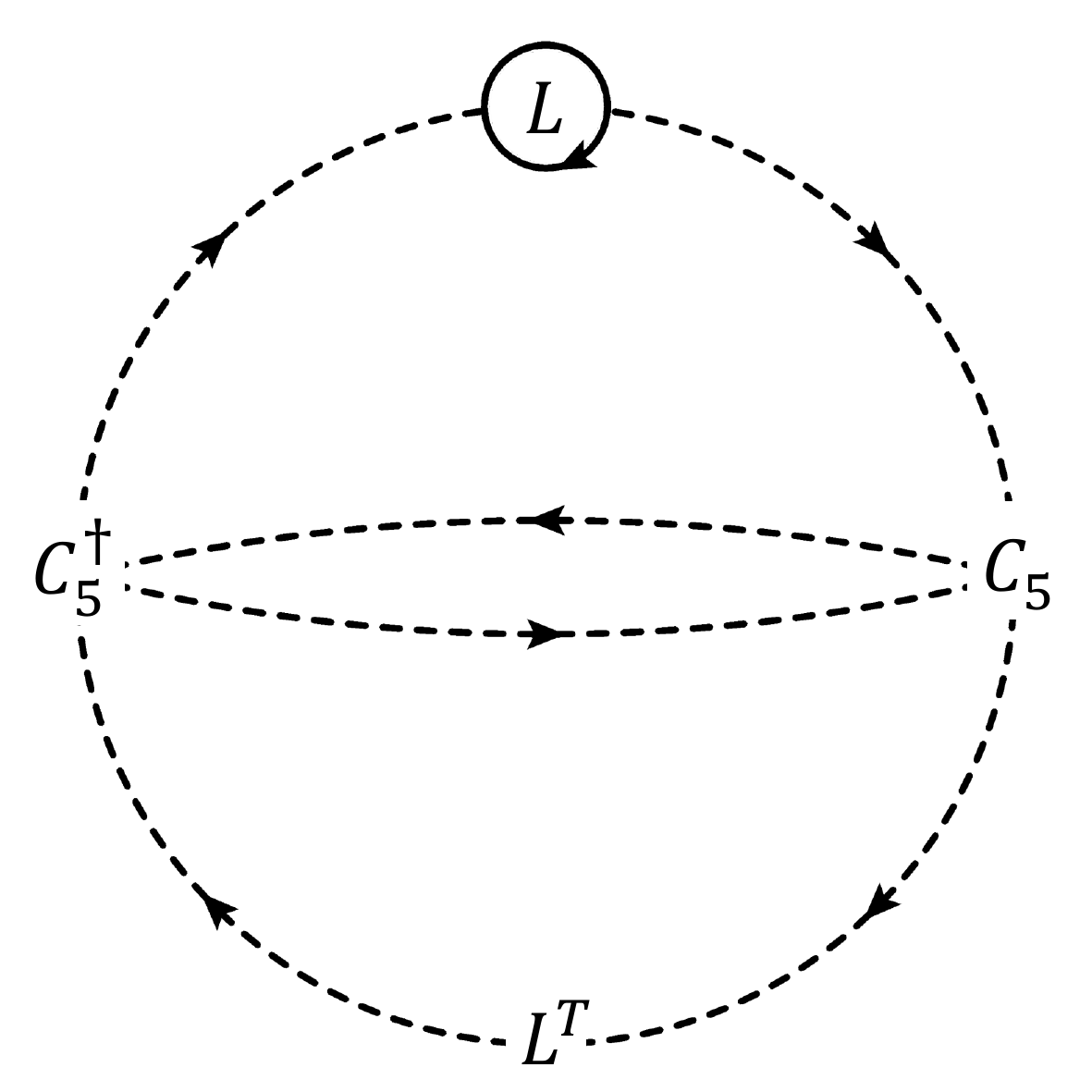}
\end{center}
\caption{The Ring-diagram for $\nu$SMEFT with one Wilson coefficient $C_5$.}
\label{RD}
\end{figure} 
Accordingly, the building blocks become
\begin{eqnarray}\small
 M_L &\equiv & \text{diag} \begin{pmatrix} L & 0 & 0 & 0 \end{pmatrix}\equiv \hspace{0.2cm}{\raisebox{-0.7em}{\begin{tikzpicture}
\def \radius {0.4cm}
 \node[draw, circle,style={thick},-latex] {$L$};
 \draw[-latex,->,thick] ({180}:\radius);
\end{tikzpicture}}}\,,
\nonumber \\
 {M_{c}} &\equiv & \sigma_0 \otimes \begin{pmatrix}
C_5\,C_5^{\dagger} & 0 \\
0 & \, C_5^{\dagger}\,C_5 \\
\end{pmatrix}~ \equiv \raisebox{-1em}{ 
\begin{tikzpicture}[transform shape,line width=0.9pt]
\node (A) at (0,0) {$C_5$}; 
\node (B) at (1.2,0) { $C_5^{\dagger}$ }; 
\path[draw] 
(A) edge[bend right=-40,black,dashed,->] (B)
(B) edge[bend right=-40,black,-latex,dashed,->] (A);
\end{tikzpicture}}\equiv \hspace{0.2cm}{\raisebox{-0.7em}{\begin{tikzpicture}
\def \radius {0.4cm}
 \node[draw, circle,style={thick},-latex] {$C$};
 \draw[-latex,->,thick] ({180}:\radius);
\end{tikzpicture}}}\,,
\nonumber \\
M_{lk_m}^\pm &\equiv & \sigma_0 \otimes \begin{pmatrix}
L \,C_5\,{L^{\sf T}}^m\,C_5^{\dagger}& 0\\
0 &\, \pm C_5 \,{L^{\sf T}}^m\,C_5^{\dagger}\,L \\
\end{pmatrix}~ \equiv \raisebox{-0.1em}{ 
\raisebox{-2.8em}{ \begin{tikzpicture}[transform shape,line width=1pt]
\node (A) at (0.9,1) {$C_5^{\dagger}$}; 
\node (B) at (2,2) {$L$}; 
\node (C) at (3,1) {$C_5$};
\node (D) at (2,0) {${L^{\sf T}}^m$}; 
\node (E) at (1,0.8) {}; 
\path[draw] 
(A) edge[bend right=-30,black,dashed,->] (B) 
(B) edge[bend right=-30,black,dashed,->] (C)
(C) edge[bend right=-30,black,dashed,->] (D) 
(D) edge[bend right=-30,black,dashed,->] (E);
\end{tikzpicture}}}\,
\nonumber \\
&\equiv & \sigma_0 \otimes \begin{pmatrix}
L \,K_m& 0\\
0 &\, \pm K_m\,L \\
\end{pmatrix}~ \equiv \raisebox{-1em}{ 
\begin{tikzpicture}[transform shape,line width=0.9pt]
\node (A) at (0,0) {$L$}; 
\node (B) at (1.2,0) { $K_m$ }; 
\path[draw] 
(A) edge[bend right=-40,black,dashed,->] (B)
(B) edge[bend right=-40,black,-latex,dashed,->] (A);
\end{tikzpicture}} \,,
 \label{blocks-dim-6-2}
\end{eqnarray}
with $K_m=C_5 \,{L^{\sf T}}^m\,C_5^{\dagger}$ and $m=1,2$. Therefore, blocks $C$, $K_1$, and $K_2$ correspond to orders two, three, and four, respectively. However, $K_1$ and $K_2$ are derived from elements that also contribute to lower-order blocks. Consequently, blocks denoted by ${K_m^n}$ (where $n>1$) do not represent standalone invariants but, through their integration with other blocks, yield unique invariants $M_{lk_m}^\pm$. The roles of $C$, $K_1$, and $K_2$ in the petal framework are indicated by green, gray, and purple colors, respectively. These colors guide the assembly of higher-order invariants by marking the corresponding blocks.

Basic invariants can thus be derived in the subsequent manner:
\begin{itemize}
\item Order-1:
\begin{center}\includegraphics[width=0.9\textwidth]{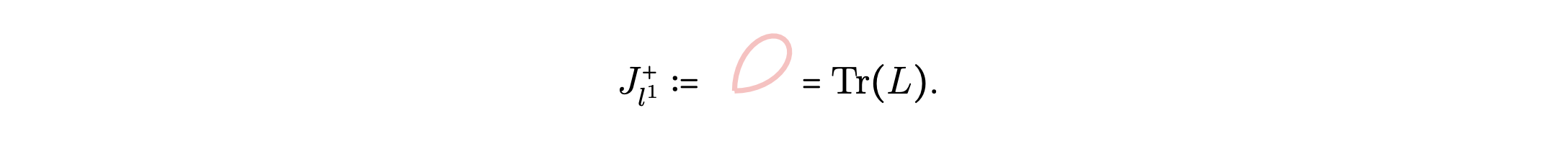}\end{center}
\item Order-2:
\begin{center}\includegraphics[width=0.9\textwidth]{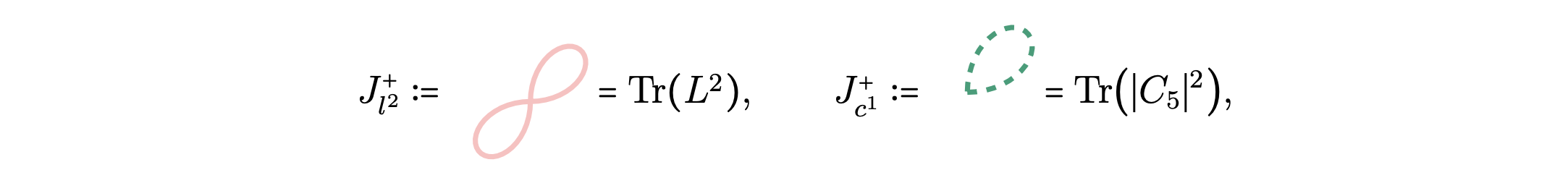}\end{center}
where $\hspace{0.0cm}{\raisebox{0.2em}{ \begin{tikzpicture}\begin{polaraxis}[width=3.6cm,grid=none, axis lines=none]	\addplot+[mark=none,domain=0:90, samples=30,color=darkgreen,style=dashed,style={ultra thick}] {sin(2*x)};\end{polaraxis} \end{tikzpicture}}}$ stands for order two blocks. 
 
\item Order-3:
\begin{center}\includegraphics[width=0.9\textwidth]{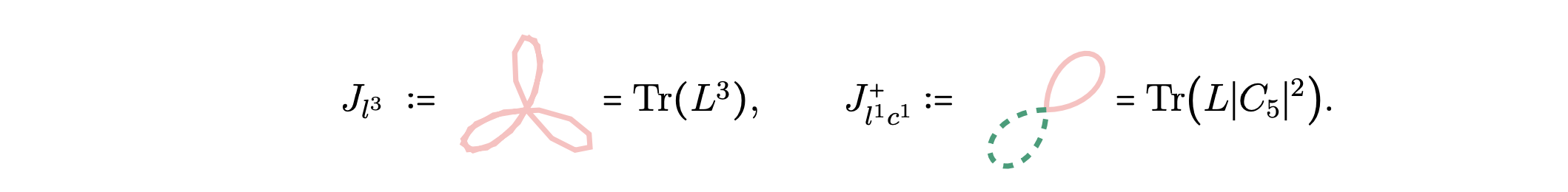}\end{center}
\item Order-4:
\begin{center}\includegraphics[width=0.9\textwidth]{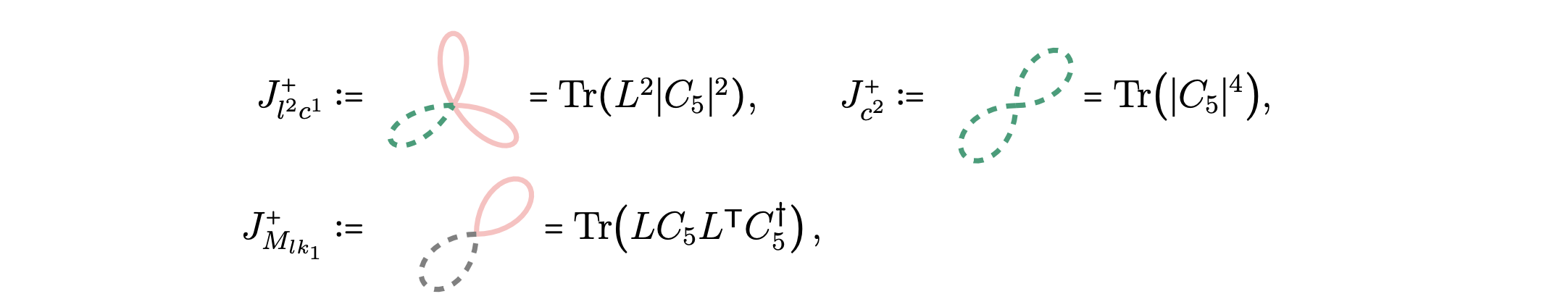}\end{center}
where the dashed gray petal presenting a order three ring $K_1 \equiv C_5 {L^{\sf T}} C_5^\dagger$.
\item Order-5:
\begin{center}\includegraphics[width=0.9\textwidth]{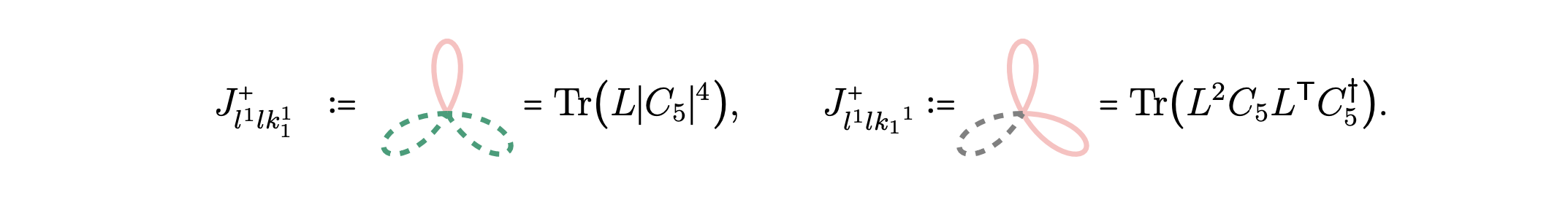}\end{center}
\item Order-6:
\begin{center}\includegraphics[width=0.9\textwidth]{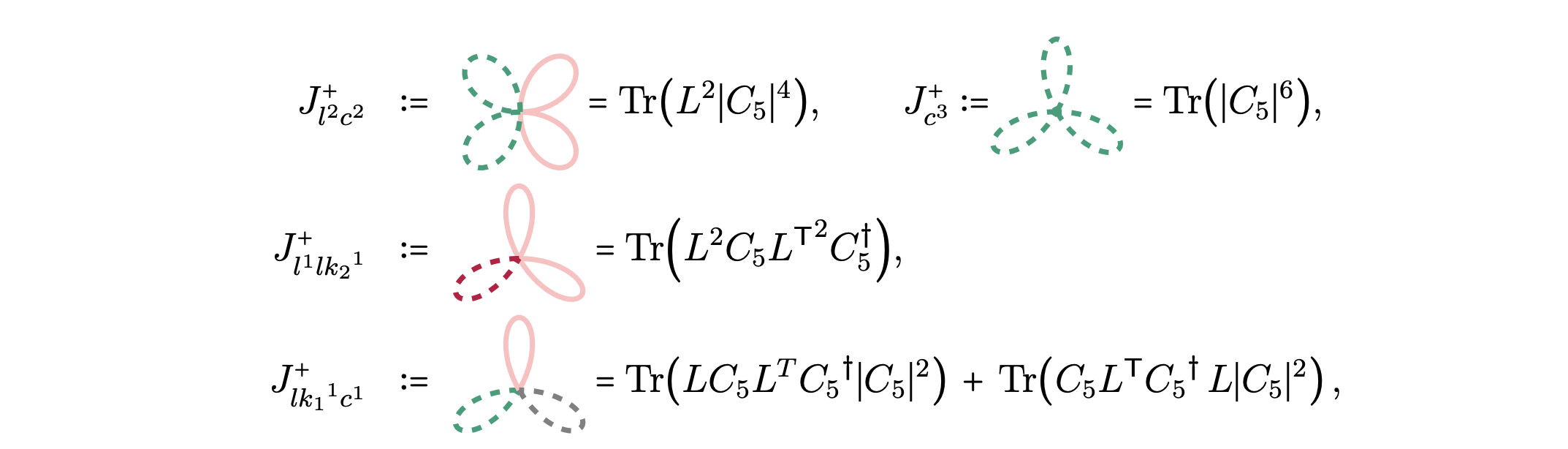}\end{center}
where the dashed gray petal presenting a order four $K_2 \equiv C_5 {L^{\sf T}}^2 C_5^\dagger$.
\end{itemize}
Additionally, the following non-reducible CP-odd invariants can be constructed
 {\allowdisplaybreaks \ytableausetup{smalltableaux}
\bea
 J^-_{{lk_1}^1c^1} &:=&  \hspace{-0.1cm}{\raisebox{-1.1em}{\includegraphics[width=0.09\textwidth]{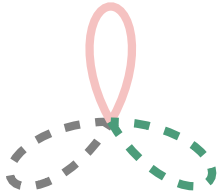}}}= \Tr({L C_5 {L}^T{C_5}^\dagger |C_5|^2})\,-\,\Tr({ C_5 {L}^{\sf T}{C_5}^\dagger\, L |C_5|^2})\,,
 \\
 J^-_{l^1 {lk_1}^1c^1} &:=&  \hspace{-0.1cm}{\raisebox{-1.1em}{\includegraphics[width=0.09\textwidth]{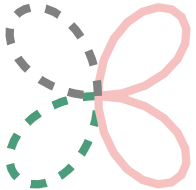}}}= \Tr({L^2 C_5 L^{\sf T} {C_5}^\dagger |C_5|^2})\,-\,\Tr({ C_5 L^{\sf T}{C_5}^\dagger\, L^2 |C_5|^2})\,,
\label{sec:d-5o7} \\ 
J^-_{ {lk_1}^1c^2} &:=& \hspace{-0.1cm}{\raisebox{-1.1em}{\includegraphics[width=0.09\textwidth]{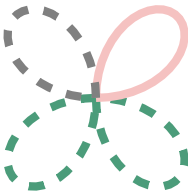}}}=\Tr({L C_5 {L}^{\sf T}{C_5}^\dagger |C_5|^4 }) \;-\; \Tr({ C_5 {L}^{\sf T}{C_5}^\dagger\, L |C_5|^4} )\,,
\label{sec:d-5o8} \\ 
 J^-_{l^1 {lk_2}^1c^1}&:=&  \hspace{-0.1cm}{\raisebox{-1.1em}{\includegraphics[width=0.09\textwidth]{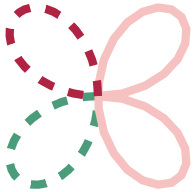}}}= \Tr({C_5 {L^{\sf T}}^2{C_5}^\dagger |C_5|^2 L^2 }) \;-\; \Tr({ C_5 {L^{\sf T}}^2 {C_5}^\dagger L^2 |C_5|^2 })\,,
\label{sec:d-5o8-2} \\ 
 J^-_{l^1 {lk_1}^1c^1}&:=&  \hspace{-0.1cm}{\raisebox{-1.1em}{\includegraphics[width=0.09\textwidth]{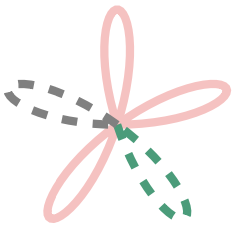}}}= \Tr({L^2 C_5 {L^{\sf T}} {C_5}^\dagger L |C_5|^2 })
 \;-\; \Tr({L^2 |C_5|^2 L C_5 {L^{\sf T}} {C_5}^\dagger })\,.
 \label{sec:d-5o8-3}
\eea}
From the aforementioned invariant, CP violation occurs \textit{iff} ${\rm Im}J^- \neq 0$. The first CP-odd invariant emerges at order six. Moreover, the initial joint invariant appears at order seven, denoted as $J^+$, which correlates with Eq. \eqref{sec:d-5o7} and contributes two invariants at this level. Notably, at order eight, there are two joint invariants linked to Eqs. \eqref{sec:d-5o8} and \eqref{sec:d-5o8-2}, alongside $J^+$, cumulatively resulting in five invariants at this tier. 

Given the assignments $z_1+z_2+z_3$ for ${\bf 3}$ and $z_1^{-1}+z_2^{-1}+z_3^{-1}$ for ${\bf 3}^*$, the character functions of the primary elements are delineated as:
\begin{eqnarray}
\small
\chi_l (z_1,z_2,z_3) &=& (z_1 + z_2 + z_3)(z_1^{-1} + z_2^{-1} + z_3^{-1}),\nonumber\\
\chi_5 (z_1,z_2,z_3) &=& z_1^2 + z_2^2 + z_3^2 + z_1 z_2 + z_1 z_3 + z_2 z_3 \nonumber\\
&& + z_1^{-2} + z_2^{-2} + z_3^{-2} + z_1^{-1}z_2^{-1} + z_1^{-1}z_3^{-1} + z_2^{-1}z_3^{-1},
\end{eqnarray}
where $z_i $ (for $i=1,2,3$) denote the coordinates on the maximum torus of ${\rm U}(3)$. Then the subsequent PE function can be written as 
\begin{eqnarray}\small
	\label{eq:PE eff5}
	{\rm PE}\left(z_1,z_2,z_3 ;q\right)&=& {\rm exp}\left(\sum_{k=1}^\infty\frac{\chi_l\left(z_1^k,z_2^k,z_3^k\right)q^k+\chi_5\left(z_1^k,z_2^k,z_3^k\right)q^k}{k}\right)\nonumber\\
	&=&\left[\left(1-q\right)^3\left(1-q z_1 z_2^{-1}\right) \left(1-q z_2 z_1^{-1}\right) \left(1-q z_1 z_3^{-1}\right) \left(1-q z_3 z_1^{-1}\right) \right.\nonumber\\
	&&\left.\times \left(1-q z_2 z_3^{-1}\right) \left(1-q z_3 z_2^{-1}\right) \left(1-q z_1^2\right)\left(1-q z_2^2\right)\left(1-q z_3^2\right)\left(1-q z_1 z_2 \right)\right.\nonumber\\
	&&\left.\times \left(1-q z_1 z_3 \right)\left(1-q z_2 z_3 \right)
	\left(1-q z_1^{-2}\right)\left(1-q z_2^{-2}\right)\left(1-q z_3^{-2}\right)\left(1-q z_1^{-1}z_2^{-1}\right)\right.\nonumber\\
	&& \left.\times \left(1-q z_1^{-1}z_3^{-1}\right)\left(1-q z_2^{-1}z_3^{-1}\right)\right]^{-1}\;.
\end{eqnarray}
Using the MW formula, one obtains the HS in the $\nu$SMEFT for the three-generation case
\begin{eqnarray}\small
	\label{eq:HS5}
	{ \cal H} (q)&=&\int \left[{\rm d}\mu\right]_{\rm U (3)} {\rm PE}\left(z_1,z_2,z_3 ;q\right)\nonumber\\
	&=&\frac{1}{6}\frac{1}{\left(2\pi {\rm i}\right)^3}\oint_{\left|z_1\right|=1}\oint_{\left|z_2\right|=1}\oint_{\left|z_3\right|=1}\left[-\frac{\left(z_2-z_1\right)^2\left(z_3-z_1\right)^2\left(z_3-z_2\right)^2}{z_1^2z_2^2z_3^2}\right]{\rm PE}\left(z_1,z_2,z_3 ;q\right)\;,\nonumber\\
\end{eqnarray}
where the Haar measure of ${\rm U}(3)$ group has been inserted and one may obtains
\begin{eqnarray}\small
	\label{eq:HS5}
	{ \cal H} (q)_5=\frac{{ \cal N} (q)_5}{{ \cal D} (q)_5}\;,
\end{eqnarray}
with
\begin{eqnarray}\small
	\label{eq:numerator2}
{ \cal N} (q)_5&=&1 + 3 q^6 + 2 q^7 + 5 q^8 + 4 q^9 + 5 q^{10} + 2 q^{11} + 6 q^{12} + 2 q^{13} + 5 q^{14} \nonumber\\
	&&+ 4 q^{15} + 5 q^{16} + 2 q^{17} + 3 q^{18} + q^{24 },
\end{eqnarray}
and
\begin{eqnarray}\small
	\label{eq:denominator5}
		{ \cal D} (q)_5=\left(1 - q\right) \left(1 - q^2\right)^2 \left(1 - q^3\right)^2 \left(1 - q^4\right)^3 \left(1 - q^5\right)^2 \left(1 - q^6\right)^2\;.
\end{eqnarray}
Therefore, the total number of invariants can be given by the leading positive terms of PL as
\begin{eqnarray}\small
	{\rm PL}\left[{\cal H}(q)_5\right]&=&q + 2 q^2 + 2 q^3 + 3 q^4 + 2 q^5 + 5 q^6 + 2 q^7 + 5 q^8 +4 q^9 
	\\\nonumber 
	&&+ 5 q^{10} + 2 q^{11}-{\cal O}\left(q^{13}\right)\;,
\end{eqnarray}
where the first CP-odd invariants as discussed earlier show up at order six and we list them up to order eight. Additionally, the first joint invariant appears in order seven. 
For the $\nu$SMEFT dim-5 with three generations of neutrinos, the invariants identified using our method correspond entirely with those observed in~\cite{Jenkins:2009dy}.

In the next section, we show identification of Ring-diagram in the presence of the dim-5 and one dim-6 operators.

\subsection{Two Dim-5 and -6 Operators}
\label{sec:dim-6}
Here, the Lagrangian \eqref{eq:EFT-lag5} is extended to the order of dim-6 as
\begin{eqnarray}\small
	\label{eq:EFT-lag6}
\mathcal{L}= {\mathcal L}_{\rm SM} -\left(\frac{C_5 }{2\Lambda} {\cal O}_5 + {\rm h.c.}\right)+\frac{C_6 }{\Lambda^2}{\cal O}_6 \;,
\end{eqnarray}
with the following dim-6 operator
\begin{eqnarray}\small
{\cal O}_5 =\overline L_{\sf L}\widetilde{H}\widetilde{H}^{\sf T}L_{\sf L}^{\sf C}\;,\quad
{\cal O}_6 =\left(\widetilde{H}^\dag i\overleftrightarrow{\partial}_\mu \widetilde{H}\right)\left(\overline L_i \gamma^\mu L_j\right)\;.	
\end{eqnarray}
Thus the corresponding Wilson coefficients read
\begin{eqnarray}\small
	\label{eq:Neutino-wilson}
	C_5 =-Y_\nu {\Lambda\over M_{\sf R}} Y_\nu^{\sf T}\;, \quad
	C_6 =Y_\nu {\Lambda^2\over M_{\sf R}^\dagger M_{\sf R}} Y_\nu^\dagger\;.
\end{eqnarray}
The general unitary transformations in the flavor space in the leptonic sector are given in \eqref{eq:field}. These induce the transformation of Yukawa matrices, the Majorana masses and the Wilson coefficient as shown in Table~\ref{tab:tab2}.
The additional entry for the transformation of $C_6$ is given in Table~\ref{tab:tab2p}.
\begin{table}[H]
\begin{center}
	\begin{tabular}{l|c|c|c}
		& \; ${\rm U}(3)_{L}$ \; & \; ${\rm U}(3)_e$ \; &\; ${\rm U}(3)_{\sf R}$ \; \\\hline
		$C_6$ &$\mathbf{3}\times\mathbf{3}^{\dagger}$ & $\mathbf{1}$ & $\mathbf{1}$ \\[0.1cm]
	\end{tabular}
	\end{center}
\caption{\it Flavor transformation under ${\rm U}(3)_{L}$, ${\rm U}(3)_e$ and ${\rm U}(3)_{\sf R}$ of the Wilson coefficient $C_{6}$. }
	\label{tab:tab2p}
\end{table}
Henceforth, the Lagrangian in Eq.~\eqref{eq:EFT-lag6} remains unchanged under the above transformation. 
Therefore, the building blocks for the construction of flavor invariants are $\left\{L \equiv Y_eY_e^\dagger, C_5, C_6 \right\}$ as can be seen in Ring-diagram ~\ref{RD7}.
\begin{figure}[H]
\begin{center}
\includegraphics[width=0.28\textwidth]{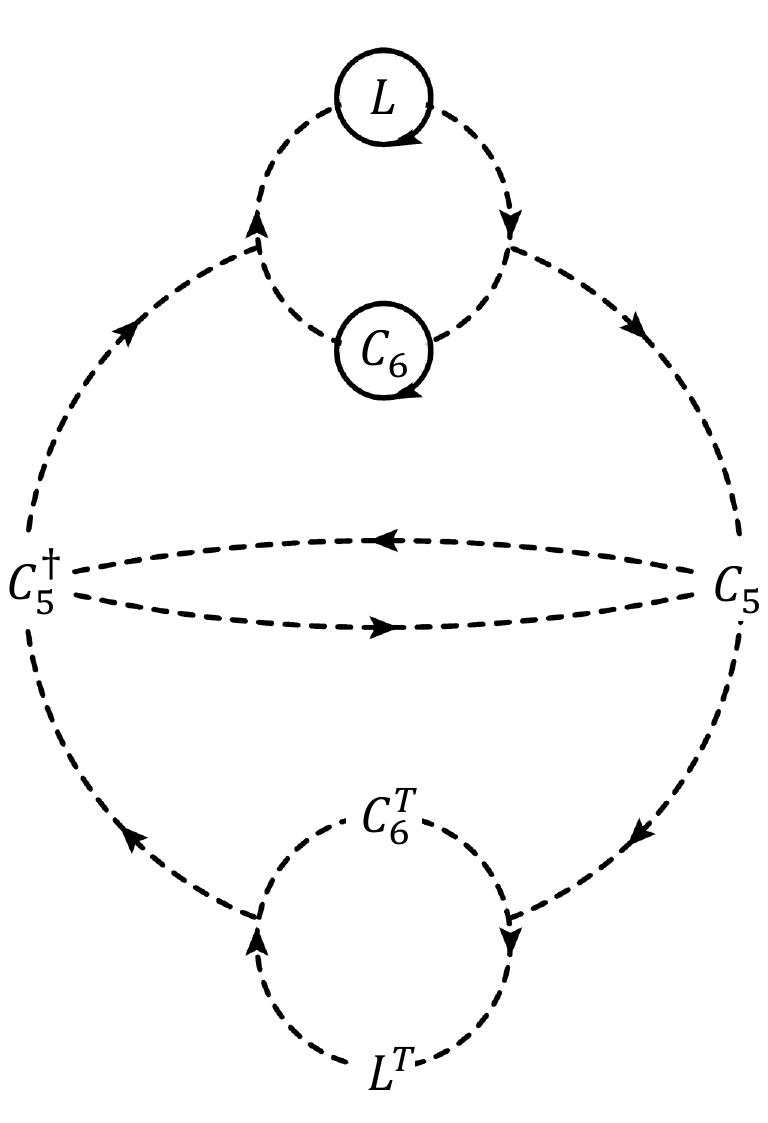}
\end{center}
\caption{The Ring-diagram for $\nu$SMEFT with Wilson coefficients $C_{5,6}$.}
\label{RD7}
\end{figure} 

Accordingly, the building blocks and their representatives rings are
{\allowdisplaybreaks
\begin{eqnarray}\small
 M_l &\equiv & \text{diag} \begin{pmatrix} L & 0 \end{pmatrix}\equiv \hspace{0.2cm}{\raisebox{-0.7em}{\begin{tikzpicture}
\def \radius {0.4cm}
 \node[draw, circle,style={thick},-latex] {$L$};
 \draw[-latex,->,thick] ({180}:\radius);
\end{tikzpicture}}}\,,
\qquad \qquad \qquad 
M_{c_6} \equiv \text{diag} \begin{pmatrix} 0 & C_6 \end{pmatrix}\equiv \hspace{0.2cm}{\raisebox{-0.7em}{\begin{tikzpicture}
\def \radius {0.4cm}
 \node[draw, circle,style={thick},-latex] {$\scriptstyle{C_6}$};
 \draw[-latex,->,thick] ({180}:\radius);
\end{tikzpicture}}}\,,
\nonumber \\
 {M_{c}} &\equiv & \begin{pmatrix}
C_5\,C_5^{\dagger} & 0 \\
0 & \, C_5^{\dagger}\,C_5 \\
\end{pmatrix}~ \equiv \raisebox{-1em}{ 
\begin{tikzpicture}[transform shape,line width=0.9pt]
\node (A) at (0,0) {$C_5$}; 
\node (B) at (1.2,0) {$C_5^{\dagger}$}; 
\path[draw] 
(A) edge[bend right=-40,black,dashed,->] (B)
(B) edge[bend right=-40,black,-latex,dashed,->] (A);
\end{tikzpicture}}\equiv \hspace{0.2cm}{\raisebox{-0.7em}{\begin{tikzpicture}
\def \radius {0.4cm}
 \node[draw, circle,style={thick},-latex] {$C$};
 \draw[-latex,->,thick] ({180}:\radius);
\end{tikzpicture}}} \,,
\nonumber \\
 {M_{lc_6}}^\pm &\equiv & \begin{pmatrix}
L\,C_6 & 0 \\
0 & \,\pm C_6 L \\
\end{pmatrix}~ \equiv \raisebox{-1em}{ 
\begin{tikzpicture}[transform shape,line width=0.9pt]
\node (A) at (0,0) {$L$}; 
\node (B) at (1.2,0) { $C_6$ }; 
\path[draw] 
(A) edge[bend right=-40,black,dashed,->] (B)
(B) edge[bend right=-40,black,-latex,dashed,->] (A);
\end{tikzpicture}}\,,
\nonumber \\
M_{lk_m}^\pm &\equiv & \begin{pmatrix}
L \,K_m& 0\\
0 &\, \pm K_m\,L \\
\end{pmatrix}~ \equiv \raisebox{-1em}{ 
\begin{tikzpicture}[transform shape,line width=0.9pt]
\node (A) at (0,0) {$L$}; 
\node (B) at (1.2,0) { $K_m$ }; 
\path[draw] 
(A) edge[bend right=-40,black,dashed,->] (B)
(B) edge[bend right=-40,black,-latex,dashed,->] (A);
\end{tikzpicture}} \equiv \raisebox{-0.1em}{ 
\raisebox{-2.8em}{ \begin{tikzpicture}[transform shape,line width=1pt]
\node (A) at (0.9,1) {$C_5^{\dagger}$}; 
\node (B) at (2,2) {$L$}; 
\node (C) at (3,1) {$C_5$};
\node (D) at (2,0) {${L^{\sf T}}^m$}; 
\node (E) at (1,0.8) {}; 
\path[draw] 
(A) edge[bend right=-30,black,dashed,->] (B) 
(B) edge[bend right=-30,black,dashed,->] (C)
(C) edge[bend right=-30,black,dashed,->] (D) 
(D) edge[bend right=-30,black,dashed,->] (E);
\end{tikzpicture}}}\,,
\nonumber \\
M_{c_6k_m}^\pm &\equiv & \begin{pmatrix}
C_6\,K_m& 0\\
0 &\, \pm K_m\,C_6 \\
\end{pmatrix}~ \equiv \raisebox{-1em}{ 
\begin{tikzpicture}[transform shape,line width=0.9pt]
\node (A) at (0,0) {$C_6$}; 
\node (B) at (1.2,0) { $K_m$ }; 
\path[draw] 
(A) edge[bend right=-40,black,dashed,->] (B)
(B) edge[bend right=-40,black,-latex,dashed,->] (A);
\end{tikzpicture}} \,\equiv \raisebox{-0.1em}{ 
\raisebox{-2.8em}{ \begin{tikzpicture}[transform shape,line width=1pt]
\node (A) at (0.9,1) {$C_5^{\dagger}$}; 
\node (B) at (2,2) {$C_6$}; 
\node (C) at (3,1) {$C_5$};
\node (D) at (2,0) {${L^{\sf T}}^m$}; 
\node (E) at (1,0.8) {}; 
\path[draw] 
(A) edge[bend right=-30,black,dashed,->] (B) 
(B) edge[bend right=-30,black,dashed,->] (C)
(C) edge[bend right=-30,black,dashed,->] (D) 
(D) edge[bend right=-30,black,dashed,->] (E);
\end{tikzpicture}}}\,,
\nonumber \\
M_{lh_m}^\pm &\equiv & \begin{pmatrix}
L \,H_m& 0\\
0 &\, \pm H_m\,L \\
\end{pmatrix}~ \equiv \raisebox{-1em}{ 
\begin{tikzpicture}[transform shape,line width=0.9pt]
\node (A) at (0,0) {$L$}; 
\node (B) at (1.2,0) { $H_m$ }; 
\path[draw] 
(A) edge[bend right=-40,black,dashed,->] (B)
(B) edge[bend right=-40,black,-latex,dashed,->] (A);
\end{tikzpicture}} ~ \equiv \raisebox{-0.1em}{ 
\raisebox{-2.8em}{ \begin{tikzpicture}[transform shape,line width=1pt]
\node (A) at (0.9,1) {$C_5^{\dagger}$}; 
\node (B) at (2,2) {$L$}; 
\node (C) at (3,1) {$C_5$};
\node (D) at (2,0) {${{C_6}^{\sf T}}^m$}; 
\node (E) at (1,0.8) {}; 
\path[draw] 
(A) edge[bend right=-30,black,dashed,->] (B) 
(B) edge[bend right=-30,black,dashed,->] (C)
(C) edge[bend right=-30,black,dashed,->] (D) 
(D) edge[bend right=-30,black,dashed,->] (E);
\end{tikzpicture}}}\,,
 \nonumber \\
M_{c_6H_m}^\pm &\equiv & \begin{pmatrix}
C_6 \,H_m& 0\\
0 &\, \pm H_m\,C_6 \\
\end{pmatrix}~ \equiv \raisebox{-1em}{ 
\begin{tikzpicture}[transform shape,line width=0.9pt]
\node (A) at (0,0) {$C_6$}; 
\node (B) at (1.2,0) { $H_m$ }; 
\path[draw] 
(A) edge[bend right=-40,black,dashed,->] (B)
(B) edge[bend right=-40,black,-latex,dashed,->] (A);
\end{tikzpicture}} \, \equiv \raisebox{-0.1em}{ 
\raisebox{-2.8em}{ \begin{tikzpicture}[transform shape,line width=1pt]
\node (A) at (0.9,1) {$C_5^{\dagger}$}; 
\node (B) at (2,2) {$C_6$}; 
\node (C) at (3,1) {$C_5$};
\node (D) at (2,0) {${{C_6}^{\sf T}}^m$}; 
\node (E) at (1,0.8) {}; 
\path[draw] 
(A) edge[bend right=-30,black,dashed,->] (B) 
(B) edge[bend right=-30,black,dashed,->] (C)
(C) edge[bend right=-30,black,dashed,->] (D) 
(D) edge[bend right=-30,black,dashed,->] (E);
\end{tikzpicture}}}\,,
 \nonumber \\
M_{l{f_{mn}}^\pm}^\pm &\equiv & \begin{pmatrix}
L \,{F_{mn}}^\pm& 0\\
0 &\, \pm {F_{mn}}^\pm\,L \\
\end{pmatrix}~ \equiv \raisebox{-1em}{ 
\begin{tikzpicture}[transform shape,line width=0.9pt]
\node (A) at (0,0) {$L$}; 
\node (B) at (1.2,0) { ${F_{mn}}^\pm$ }; 
\path[draw] 
(A) edge[bend right=-40,black,dashed,->] (B)
(B) edge[bend right=-40,black,-latex,dashed,->] (A);
\end{tikzpicture}} \, \equiv \raisebox{-0.1em}{ 
\raisebox{-2.8em}{ \begin{tikzpicture}[transform shape,line width=1pt]
\node (A) at (0.9,1) {$C_5^{\dagger}$}; 
\node (B) at (2.2,2.2) {$L$}; 
\node (C) at (3.3,1) {$C_5$};
\node (D) at (2.2,0) {${{X}^{mn}}^\pm$}; 
\node (E) at (1,0.7) {}; 
\path[draw] 
(A) edge[bend right=-30,black,dashed,->] (B) 
(B) edge[bend right=-30,black,dashed,->] (C)
(C) edge[bend right=-30,black,dashed,->] (D) 
(D) edge[bend right=-30,black,dashed,->] (E);
\end{tikzpicture}}}\,,
 \label{blocks-dim-6-2}
\end{eqnarray}}
where $K_m=C_5 \,{L^{\sf T}}^m\,C_5^{\dagger}$, $H_m=C_5 \,{C_6^{\sf T}}^m\,C_5^{\dagger}$ and $F_{mn}^\pm=C_5 \,{X^{mn}}^\pm\,C_5^{\dagger}$ with $X^{mn\pm}={L^{\sf T}}^m {C_6^{\sf T}}^n \pm{C_6^{\sf T}}^n{L^{\sf T}}^m$ running over $m,n=1,2$. Additionally, the blocks $M_{c_6H_m}^\pm$ out of ring $ \raisebox{-1em}{ 
\begin{tikzpicture}[transform shape,line width=0.9pt]
\node (A) at (0,0) {$C_6$}; 
\node (B) at (1.2,0) { $H_m$ }; 
\path[draw] 
(A) edge[bend right=-40,black,dashed,->] (B)
(B) edge[bend right=-40,black,-latex,dashed,->] (A);
\end{tikzpicture}}$ can be skipped as produces dim-10. However, we can keep this block for comparison with the number derived via the HS. One can organise the basic invariants beginning with the lowest order blocks as follows:
\begin{itemize}
\item[$\bullet$] Order-1:\hfill
\begin{center}\includegraphics[width=0.9\textwidth]{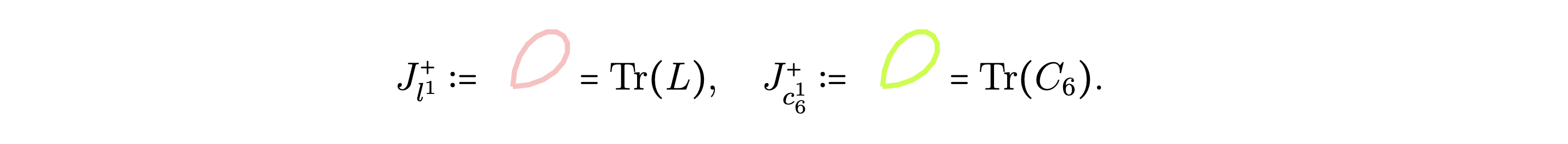}\end{center}
\item[$\bullet$] Order-2:\hfill
\begin{center}\includegraphics[width=0.9\textwidth]{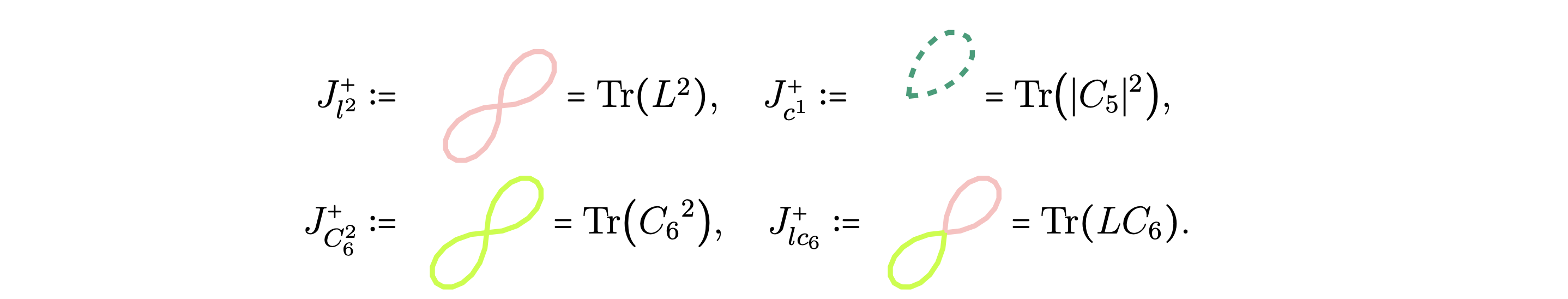}\end{center}
\item[$\bullet$] Order-3:\hfill
\begin{center}\includegraphics[width=0.9\textwidth]{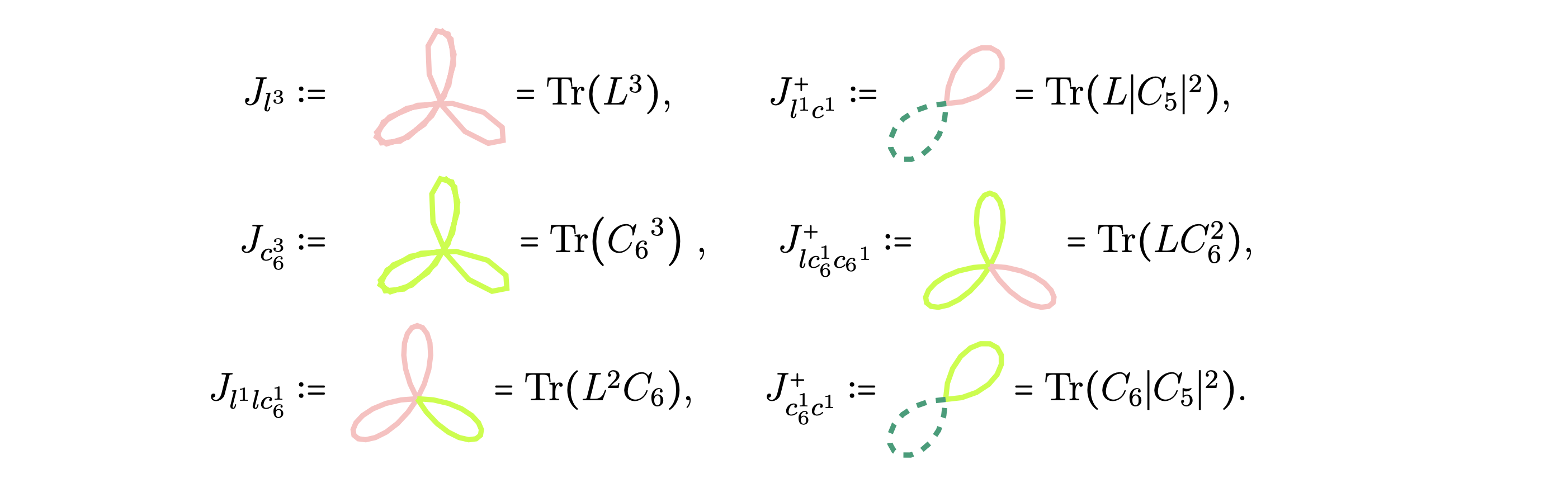}\end{center}
\item[$\bullet$] Order-4:\hfill
{\begin{center}\includegraphics[width=0.9\textwidth]{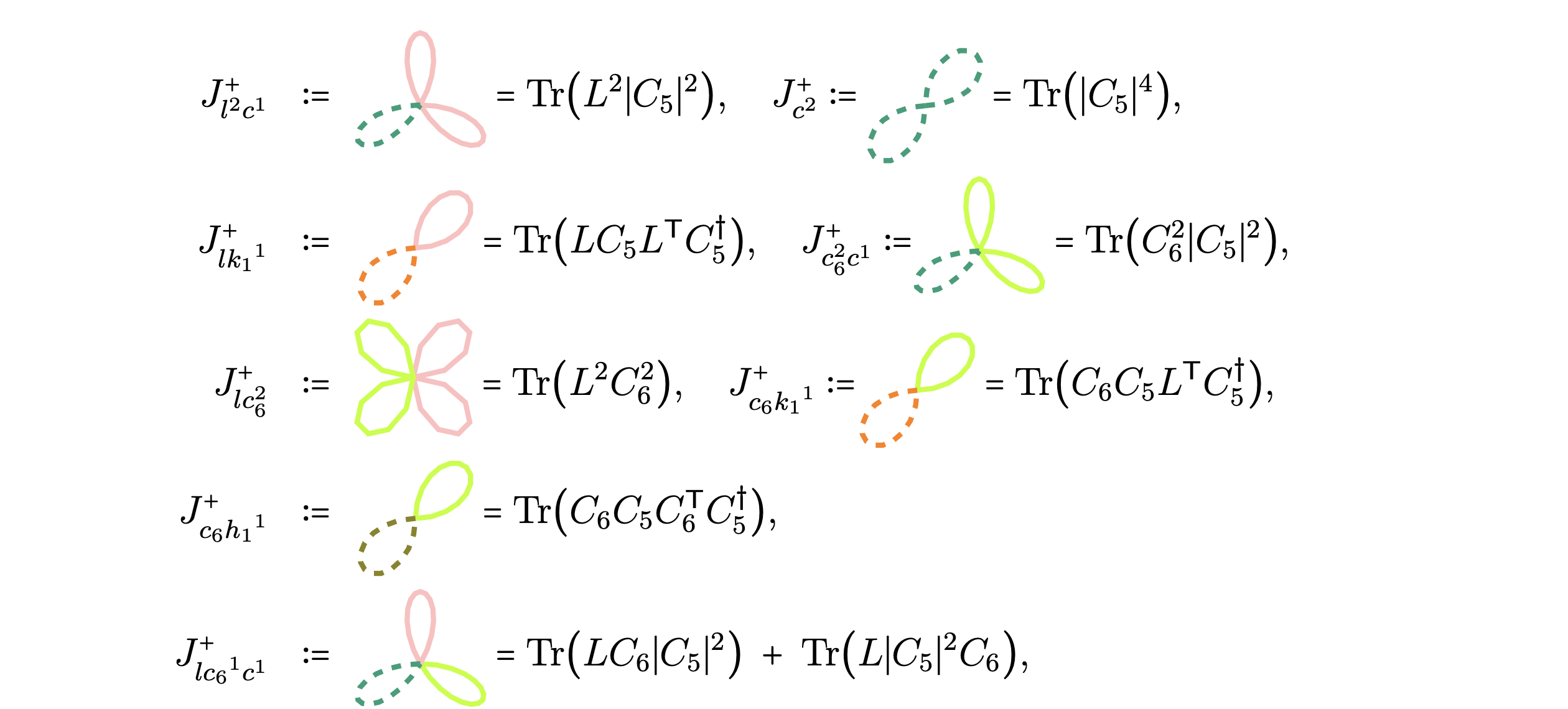}\end{center}}
where the dashed orange and olive petals presenting order three blocks $K_1 \equiv C_5 {L^{\sf T}} C_5^\dagger$ and $H_1 \equiv C_5 {C_6^{\sf T}} C_5^\dagger$, respectively.

\item[$\bullet$] Order-5:
\hfill
\begin{center}\includegraphics[width=0.9\textwidth]{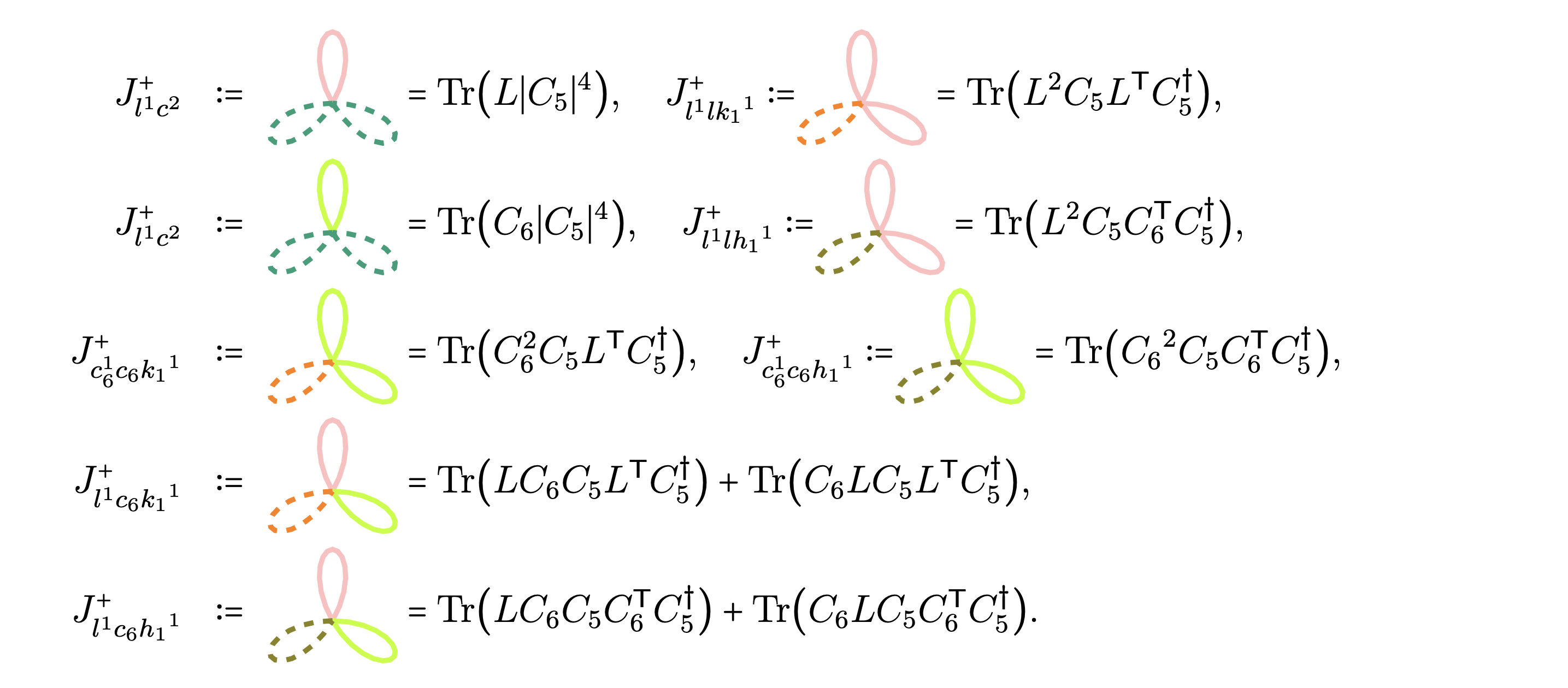}\end{center}
$\bullet$ Order-6:
\hfill
{\begin{center}\includegraphics[width=0.9\textwidth]{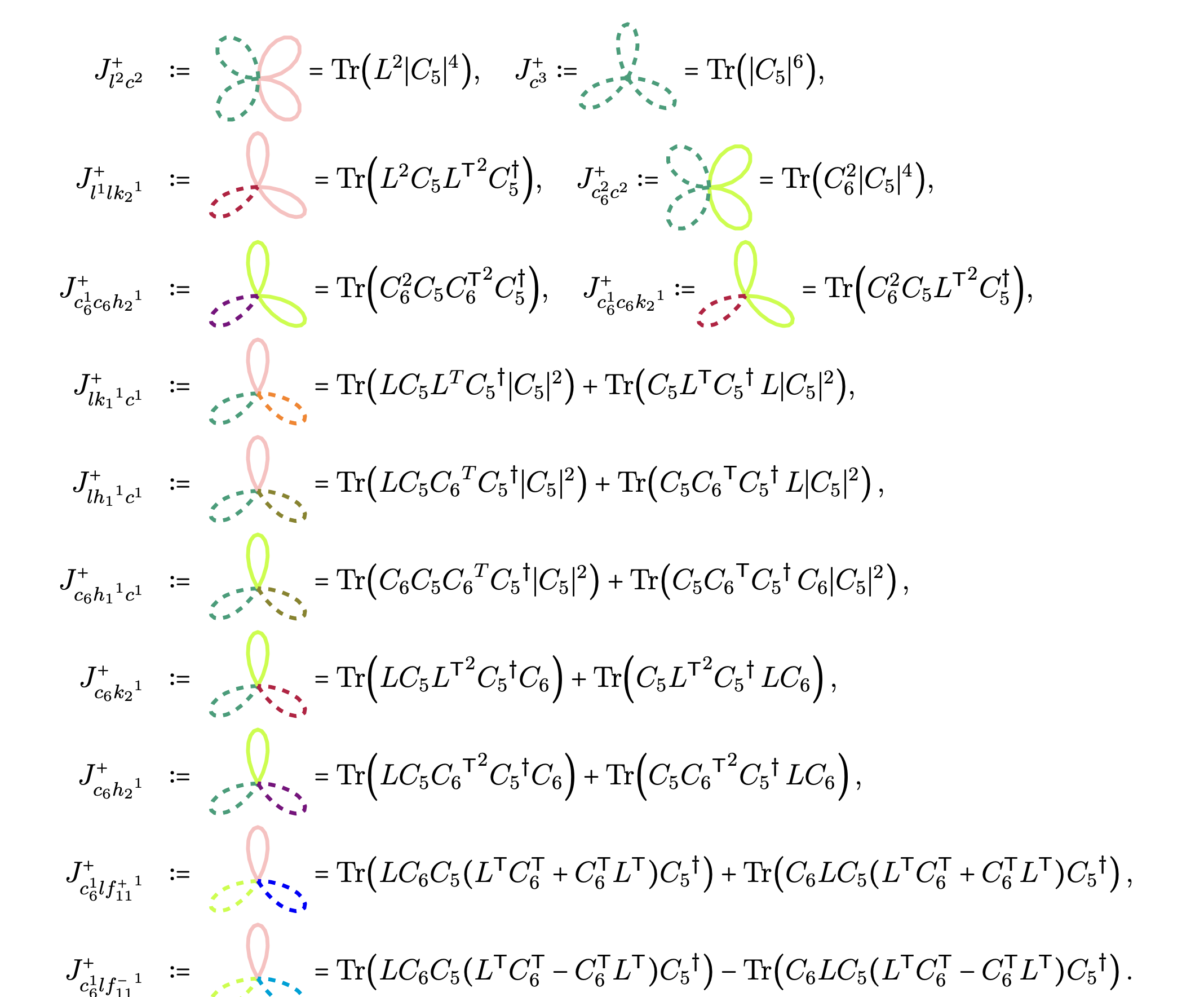}\end{center}}
The dashed petals in violet, purple, blue, and cyan represent the fourth-order invariants $K_2 \equiv C_5 {L^{\sf T}}^2 C_5^\dagger$, $H_2 \equiv C_5 {C_6^{\sf T}}^2 C_5^\dagger$, and $F_{11}^\pm\equiv C_5 ({L^{\sf T}}{C_6^{\sf T}} \;\pm\; {C_6^{\sf T}}{L^{\sf T}}) C_5^\dagger$.
\end{itemize}

Besides these, there are CP-odd and joint invariants. The first CP-odd invariant appears at fourth order and consists of nine invariants. At fifth order, the first joint invariant emerges, and in combination with four CP-odd and basic invariants, accounts for 14 invariants. Similarly, at sixth order, there are 20 additional CP-odd and joint invariants, on top of the previous 13. Up to the fifth order, all invariants are basic. From this order onward, joint invariants can also be recognised.

Note that for each $J^+$ invariant (containing two terms), there are CP-odd $J^-$ invariants relevant only for their subtractions. However, only one $J^-$ invariant corresponds to the last two relations.

The following list presents CP-odd invariants, starting from the lowest conceivable order:
{\allowdisplaybreaks \ytableausetup{smalltableaux}
\bea
J^-_{{lc_6}^1c^1} &:=&  \ \hspace{-0.1cm}{\raisebox{-1.1em}{\includegraphics[width=0.09\textwidth]{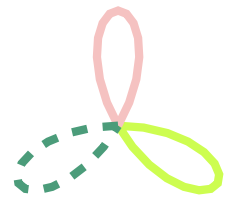}}} =\Tr(L C_6 |C_5|^2 )\;- \Tr( L |C_5|^2 C_6 ),
  \\
 J^-_{l^1{lc_6}^1c^1} &:=&  \hspace{-0.1cm}{\raisebox{-1.1em}{\includegraphics[width=0.09\textwidth]{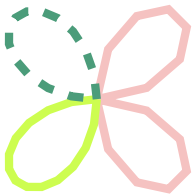}}} = \Tr(L^2 C_6 |C_5|^2 )- \Tr( L^2 |C_5|^2 C_6 ),
  \\
  J^-_{c_6^1{lc_6}^1c^1} &:=&   \hspace{-0.1cm}{\raisebox{-1.1em}{\includegraphics[width=0.09\textwidth]{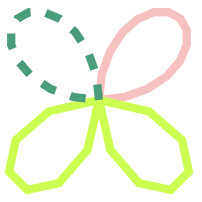}}} =\Tr(L C_6^2 |C_5|^2 ) - \Tr( L |C_5|^2 C_6^2 ),
  \\
 J^-_{l^1{c_6k_1}^1}&:=&   \hspace{-0.1cm}{\raisebox{-1.1em}{\includegraphics[width=0.09\textwidth]{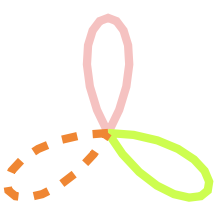}}}= \Tr (L C_6 C_5 L^{\sf T} C_5^\dagger)
 - \Tr (C_6 L C_5 L^{\sf T} C_5^\dagger),
 \\ 
  J^-_{l^1{c_6h_1}^1}&:=& \hspace{-0.1cm}{\raisebox{-1.1em}{\includegraphics[width=0.09\textwidth]{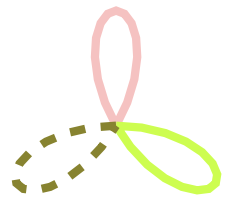}}} = \Tr (L C_6 C_5 C_6^{\sf T} C_5^\dagger)
 -\Tr (C_6 L C_5 C_6^{\sf T} C_5^\dagger),
\\
 J^-_{{lc_6}^1c^2} &:=&  \hspace{-0.1cm}{\raisebox{-1.1em}{\includegraphics[width=0.09\textwidth]{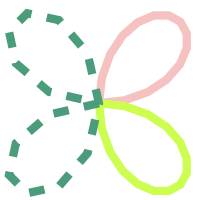}}} =\Tr(L C_6 |C_5|^4 )- \Tr( L |C_5|^4 C_6)\,,
\eea }
where certain invariants involving higher orders of $C_{5,6}$ can be disregarded. From the aforementioned invariants, it is evident that CP violation occurs \textit{if and only if} ${\rm Im}J^-\neq 0$. 

In scenarios involving degeneracies, the evaluation of higher-order CP-violating invariants becomes essential. In addition to $J^-$, which corresponds to the petals and relations for $J^+$ (with reversed signs between terms) as shown in the fundamental invariant at the sixth order, a detailed list of other higher-order CP-violating invariants can be found in Reference~\cite{Darvishi:2023ckq}.

A useful exercise is to compare the magnitude of these invariants with predictions from the Hilbert series and the results obtained in References~\cite{Yu:2021cco,Yu:2022ttm}. We begin by defining the character functions in terms of ${\bf 3}$ and ${\bf 3}^*$ as $z_1 + z_2 + z_3$ and $z_1^{-1} + z_2^{-1} + z_3^{-1}$, respectively, where $z_{1,2}$ and $z_3$ are coordinates on the maximal torus of $U(3)$. The character functions of the flavor invariants, $\left\{L \equiv Y_eY_e^\dagger, C_5, C_6 \right\}$, analogous to Tables \ref{tab:tab2} and \ref{tab:tab2p}, are expressed as:
\begin{eqnarray}
\chi_{l,6} (z_1, z_2, z_3) &=& (z_1 + z_2 + z_3)(z_1^{-1} + z_2^{-1} + z_3^{-1}),\nonumber\\
&& 
\\
\chi_5 (z_1, z_2, z_3) &=& z_1^2 + z_2^2 + z_3^2 + z_1 z_2 + z_1 z_3 + z_2 z_3 + z_1^{-2} 
\nonumber\\&+&  z_2^{-2} + z_3^{-2} + z_1^{-1}z_2^{-1} + z_1^{-1}z_3^{-1} + z_2^{-1}z_3^{-1}.
\nonumber
\end{eqnarray}
Thus Hilbert series in the $\nu$SMEFT with operators of dimensions 5 and 6 for the $3\times3$ matrices becomes
\begin{eqnarray}
{\cal H}_6 (q) &=& \frac{{ \cal N}_6 (q)}{{ \cal D}_6 (q)},
\end{eqnarray}
where ${\cal N}_6 (q)$ and the denominator ${\cal D}_6 (q)$ read
\begin{eqnarray}
\label{eq:numerator2}
{\cal N}_6 (q) & = & 1 + 2 q^3 + 4 q^4 + 11 q^5 + 33 q^6 + 52 q^7  +  104 q^8 + 182 q^9 + 307 q^{10} + 495 q^{11} + 808 q^{12} \nonumber \\
& + & 1176 q^{13} + 1692 q^{14} + 2307 q^{15} + 2995 q^{16}  +  3736 q^{17} + 4546 q^{18} + 5246 q^{19} + 5902 q^{20} \nonumber \\
& + & 6401 q^{21} + 6632 q^{22} + 6632 q^{23} + 6401 q^{24} + 5902 q^{25} + 5246 q^{26} + 4546 q^{27} + 3736 q^{28} \nonumber \\
& + & 2995 q^{29} + 2307 q^{30} + 1692 q^{31} + 1176 q^{32} + 808 q^{33} + 495 q^{34} + 307 q^{35}  
\nonumber \\
& + & 182 q^{36} + 104 q^{37}  + 52 q^{38} + 33 q^{39} + 11 q^{40} + 4 q^{41} + 2 q^{42} + q^{45},
\end{eqnarray}
and 
\begin{eqnarray}
\label{eq:denominator}
{\cal D}_6 (q) &=& \left(1 - q\right)^2 \left(1 - q^2\right)^4 \left(1 - q^3\right)^4 \left(1 - q^4\right)^5  
 \left(1 - q^5\right)^3  \left(1 -  q^6\right)^3\;.
\end{eqnarray}
Finally, the associated Poincaré series for the Hilbert series can be derived as follows:
\begin{eqnarray}
	\label{eq:PL}
	{\rm PL}\left[{\cal H}_6(q)\right] &=& 2q+4q^2+6q^3+9q^4+14q^5+33q^6+44q^7
 \nonumber \\
&+& 72q^8+74q^9+21q^{10}-{\cal O}\left(q ^{11}\right),
\end{eqnarray}
which aligns with the results obtained in References~\cite{Yu:2021cco,Yu:2022ttm}. 
Clearly, the presentation of ${\cal N}(q)$ and ${\cal D}(q)$ is not unique since these are elements of a ratio; therefore, defining primary invariants based on the denominator ${\cal D}(q)$ is not a viable method. Instead, the current formalism in this article get the match number of invariants as the above ${\rm PL}$, with primary invariants clearly identified up to the sixth order. The starting point for CP-odd invariants is at fourth order, and for joint invariants, it is at fifth order.

\subsection{Two Dim-5 and -7 Operators}
\label{sec:dim-7}
Here, the Lagrangian of $\nu$SMEFT is considered as the extension of \eqref{eq:EFT-lag5} to the order of dim-7 
\begin{eqnarray}\small
	\label{eq:EFT-lag7}
\mathcal{L}= {\mathcal L}_{\rm SM}-\left(\frac{C_5 }{2\Lambda} {\cal O}_5 +{\rm h.c.}\right)\,+\,\left(\frac{C_7 }{\Lambda^3}{\cal O}_7+{\rm h.c.}\right) \;,
\end{eqnarray}
with
\begin{eqnarray}\small
{\cal O}_5 =\overline L_{\sf L}\widetilde{H}\widetilde{H}^{\sf T}L_{\sf L}^{\sf C}\;,\quad
{\cal O}_7 =\overline L_{\sf L}\widetilde{H}\widetilde{H}^{\sf T}L_{\sf L}^{\sf C} {H}^{\rm \dagger} H\;,
\end{eqnarray}
and the corresponding Wilson coefficients read
\begin{eqnarray}\small
	\label{eq:Neutino-wilson}
	C_5 =-Y_\nu {\Lambda\over M_{\sf R}} Y_\nu^{\sf T}\;, \quad
	C_7 =Y_\nu {\Lambda^3 \over M_{\sf R} v^2} Y_\nu^{\sf T}\;.
\end{eqnarray}

The general unitary transformation in the flavor space in the leptonic sector induces the transformation of Yukawa matrices, the Majorana masses and the Wilson coefficient as shown in the previous section, where $C_7$ is replaced with $C_6$ but with the same transformation as $C_5$ as shown in Table~\ref{tab:tabIV}.
\begin{table}[H]
\begin{center}
	\begin{tabular}{l|c|c|c}
		& \; ${\rm U}(3)_{L}$ \; & \; ${\rm U}(3)_e$ \; &\; ${\rm U}(3)_{\sf R}$ \; \\\hline
		$C_7$ &$\mathbf{3}\times\mathbf{3}^{\sf T}$ & $\mathbf{1}$ & $\mathbf{1}$ \\[0.1cm]
	\end{tabular}
	\end{center}
\caption{\it Flavor transformation under ${\rm U}(3)_{L}$, ${\rm U}(3)_e$ and ${\rm U}(3)_{\sf R}$ of Wilson coefficient $C_{7}$. The rest of the transformations are the same as Table \ref{tab:tab2}.}
	\label{tab:tabIV}
\end{table}

Therefore, the ring diagram can be sketched including flavor invariants $\left\{L \equiv Y_eY_e^\dagger,C_5,C_7\right\}$ as shown in Ring-diagram \ref{RD8}.
\begin{figure}[H]
\begin{center}
\includegraphics[width=0.28\textwidth]{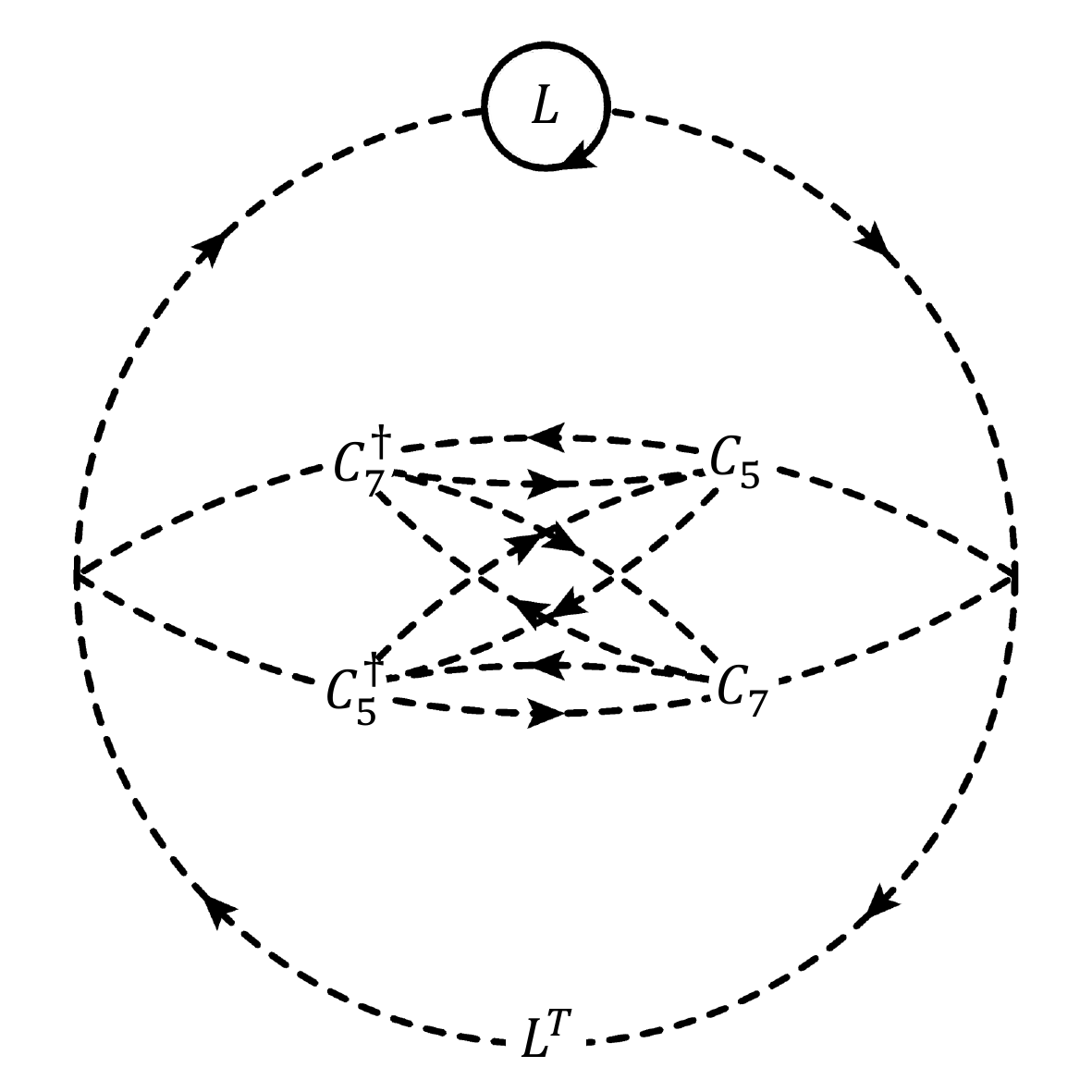}
\end{center}
\caption{
The Ring-diagram for $\nu$SMEFT with Wilson coefficients $C_{5,7}$.}
\label{RD8}
\end{figure}

Thus, the building blocks take the following form
{\allowdisplaybreaks
\begin{eqnarray}\small
 M_l &\equiv & \text{diag} \begin{pmatrix} L & 0 \end{pmatrix}\equiv \hspace{0.2cm}{\raisebox{-0.7em}{\begin{tikzpicture}
\def \radius {0.4cm}
 \node[draw, circle,style={thick},-latex] {$L$};
 \draw[-latex,->,thick] ({180}:\radius);
\end{tikzpicture}}}\,,
\nonumber \\
 {M_{c_5}} &\equiv & \begin{pmatrix}
C_5\,C_5^{\dagger} & 0 \\
0 & \, C_5^{\dagger}\,C_5 \\
\end{pmatrix}~ \equiv \raisebox{-1em}{ 
\begin{tikzpicture}[transform shape,line width=0.9pt]
\node (A) at (0,0) {$C_5$}; 
\node (B) at (1.2,0) {$C_5^{\dagger}$}; 
\path[draw] 
(A) edge[bend right=-40,black,dashed,->] (B)
(B) edge[bend right=-40,black,-latex,dashed,->] (A);
\end{tikzpicture}}\equiv \hspace{0.2cm}{\raisebox{-0.7em}{\begin{tikzpicture}
\def \radius {0.4cm}
 \node[draw, circle,style={thick},-latex] {$\scriptstyle{C_5}$};
 \draw[-latex,->,thick] ({180}:\radius);
\end{tikzpicture}}} \,,
\nonumber \\
 {M_{c_7}} &\equiv & \begin{pmatrix}
C_7\,C_7^{\dagger} & 0 \\
0 & \, C_7^{\dagger}\,C_7 \\
\end{pmatrix}~ \equiv \raisebox{-1em}{ 
\begin{tikzpicture}[transform shape,line width=0.9pt]
\node (A) at (0,0) {$C_7$}; 
\node (B) at (1.2,0) {$C_7^{\dagger}$}; 
\path[draw] 
(A) edge[bend right=-40,black,dashed,->] (B)
(B) edge[bend right=-40,black,-latex,dashed,->] (A);
\end{tikzpicture}}\equiv \hspace{0.2cm}{\raisebox{-0.7em}{\begin{tikzpicture}
\def \radius {0.4cm}
 \node[draw, circle,style={thick},-latex] {$\scriptstyle{C_7}$};
 \draw[-latex,->,thick] ({180}:\radius);
\end{tikzpicture}}} \,,
\nonumber \\
 M_{c} &\equiv & \begin{pmatrix}
C_5\,C_7^{\dagger} & 0 \\
0 & \, C_7^{\dagger}\,C_5 \\
\end{pmatrix}~ \equiv \raisebox{-1em}{ 
\begin{tikzpicture}[transform shape,line width=0.9pt]
\node (A) at (0,0) {$C_5$}; 
\node (B) at (1.2,0) {$C_7^{\dagger}$}; 
\path[draw] 
(A) edge[bend right=-40,black,dashed,->] (B)
(B) edge[bend right=-40,black,-latex,dashed,->] (A);
\end{tikzpicture}}\equiv \hspace{0.2cm}{\raisebox{-0.7em}{\begin{tikzpicture}
\def \radius {0.4cm}
 \node[draw, circle,style={thick},-latex] {$\scriptstyle{\,C~}$};
 \draw[-latex,->,thick] ({180}:\radius);
\end{tikzpicture}}} \,,
\nonumber \\
 M_{c^\dagger} &\equiv & \begin{pmatrix}
C_7\,C_5^{\dagger} & 0 \\
0 & \, C_5^{\dagger}\,C_7 \\
\end{pmatrix}~ \equiv \raisebox{-1em}{ 
\begin{tikzpicture}[transform shape,line width=0.9pt]
\node (A) at (0,0) {$C_7$}; 
\node (B) at (1.2,0) {$C_5^{\dagger}$}; 
\path[draw] 
(A) edge[bend right=-40,black,dashed,->] (B)
(B) edge[bend right=-40,black,-latex,dashed,->] (A);
\end{tikzpicture}}\equiv \hspace{0.2cm}{\raisebox{-0.7em}{\begin{tikzpicture}
\def \radius {0.4cm}
 \node[draw, circle,style={thick},-latex] {$\scriptstyle{C^\dagger}$};
 \draw[-latex,->,thick] ({180}:\radius);
\end{tikzpicture}}} \,,
\nonumber \\
M_{lk_m}^\pm &\equiv & \begin{pmatrix}
L \,K_m& 0\\
0 &\, \pm K_m\,L \\
\end{pmatrix}~ \equiv \raisebox{-1em}{ 
\begin{tikzpicture}[transform shape,line width=0.9pt]
\node (A) at (0,0) {$L$}; 
\node (B) at (1.2,0) { $K_m$ }; 
\path[draw] 
(A) edge[bend right=-40,black,dashed,->] (B)
(B) edge[bend right=-40,black,-latex,dashed,->] (A);
\end{tikzpicture}} \equiv \raisebox{-0.1em}{ 
\raisebox{-2.8em}{ \begin{tikzpicture}[transform shape,line width=1pt]
\node (A) at (0.9,1) {$C_5^{\dagger}$}; 
\node (B) at (2,2) {$L$}; 
\node (C) at (3,1) {$C_5$};
\node (D) at (2,0) {${L^{\sf T}}^m$}; 
\node (E) at (1,0.8) {}; 
\path[draw] 
(A) edge[bend right=-30,black,dashed,->] (B) 
(B) edge[bend right=-30,black,dashed,->] (C)
(C) edge[bend right=-30,black,dashed,->] (D) 
(D) edge[bend right=-30,black,dashed,->] (E);
\end{tikzpicture}}}\,,
\nonumber \\
M_{lp_m}^\pm &\equiv & \begin{pmatrix}
L \,P_m& 0\\
0 &\, \pm P_m\,L \\
\end{pmatrix}~ \equiv \raisebox{-1em}{ 
\begin{tikzpicture}[transform shape,line width=0.9pt]
\node (A) at (0,0) {$L$}; 
\node (B) at (1.2,0) { $P_m$ }; 
\path[draw] 
(A) edge[bend right=-40,black,dashed,->] (B)
(B) edge[bend right=-40,black,-latex,dashed,->] (A);
\end{tikzpicture}} \equiv \raisebox{-0.1em}{ 
\raisebox{-2.8em}{ \begin{tikzpicture}[transform shape,line width=1pt]
\node (A) at (0.9,1) {$C_7^{\dagger}$}; 
\node (B) at (2,2) {$L$}; 
\node (C) at (3,1) {$C_7$};
\node (D) at (2,0) {${L^{\sf T}}^m$}; 
\node (E) at (1,0.8) {}; 
\path[draw] 
(A) edge[bend right=-30,black,dashed,->] (B) 
(B) edge[bend right=-30,black,dashed,->] (C)
(C) edge[bend right=-30,black,dashed,->] (D) 
(D) edge[bend right=-30,black,dashed,->] (E);
\end{tikzpicture}}}\,,
\nonumber \\
M_{lx_m}^\pm &\equiv & \begin{pmatrix}
L \,X_m& 0\\
0 &\, \pm X_m\,L \\
\end{pmatrix}~ \equiv \raisebox{-1em}{ 
\begin{tikzpicture}[transform shape,line width=0.9pt]
\node (A) at (0,0) {$L$}; 
\node (B) at (1.2,0) { $X_m$ }; 
\path[draw] 
(A) edge[bend right=-40,black,dashed,->] (B)
(B) edge[bend right=-40,black,-latex,dashed,->] (A);
\end{tikzpicture}} \equiv \raisebox{-0.1em}{ 
\raisebox{-2.8em}{ \begin{tikzpicture}[transform shape,line width=1pt]
\node (A) at (0.9,1) {$C_7^{\dagger}$}; 
\node (B) at (2,2) {$L$}; 
\node (C) at (3,1) {$C_5$};
\node (D) at (2,0) {${L^{\sf T}}^m$}; 
\node (E) at (1,0.8) {}; 
\path[draw] 
(A) edge[bend right=-30,black,dashed,->] (B) 
(B) edge[bend right=-30,black,dashed,->] (C)
(C) edge[bend right=-30,black,dashed,->] (D) 
(D) edge[bend right=-30,black,dashed,->] (E);
\end{tikzpicture}}}\,,
\nonumber \\
M_{ly_m}^\pm &\equiv & \begin{pmatrix}
L \,Y_m& 0\\
0 &\, \pm Y_m\,L \\
\end{pmatrix}~ \equiv \raisebox{-1em}{ 
\begin{tikzpicture}[transform shape,line width=0.9pt]
\node (A) at (0,0) {$L$}; 
\node (B) at (1.2,0) { $Y_m$ }; 
\path[draw] 
(A) edge[bend right=-40,black,dashed,->] (B)
(B) edge[bend right=-40,black,-latex,dashed,->] (A);
\end{tikzpicture}} \equiv \raisebox{-0.1em}{ 
\raisebox{-2.8em}{ \begin{tikzpicture}[transform shape,line width=1pt]
\node (A) at (0.9,1) {$C_5^{\dagger}$}; 
\node (B) at (2,2) {$L$}; 
\node (C) at (3,1) {$C_7$};
\node (D) at (2,0) {${L^{\sf T}}^m$}; 
\node (E) at (1,0.8) {}; 
\path[draw] 
(A) edge[bend right=-30,black,dashed,->] (B) 
(B) edge[bend right=-30,black,dashed,->] (C)
(C) edge[bend right=-30,black,dashed,->] (D) 
(D) edge[bend right=-30,black,dashed,->] (E);
\end{tikzpicture}}}\,,
\end{eqnarray}}
with $K_m=C_5 \,{L^{\sf T}}^m\,C_5^{\dagger}$, $P_m=C_7 \,{L^{\sf T}}^m\,C_7^{\dagger}$, $X_m=C_5 \,{L^{\sf T}}^m\,C_7^{\dagger}$ and $Y_m=C_7 \,{L^{\sf T}}^m\,C_5^{\dagger}$ where running over $m=1,2$. 
Also, it would be more convenient to redefine $(M_{c} $ and $M_{c^\dagger}$ into their real and imaginary components i.e. $M_{c_1}^+= (M_{c} + M_{c^\dagger} ) /2$ and $M_{c_2}^-= (M_{c} - M_{c^\dagger} ) /2$. Similarly, we can re-express $X_m$ and $Y_m$ with $R_m^+= (X_m + Y_m ) /2$ and $S_m^-= (X_m - Y_m ) /2$.

In this framework, the list of basic invariants is long and similar to the previous case extended to dim-6. Here, we classify these invariants to order 5 and provide only a couple of examples from order 6:
\begin{itemize}
\item[$\bullet$] Order-1:\hfill
\begin{center}\includegraphics[width=0.9\textwidth]{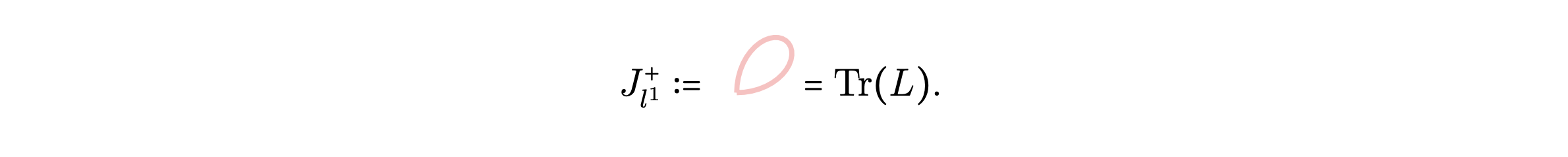}
\end{center}
\item[$\bullet$] Order-2:\hfill
\begin{center}
\includegraphics[width=1\textwidth]{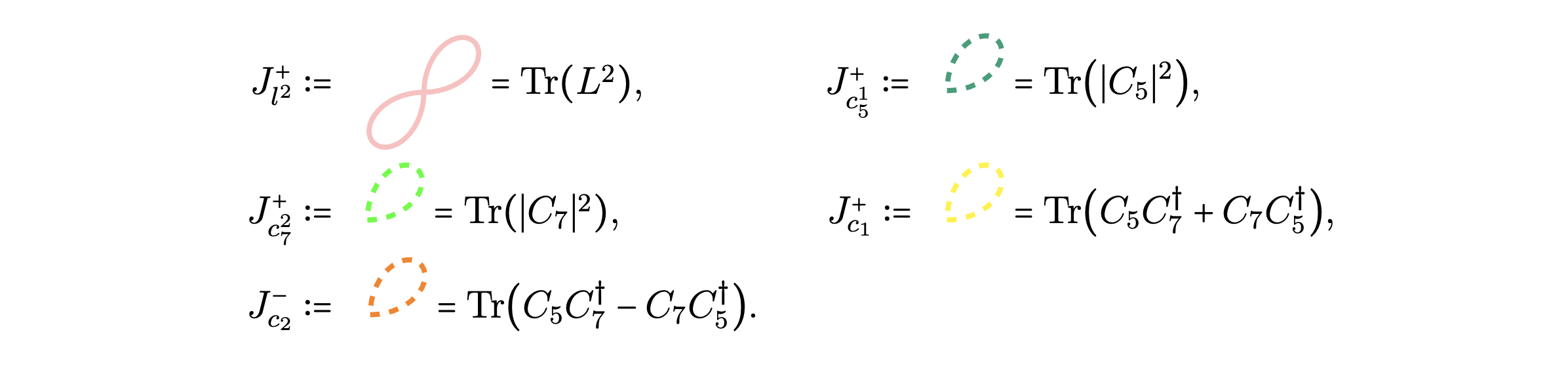}
\end{center}
\item[$\bullet$] Order-3:\hfill
\begin{center}
\includegraphics[width=1\textwidth]{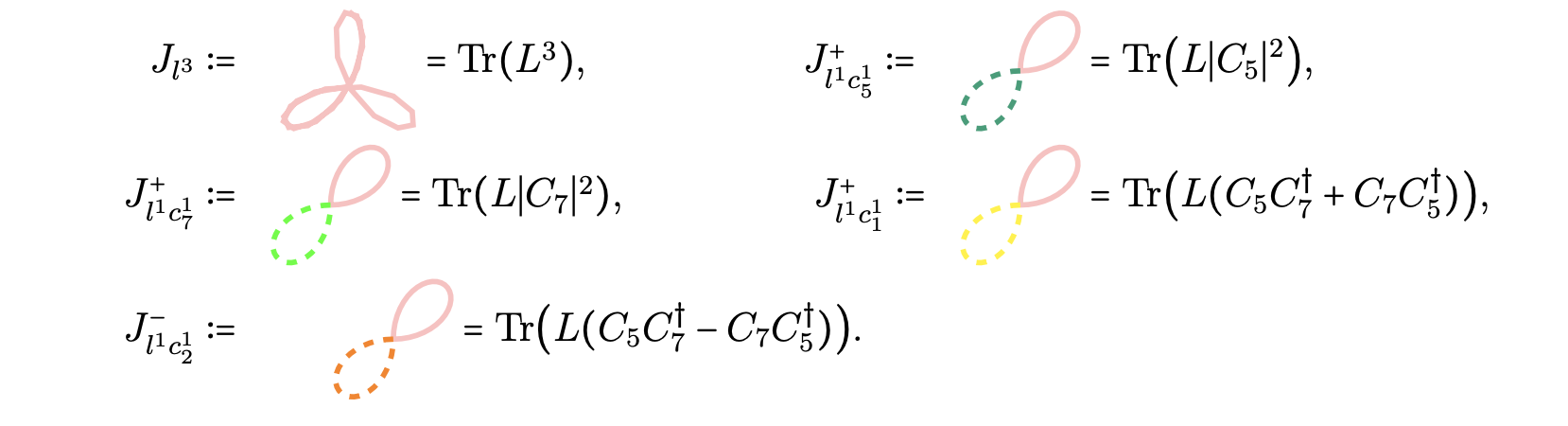}
\end{center}
\item[$\bullet$] Order-4:\hfill
\begin{center}\includegraphics[width=0.9\textwidth]{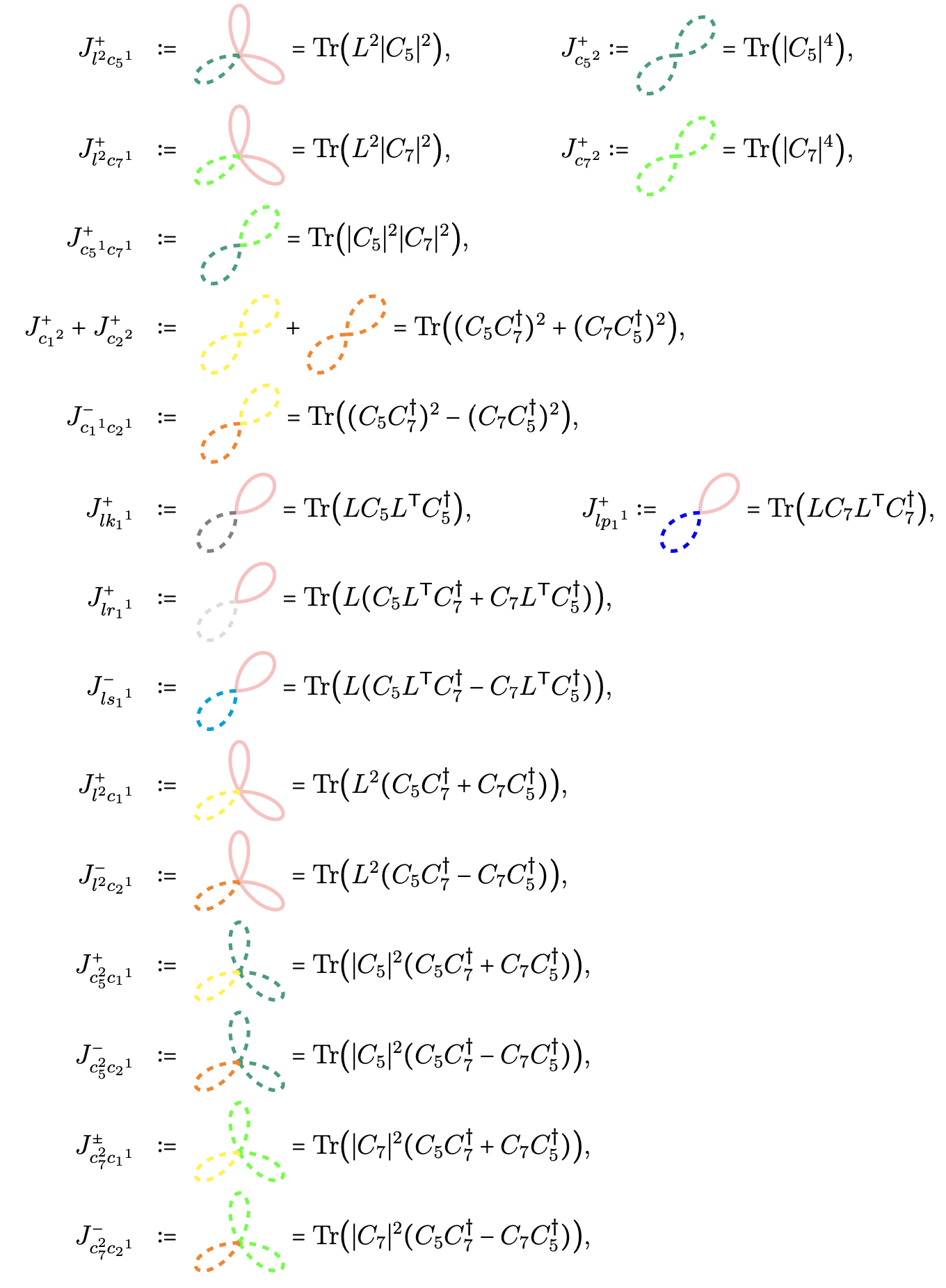}
\end{center}
where the dashed grey, blue, light grey and cyan petals presenting order three blocks $K_1 \equiv C_5 {L^{\sf T}} C_5^\dagger$, $P_1 \equiv C_7 {L^{\sf T}} C_7^\dagger$, $R_1 \equiv C_5 {L^{\sf T}} C_7^\dagger$ and $S_1 \equiv C_7 {L^{\sf T}} C_5^\dagger$, respectively.
\item[$\bullet$] Order-5:\hfill
\begin{center}\includegraphics[width=0.9\textwidth]{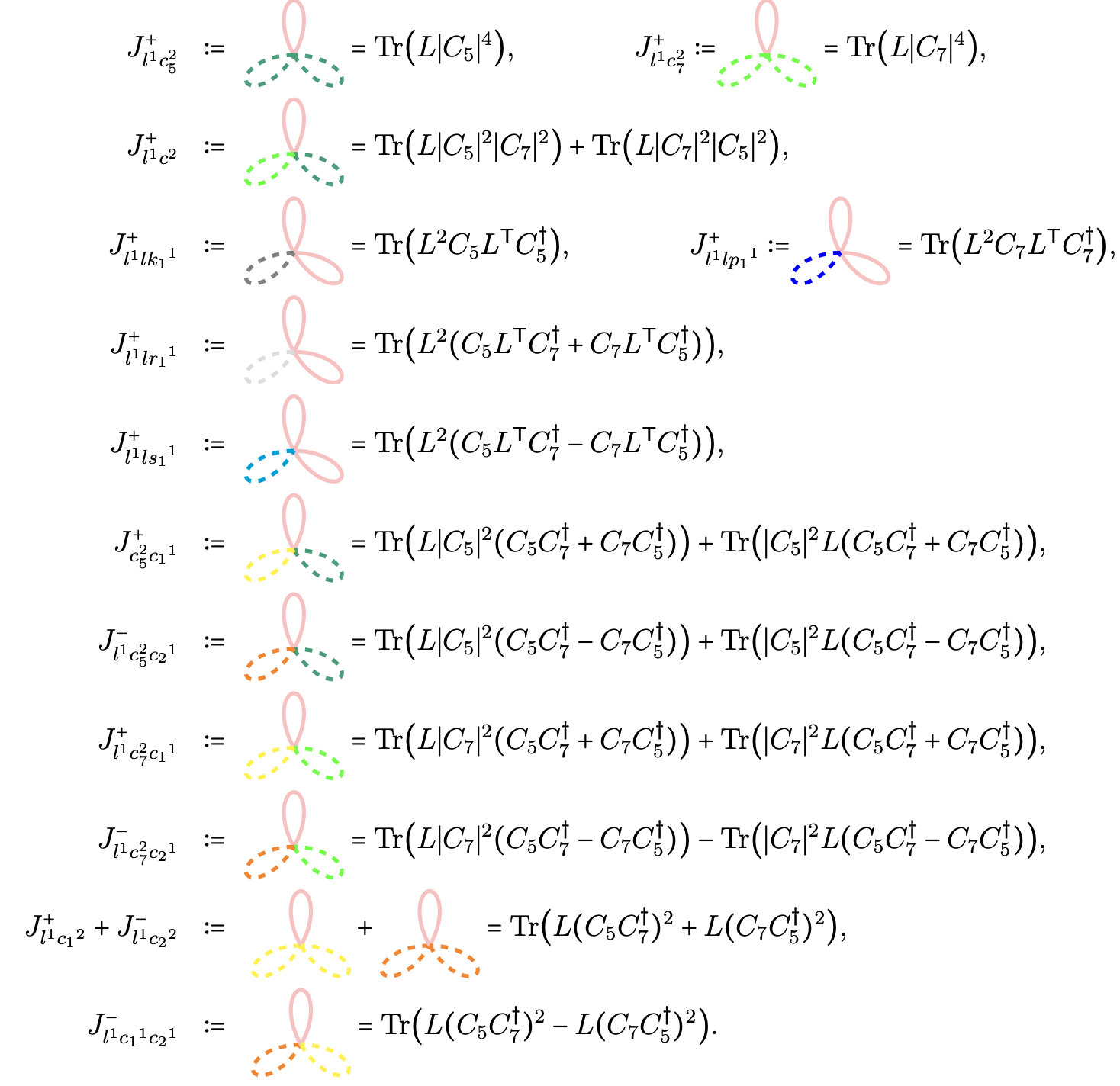}
\end{center}
\item[$\bullet$] Order-6:\hfill
\begin{center}\includegraphics[width=0.9\textwidth]{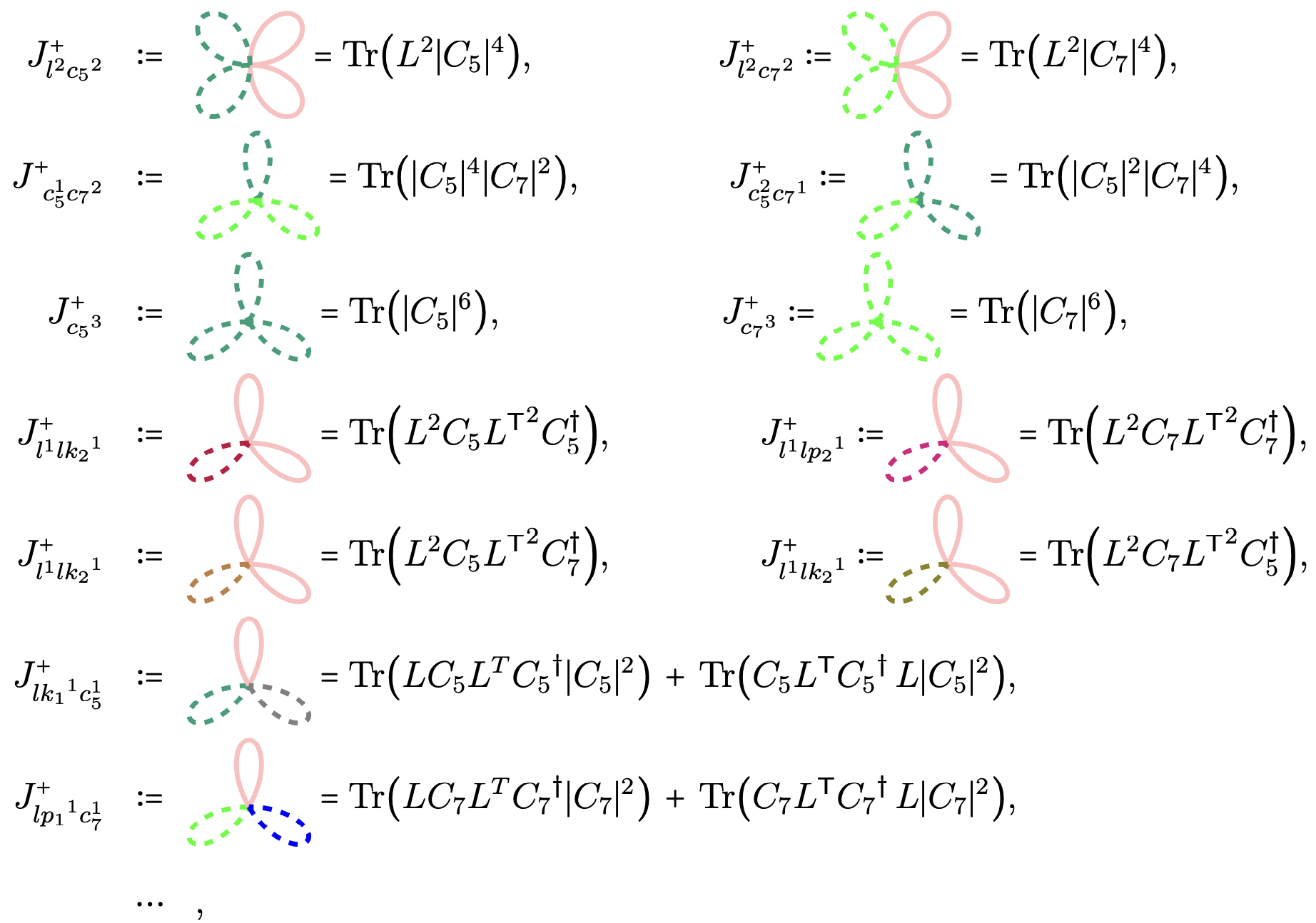}
\end{center}
where the dashed petals in grey, purple, brown and olive present order four $K_2 \equiv C_5 {L^{\sf T}}^2 C_5^\dagger$, $H_2 \equiv C_5 {L^{\sf T}}^2 C_5^\dagger$, $R_2^+ \equiv C_5 {L^{\sf T}}^2 C_7^\dagger +C_7 {L^{\sf T}}^2 C_5^\dagger$ and $R_2^- \equiv C_5 {L^{\sf T}}^2 C_7^\dagger - C_7 {L^{\sf T}}^2 C_5^\dagger$ in the ring diagram. 
\end{itemize}
In the above list, there are already CP-odd invariants $J^-$, where their non-zero values account for CP violation. 

Taking the character functions of ${\bf 3}$ and ${\bf 3}^*$ to be $z_1 +z_2 +z_3 $ and $z_1^{-1}+z_2^{-1}+z_3^{-1}$ respectively, one can calculate the character functions of the flavor invariants
 \small{\begin{eqnarray}
	\chi_l \left(z_1,z_2,z_3 \right)&=&\left(z_1 +z_2 +z_3 \right)\left(z_1^{-1}+z_2^{-1}+z_3^{-1}\right)\;,\nonumber\\
	\chi_5 \left(z_1,z_2,z_3 \right)&=&z_1^2+z_2^2+z_3^2+z_1 z_2 +z_1 z_3 +z_2 z_3 \\
	&&+z_1^{-2}+z_2^{-2}+z_3^{-2}+z_1^{-1}z_2^{-1}+z_1^{-1}z_3^{-1}+z_2^{-1}z_3^{-1}\;,\nonumber\\
	\chi_7 \left(z_1,z_2,z_3 \right)&=&z_1^2+z_2^2+z_3^2+z_1 z_2 +z_1 z_3 +z_2 z_3 \\
	&&+z_1^{-2}+z_2^{-2}+z_3^{-2}+z_1^{-1}z_2^{-1}+z_1^{-1}z_3^{-1}+z_2^{-1}z_3^{-1}\;.
\end{eqnarray}}
 Then the PE function can be written as 
 \small{\begin{eqnarray}\small
	\label{eq:PE eff 3g}
	{\rm PE}\left(z_1,z_2,z_3 ;q\right)&=& {\rm exp}(\sum_{k=1}^\infty\frac{\chi_l(z_1^k,z_2^k,z_3^k)q^k+\chi_5(z_1^k,z_2^k,z_3^k)q^k+\chi_7(z_1^k,z_2^k,z_3^k)q^k}{k})\nonumber\\
		&=&\Big[\left(1-q\right)^3\left(1-q z_1 z_2^{-1}\right) \left(1-q z_2 z_1^{-1}\right) \left(1-q z_1 z_3^{-1}\right) \left(1-q z_3 z_1^{-1}\right) \nonumber\\
	& & \left(1-q z_2 z_3^{-1}\right) \left(1-q z_3 z_2^{-1}\right) \left(1-q z_1^2\right)^2\left(1-q z_2^2\right)^2\left(1-q z_3^2\right)^2\nonumber\\
	& &\left(1-q z_1 z_2 \right)^2 \left(1-q z_1 z_3 \right)^2\left(1-q z_2 z_3 \right)^2
	\left(1-q z_1^{-2}\right)^2\left(1-q z_2^{-2}\right)^2\nonumber\\
	& & \left(1-q z_3^{-2}\right)^2 \left(1-q z_1^{-1}z_2^{-1}\right)^2\left(1-q z_1^{-1}z_3^{-1}\right)^2\left(1-q z_2^{-1}z_3^{-1}\right)^2\Big]^{-1}\;.
\end{eqnarray}}
Thus, one obtains the HS in the dim-7 $\nu$SMEFT where the Haar measure of ${U}(3)$ group has been inserted and thus one may obtain
 \small{\begin{eqnarray} 
	\label{eq:HS-3g}
	{ \cal H}_7 (q)=\frac{{ \cal N}_7 (q)}{{ \cal D}_7 (q)}\;,
\end{eqnarray}
where 
\small{\begin{eqnarray}
	\label{eq:numerator2}
{ \cal N}_7 (q)&=&1 + q^2 + 2 q^3 + 11 q^4 + 18 q^5 + 73 q^6 + 120 q^7 + 335 q^8 + 
 633 q^9
\nonumber\\
	&&+ 1454 q^{10}+ 2690 q^{11} + 5456 q^{12} + 9501 q^{13} + 
 17145 q^{14} + 28182 q^{15}
 \nonumber\\
	&&+ 46469 q^{16} + 71634 q^{17} + 109578 q^{18} + 
 158647 q^{19} + 226102 q^{20}
 \nonumber\\
	&&+ 308209 q^{21} + 411372 q^{22} + 
 528477 q^{23} + 663734 q^{24} + 805058 q^{25} \nonumber\\
	&&+ 953211 q^{26} + 
 1093469 q^{27} + 1222533 q^{28} + 1326782 q^{29} + 1402711 q^{30}
 \nonumber\\
	&&+ 
 1441224 q^{31} + 1441224 q^{32} + 1402711 q^{33} + 1326782 q^{34} + 
 1222533 q^{35}
 \nonumber\\
	&& + 1093469 q^{36}+ 953211 q^{37} + 805058 q^{38} + 
 663734 q^{39} + 528477 q^{40}
 \nonumber\\
	&& + 411372 q^{41} + 308209 q^{42} + 
 226102 q^{43} + 158647 q^{44}+ 109578 q^{45} 
 \nonumber\\
	&&+ 71634 q^{46} + 46469 q^{47} + 
 28182 q^{48} + 17145 q^{49} + 9501 q^{50} + 5456 q^{51} 
	 \nonumber\\
	&&+ 2690 q^{52}+ 
 1454 q^{53} + 633 q^{54} + 335 q^{55} + 120 q^{56} + 73 q^{57} + 18 q^{58} \nonumber\\
	&&+ 
 11 q^{59} + 2 q^{60} + q^{61} + q^{63},
\end{eqnarray}}
and
\begin{eqnarray}\small
	\label{eq:denominator}
		{ \cal D}_7 (q) = (1 - q )^1 (1 - q^2 )^4 (1 - q^3 )^3 (1 - q^4)^7 (1 - q^5 )^4 (1 - q^6 )^5\;.
\end{eqnarray}
Hence, the PL takes on the form
\begin{eqnarray}\small
	\label{eq:PL7}
	{\rm PL}\left[{\cal H}(q)\right]&=&q + 5 q^2 + 5 q^3 + 17 q^4 + 20 q^5 + 64 q^6 + 82 q^7 + 175 q^8 + 231 q^9 \nonumber\\
	&&+ 199 q^{10}-{\cal O}\left(q^{11}\right)\;,
\end{eqnarray}
that aligns with the results obtained earlier in this section using petals of building blocks.

Our procedure, suggests an automated way to find explicitly the structure of invariants other than providing only the number of invariants. These classifications of invariants play a crucial role in analysing theories which explain flavor by a dynamical mechanism. 

\subsection{Three-Generation Type-I Seesaw Model}
As a representative example of transformations under a product flavor symmetry group ${\rm U}(m) \otimes {\rm U}(n)$, we consider the full Type-I seesaw model. In this context, we take $m = n = 3$, corresponding to three generations of right-handed neutrinos. The fundamental building blocks for constructing flavor invariants in this model are the Yukawa matrices $Y_l$ and $Y_\nu$, along with a symmetric matrix $Y_{\rm R}$ associated with the Majorana mass term for the right-handed neutrinos. Their transformation properties under the flavor symmetry group ${\rm U}(3)_L \otimes {\rm U}(3)_e \otimes {\rm U}(3)_{\sf R}$ are summarized in Table~\ref{tab:seesaw-basic}.


\begin{table}[H]
\begin{center}
	\begin{tabular}{l|c|c|c}
		& \; ${\rm U}(3)_{L}$ \; & \; ${\rm U}(3)_e$ \; & \; ${\rm U}(3)_{\sf R}$ \; \\\hline
		$Y_l$ & $\mathbf{3}$ & $\mathbf{\overline{3}}$ & $\mathbf{1}$ \\[0.1cm]
		$Y_\nu$ & $\mathbf{3}$ & $\mathbf{1}$ & $\mathbf{\overline{3}}$ \\[0.1cm]
		$Y_{\rm R}$ & $\mathbf{1}$ & $\mathbf{1}$ & $\left(\mathbf{\overline{3}} \otimes \mathbf{\overline{3}}\right)_{\rm s}$ \\[0.1cm]
	\end{tabular}
\end{center}
\caption{\it Transformation properties of the basic Yukawa and Majorana matrices under the flavor symmetry ${\rm U}(3)_L \otimes {\rm U}(3)_e \otimes {\rm U}(3)_{\sf R}$.}
\label{tab:seesaw-basic}
\end{table}

Therefore, using the Hermitian combinations $ L \equiv Y_l Y_l^\dagger $ and $ R \equiv Y_R^\dagger Y_R $, along with the basic building blocks $ \{Y_\nu,\, Y_\nu^\dagger,\, Y_\nu^*\} $, the resulting ring of flavor invariants can be represented diagrammatically as shown in Fig.~\ref{RDss}.
\begin{figure}[H]
\begin{center}
\includegraphics[width=0.28\textwidth]{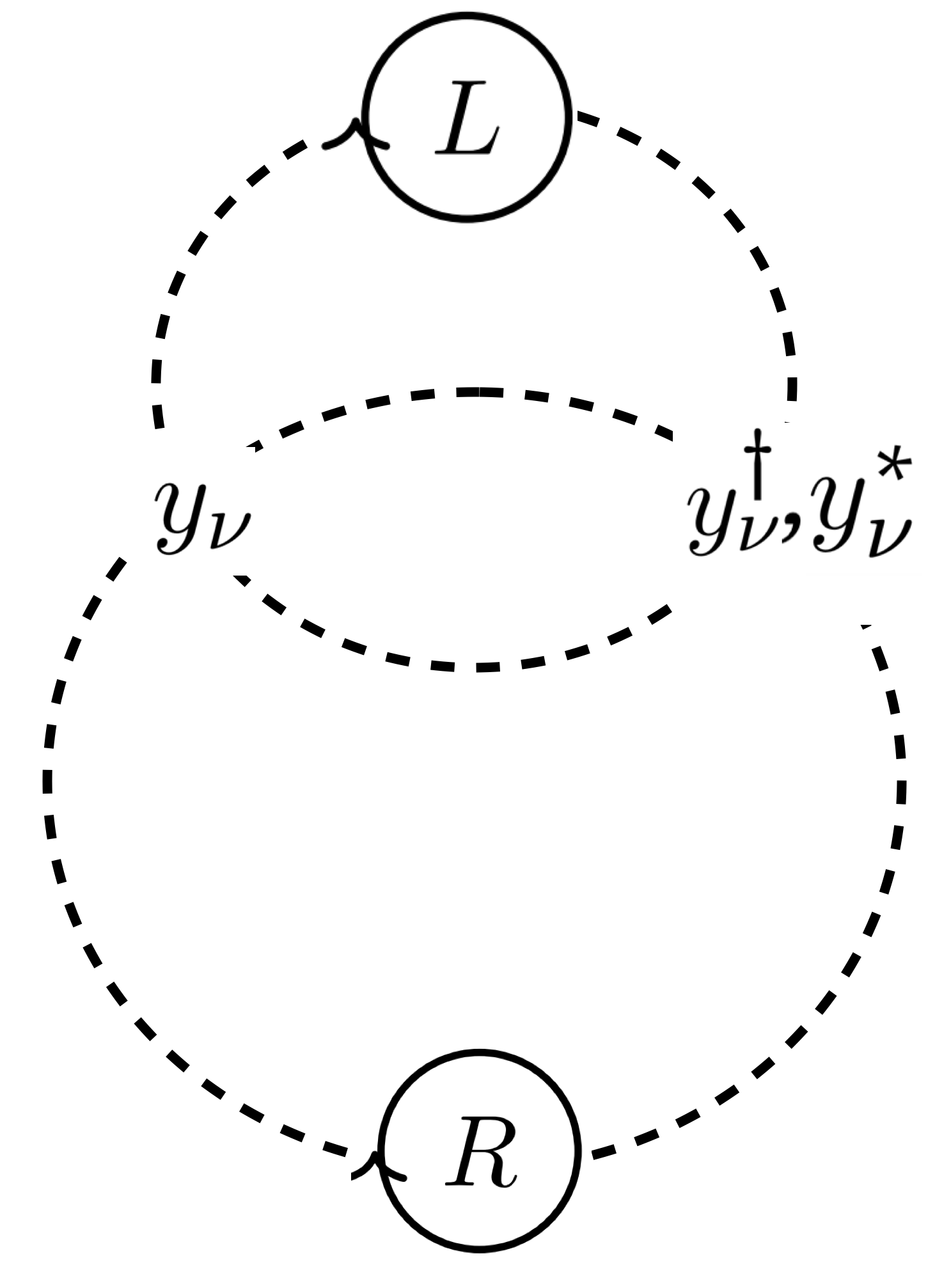}
\end{center}
\caption{The Ring-diagram for seesaw and three generations of right-handed neutrinos.}
\label{RDss}
\end{figure} 
Accordingly, the relevant building blocks can be organized as follows:
\begin{eqnarray}\small
 M_L &\equiv & \text{diag} \begin{pmatrix} L & 0 & 0 \end{pmatrix}_m\equiv \hspace{0.2cm}{\raisebox{-0.7em}{\begin{tikzpicture}
\def \radius {0.4cm}
 \node[draw, circle,style={thick},-latex] {$L$};
 \draw[-latex,->,thick] ({180}:\radius);
\end{tikzpicture}}}\,,
\nonumber \\
\small
 M_R &\equiv & \text{diag} \begin{pmatrix} 0 & R & 0 \end{pmatrix}_n\equiv \hspace{0.2cm}{\raisebox{-0.7em}{\begin{tikzpicture}
\def \radius {0.4cm}
 \node[draw, circle,style={thick},-latex] {$R$};
 \draw[-latex,->,thick] ({180}:\radius);
\end{tikzpicture}}}\,,
\nonumber \\
\small
 M_N &\equiv & \text{diag} \begin{pmatrix} 0 & 0 & N \end{pmatrix}_{mn}\equiv \hspace{0.2cm}{\raisebox{-0.7em}{\begin{tikzpicture}
\def \radius {0.4cm}
 \node[draw, circle,style={thick},-latex] {$N$};
 \draw[-latex,->,thick] ({180}:\radius);
\end{tikzpicture}}}\,,
\nonumber \\
M_{LN}^\pm &\equiv &  \begin{pmatrix}
 LN & 0\\
 0 & \pm \,NL \\
\end{pmatrix}~ \equiv \raisebox{-0.5em}{ 
\begin{tikzpicture}[transform shape,line width=0.9pt]
\node (A) at (0,0) {$L$}; 
\node (B) at (1.2,0) {$N$}; 
\path[draw] 
(A) edge[bend right=-40,black,dashed,->] (B)
(B) edge[bend right=-40,black,-latex,dashed,->] (A);
\end{tikzpicture}} \,,
\nonumber \\
M_{RN}^\pm &\equiv &  \begin{pmatrix}
 RN & 0\\
 0 & \pm \,NR \\
\end{pmatrix}~ \equiv \raisebox{-0.5em}{ 
\begin{tikzpicture}[transform shape,line width=0.9pt]
\node (A) at (0,0) {$N$}; 
\node (B) at (1.2,0) {$R$}; 
\path[draw] 
(A) edge[bend right=-40,black,dashed,->] (B)
(B) edge[bend right=-40,black,-latex,dashed,->] (A);
\end{tikzpicture}} \,,
\nonumber \\
M_{lk_m}^\pm &\equiv &\textbf{1}_3 \otimes \begin{pmatrix}
R \,y_\nu^{\dagger}\,{L}^m\,y_\nu& 0\\
0 &\, \pm y_\nu^{\dagger}\,{L}^m\,y_\nu \, R  \\
\end{pmatrix}~ \equiv \raisebox{-0.1em}{ 
\raisebox{-2.8em}{ \begin{tikzpicture}[transform shape,line width=1pt]
\node (A) at (0.9,1) {$y_\nu$}; 
\node (B) at (2,2) {$R$}; 
\node (C) at (3,1) {$y_\nu^{\dagger}$};
\node (D) at (2,0) {${L}^m$}; 
\node (E) at (1,0.8) {}; 
\path[draw] 
(A) edge[bend right=-30,black,dashed,->] (B) 
(B) edge[bend right=-30,black,dashed,->] (C)
(C) edge[bend right=-30,black,dashed,->] (D) 
(D) edge[bend right=-30,black,dashed,->] (E);
\end{tikzpicture}}}\,
\nonumber \\
&\equiv & \textbf{1}_3 \otimes \begin{pmatrix}
R\,K_m& 0\\
0 &\, \pm K_m\,R \\
\end{pmatrix}~ \equiv \raisebox{-1em}{ 
\begin{tikzpicture}[transform shape,line width=0.9pt]
\node (A) at (0,0) {$R$}; 
\node (B) at (1.2,0) { $K_m$ }; 
\path[draw] 
(A) edge[bend right=-40,black,dashed,->] (B)
(B) edge[bend right=-40,black,-latex,dashed,->] (A);
\end{tikzpicture}} \,,
\nonumber \\
M_{lk_m}^\pm &\equiv & \textbf{1}_3 \otimes \begin{pmatrix}
R \,y_\nu^{*}\,{L}^m\,y_\nu & 0\\
0 &\, \pm y_\nu^{*}\,{L}^m\,y_\nu\, R  \\
\end{pmatrix}~ \equiv \raisebox{-0.1em}{ 
\raisebox{-2.8em}{ \begin{tikzpicture}[transform shape,line width=1pt]
\node (A) at (0.9,1) {$y_\nu$}; 
\node (B) at (2,2) {$R$}; 
\node (C) at (3,1) {$y_\nu^{*}$};
\node (D) at (2,0) {${L}^m$}; 
\node (E) at (1,0.8) {}; 
\path[draw] 
(A) edge[bend right=-30,black,dashed,->] (B) 
(B) edge[bend right=-30,black,dashed,->] (C)
(C) edge[bend right=-30,black,dashed,->] (D) 
(D) edge[bend right=-30,black,dashed,->] (E);
\end{tikzpicture}}}\,
\nonumber \\
&\equiv & \textbf{1}_3 \otimes \begin{pmatrix}
R \,K_m& 0\\
0 &\, \pm K_m\,R \\
\end{pmatrix}~ \equiv \raisebox{-1em}{ 
\begin{tikzpicture}[transform shape,line width=0.9pt]
\node (A) at (0,0) {$R$}; 
\node (B) at (1.2,0) { $K_m*$ }; 
\path[draw] 
(A) edge[bend right=-40,black,dashed,->] (B)
(B) edge[bend right=-40,black,-latex,dashed,->] (A);
\end{tikzpicture}} \,,
 \label{blocks-dim-6-2}
\end{eqnarray}
where $K_m=y_\nu^{\dagger}\,{L}^m\,y_\nu$ and $K_m^*=y_\nu^{*}\,{L}^m\,y_\nu$ with $m=1,2$. However, $K_1$ and $K_2$ are derived from elements that also contribute to lower-order blocks and do not represent standalone invariants but, through their integration with other blocks, yield unique invariants $M_{lk_m}^\pm$.
The basic invariants can be combined into the following form, with color coding in the diagrams indicating the building blocks: green for $L$, gray for $R$, pink for $N$, dashed blue for $K_m$, and dashed red for $K_m^*$.
\begin{itemize}
\item Order-1:\hfill
\begin{center}\includegraphics[width=0.9\textwidth]{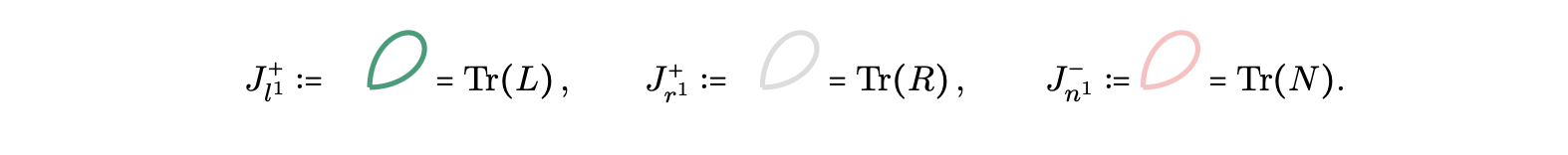}\end{center}
\item Order-2:\hfill
\begin{center}\includegraphics[width=0.9\textwidth]{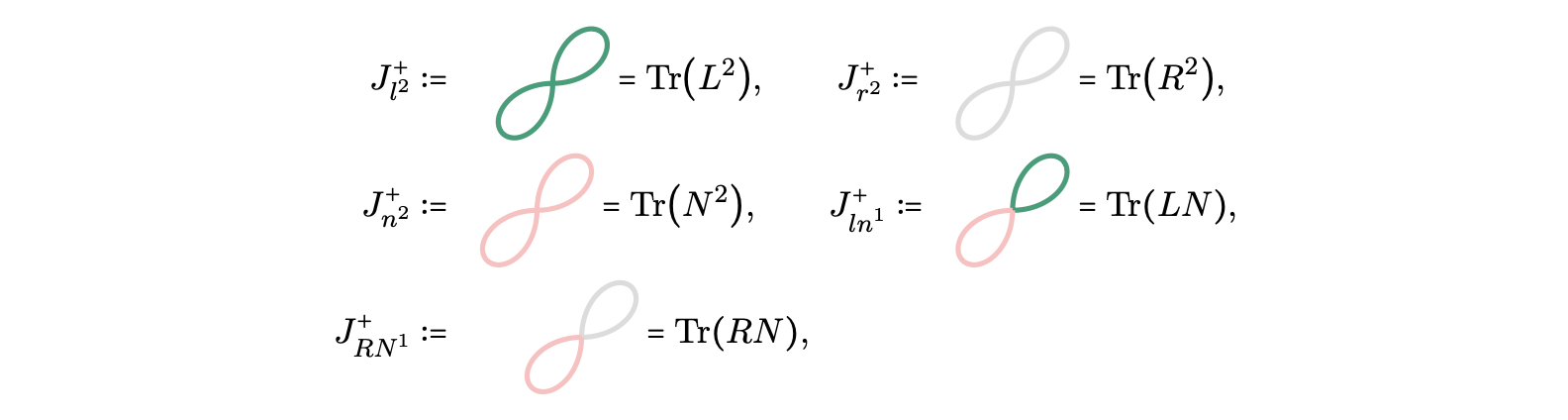}\end{center}
\item Order-3:\hfill
\begin{center}\includegraphics[width=0.9\textwidth]{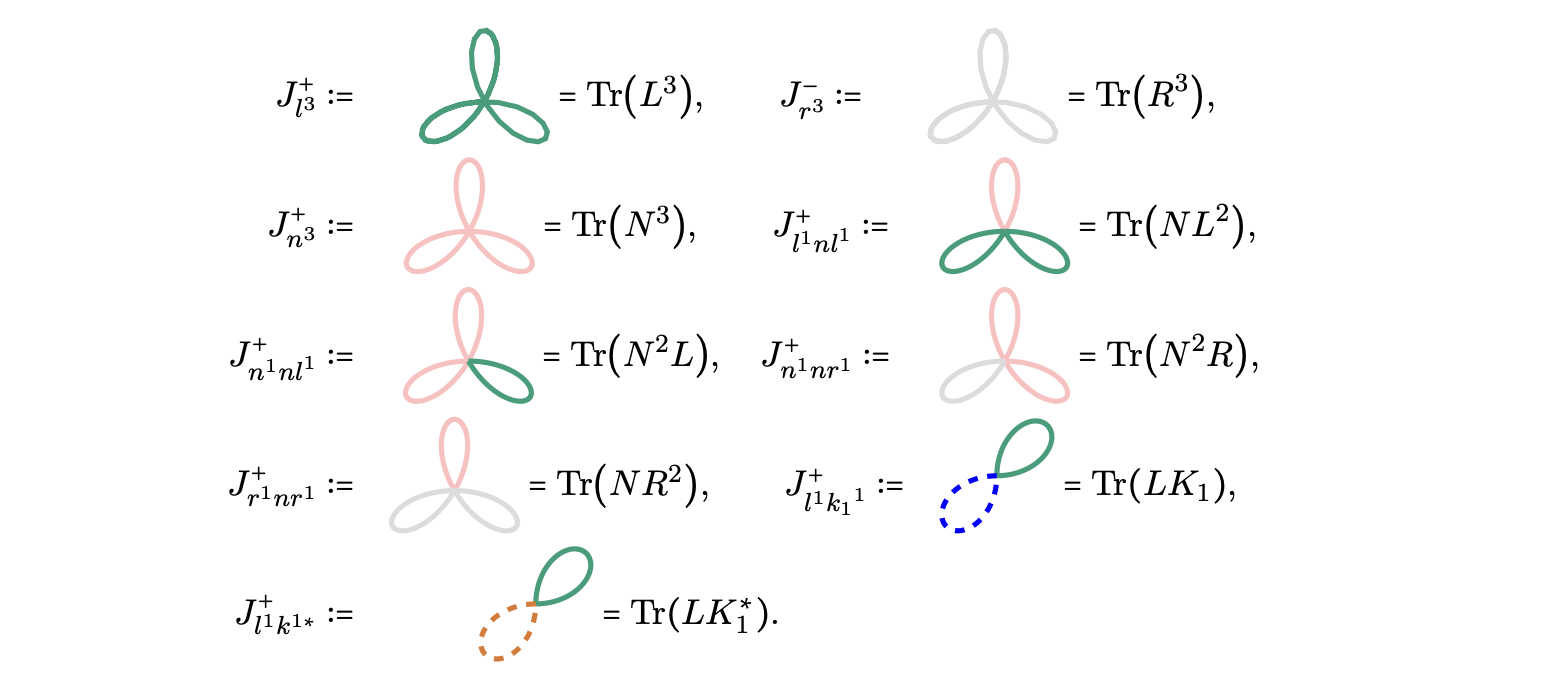}\end{center}
\end{itemize}
A similar structure can be extended to construct higher-order invariants, with the first CP-odd invariant appearing at fourth order, given by:
\bea J^-_{n^1{rk}^1} :=  \hspace{-0.1cm}{\raisebox{-1.1em}{\includegraphics[width=0.09\textwidth]{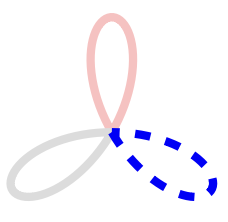}}} = \Tr({N} R K_1) -\Tr(N K_1 R).\nonumber\eea
As a result, the first joint invariant appears at order 5 as:
 \bea
J^+_{n^2{rk}^1}:= \hspace{-0.1cm}{\raisebox{-1.1em}{\includegraphics[width=0.09\textwidth]{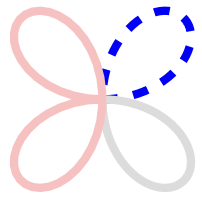}}}= \Tr({N^2} R K_1)- \Tr({N^2} K_1 R).\nonumber\eea
This suggests that the sequence of invariants is expected to terminate beyond order $4 + 5$.

To verify whether this prediction  is consistent with the Hilbert series, we analyze the character functions associated with the relevant representations. Assigning the character functions of the fundamental and anti-fundamental representations, $\mathbf{3}$ and $\mathbf{3}^*$, as $z_1 + z_2 + z_3$ and $z_1^{-1} + z_2^{-1} + z_3^{-1}$ respectively, one can compute the character functions corresponding to the flavor invariants as follows:

\begin{eqnarray}
	\chi_l\left(z_1,z_2,z_3\right)&=&\left(z_1+z_2+z_3\right)\left(z_1^{-1}+z_2^{-1}+z_3^{-1}\right)\,\nonumber\\
	\chi_\nu\left(z_1,z_2,z_3,z_4{},z_5,z_6\right)&=&\left(z_1+z_2+z_3\right)\left(z_4^{-1}+z_5^{-1}+z_6^{-1}\right)+\left(z_4+z_5+z_6\right)\left(z_1^{-1}+z_2^{-1}+z_3^{-1}\right)\;,\nonumber\\
	\chi_{\rm R}\left(z_4,z_5,z_6\right)&=&z_4^2+z_5^2+z_6^2+z_4 z_5+z_4 z_6+z_5 z_6+z_4^{-2}+z_5^{-2}+z_6^{-2}\nonumber\\
	&&+z_4^{-1}z_5^{-1}+z_4^{-1}z_6^{-1}+z_5^{-1}z_6^{-1}\;,
\end{eqnarray}
where $z_i$, for $i=1,2,3$ (or for $i=4,5,6$) denote the coordinates on the maximum torus of the ${\rm U}(3)$ group that corresponds to the flavor-basis transformation in the active neutrino (or RH neutrino) sector. Labelling the degrees of $Y_l$, $Y_\nu$ and $Y_{\rm R}$ by $q$, one can calculate the PE function
\begin{eqnarray}
	\label{eq:PE-seesaw}
	&&{\rm PE}\left(z_1,z_2,z_3,z_4,z_5,z_6;q\right)\nonumber\\
	&=&\left[\left(1-q \right)^3\left(1-q  z_1 z_2^{-1}\right)\left(1-q  z_2 z_1^{-1}\right)\left(1-q  z_1 z_3^{-1}\right)\left(1-q  z_3 z_1^{-1}\right)\left(1-q  z_2 z_3^{-1}\right)\right.\nonumber\\
	&&\left.\times\left(1-q  z_3 z_2^{-1}\right)\left(1-q z_1 z_4^{-1}\right)\left(1-q z_4 z_1^{-1}\right)\left(1-q z_1 z_5^{-1}\right)\left(1-q z_5 z_1^{-1}\right)\left(1-q z_1 z_6^{-1}\right)\right.\nonumber\\
	&&\left.\times\left(1-q z_6 z_1^{-1}\right)\left(1-q z_2 z_4^{-1}\right)\left(1-q z_4 z_2^{-1}\right)\left(1-q z_2 z_5^{-1}\right)\left(1-q z_5 z_2^{-1}\right)\left(1-q z_2 z_6^{-1}\right)\right.\nonumber\\
	&&\left.\times\left(1-q z_6 z_2^{-1}\right)\left(1-q z_3 z_4^{-1}\right)\left(1-q z_4 z_3^{-1}\right)\left(1-q z_3 z_5^{-1}\right)\left(1-q z_5 z_3^{-1}\right)\left(1-q z_3 z_6^{-1}\right)\right.\nonumber\\
	&&\left.\times\left(1-q z_6 z_3^{-1}\right)\left(1-q z_4^2\right)\left(1-q z_5^2\right)\left(1-q z_6^2\right)\left(1-q z_4 z_5\right)\left(1-q z_4 z_6\right)\left(1-q z_5 z_6\right)\right.\nonumber\\
	&&\left.\times \left(1-q z_4^{-2}\right)\left(1-q z_5^{-2}\right)\left(1-q z_6^{-2}\right)\left(1-q z_4^{-1}z_5^{-1}\right)\left(1-q z_4^{-1}z_6^{-1}\right)\left(1-q z_5^{-1}z_6^{-1}\right)
	\right]^{-1}\;.
\end{eqnarray}
Using the above equation, one gets  
\begin{eqnarray}
	\label{eq:HS-seesaw}
	{\mathcal{H}}(q)&=&\frac{{{ \cal N}}(q)}{{ \cal D}(q)} =\int \left[{\rm d}\mu\right]_{{\rm U} (3)\otimes{\rm U}(3)} {\rm PE}\left(z_1,z_2,z_3,z_4,z_5,z_6;q\right)\nonumber\\
	&=&\frac{1}{36}\frac{1}{\left(2\pi {\rm i}\right)^6}\oint_{\left|z_1\right|=1}\oint_{\left|z_2\right|=1}\oint_{\left|z_3\right|=1}\oint_{\left|z_4\right|=1}\oint_{\left|z_5\right|=1}\oint_{\left|z_6\right|=1}\left[-\frac{\left(z_2-z_1\right)^2\left(z_3-z_1\right)^2\left(z_3-z_2\right)^2}{z_1^2z_2^2z_3^2}\right]\nonumber\\
	&&\times\left[-\frac{\left(z_5-z_4\right)^2\left(z_6-z_4\right)^2\left(z_6-z_5\right)^2}{z_4^2z_5^2z_6^2}\right]{\rm PE}\left(z_1,z_2,z_3,z_4,z_5,z_6;q\right)\;,
\end{eqnarray}
where 
\begin{eqnarray}
	\label{eq:num-seesaw}
	{\cal N}(q)&=&1+q^2+5q^3+9q^4+22q^{5}+61q^{6}+126q^{7}+273q^{8}+552q^{9}+1038q^{10}+1880q^{11}\nonumber\\
	&&+3293q^{12}+5441q^{13}+8712q^{14}+13417q^{15}+19867q^{16}+28414q^{17}+39351q^{18}\nonumber\\
	&&+52604q^{19}+68220q^{20}+85783q^{21}+104588q^{22}+123852q^{23}+142559q^{24}+159328q^{25}\nonumber\\
	&&+173201q^{26}+183138q^{27}+188232q^{28}+188232q^{29}+183138q^{30}+173201q^{31}\nonumber\\
	&&+159328q^{32}+142559q^{33}+123852q^{34}+104588q^{35}+85783q^{36}+68220q^{37}+52604q^{38}\nonumber\\
	&&+39351q^{39}+28414q^{40}+19867q^{41}+13417q^{42}+8712q^{43}+5441q^{44}+3293q^{45}\nonumber\\
	&&+1880q^{46}+1038q^{47}+552q^{48}+273q^{49}+126q^{50}+61q^{51}+22q^{52}+9q^{53}+5q^{54}+q^{55}+q^{57}\;,
\end{eqnarray}
and
\begin{eqnarray}
	\label{eq:den-seesaw}
	{ \cal D}(q)=\left(1-q\right)^3\left(1-q^2\right)^4(1-q^3)^4\left(1-q^4\right)^2\left(1-q^{5}\right)^2\left(1-q^{6}\right)^3\left(1-q^{7}\right)^2\left(1-q^{8}\right)\;, \quad
\end{eqnarray}
where the PL takes on the form
\begin{eqnarray}\small
	\label{eq:PL7}
	{\rm PL}\left[{\cal H}(q)\right]&=&3q + 5 q^2 + 9 q^3 + 10 q^4+ 19 q^{5}+ 40 q^{6}+ 66 q^{7} + 92 q^{8} +70 q^{9}-{\cal O}\left(q^{10}\right)\;.
\end{eqnarray}
The result reproduces that of Ref.~\cite{Hanany:2010vu} when $q=t^2$.

\section{Summary and discussions} \label{sec:con}
This paper extends our earlier work \cite{Darvishi:2023ckq}, enhancing the Ring-Diagram framework's capability to classify CP invariants. We enrich this methodology with detailed examples and extend its application across a wider range of theoretical physics.

The Hilbert-Poincaré series (HS) and its Plethystic Logarithm (PL) are valuable tools for counting invariants in theoretical physics. In particular, the graded Hilbert series can encode important structural information, such as the spurion content of each invariant, and in some cases, it can also provide insight into their CP properties, as discussed in \cite{Henning:2017fpj,Grinstein:2023njq}. However, in practical applications, especially in theories involving high-rank tensors or multiple spurions, the graded series can become computationally demanding and challenging to interpret. Furthermore, while these algebraic tools are effective for enumeration, they often do not directly reveal the physical significance or internal structure of the invariants.

The Ring-Diagram technique provides a complementary perspective by focusing on the visual and topological organization of invariants. This method facilitates a more systematic classification and enables the identification of physically meaningful combinations. It is particularly effective in analyzing CP properties and revealing structural patterns that may not be evident through algebraic approaches. We believe this framework offers a valuable addition to the existing toolkit for studying invariant structures in complex field theories.

In this work, we have further integrated the Ring-Diagram approach with the Cayley-Hamilton theorem, resulting in an automated procedure for classifying basis invariants. This development allows us to systematically identify both basic and joint invariants and demonstrate that sums of joint invariants can always be expressed in terms of basic invariants arranged in a specific order. The formalism we present enables the following:
\begin{itemize}
\item Apply the Cayley-Hamilton theorem in practice,
\item Offer a generic method for the classification of invariants,
\item Automatically distinguish between CP-even and CP-odd invariants right from their most basic levels.
\end{itemize} 

Through this approach, we have revisited the SM invariants and demonstrated the technique's application in identifying invariants across SMEFT up to dim-$2n$ (while maintaining a dim-6 core) and low energy $\nu$SMEFT including operators of dim-{5,6,7}  as well as the type-I seesaw model. This method not only offers a new perspective on invariant identification but also provides new explanations for details in the classical HS and its PL, particularly in terms of classifying basis invariants, joint invariants, syzygies, and the negative leading numbers in PL.

The structure of invariants obtained in our paper for SM fully matches those in \cite{Jenkins:2009dy, Bonnefoy:2021tbt}. Similarly, the CP-odd invariants identified in SMEFT dim-6 align with the findings in \cite{Bonnefoy:2021tbt} using the Hilbert series. For $\nu$SMEFT dimension 5, our method's invariants correspond entirely with those observed in~\cite{Jenkins:2009dy}. While the total number of invariants for $\nu$SMEFT dim-5 and -6 matches~\cite{Yu:2022ttm} based on the Hilbert series, the number of primary invariants defined in~\cite{Yu:2022ttm} depends on the denominator of the Hilbert series, which could change with a different numerator. Our method enhances the presentation of CP-odd and primary invariants by applying the Cayley-Hamilton theorem through Ring-diagrams, providing a more robust and systematic classification which can uniformly be applied in various models.

To further evaluate the scalability of our method in more complex symmetry settings, we consider an example based on the Type-I seesaw model with three generations of neutrinos and a ${\rm U}(3) \times {\rm U}(3)$ flavor symmetry.  In such scenarios, computing the Hilbert series becomes significantly more demanding due to the need to integrate over multiple unitary groups and the rapidly growing number of residues, as initially discussed in~\cite{Jenkins:2009dy} and further developed using the ungraded Hilbert series in~\cite{Hanany:2010vu}.  We demonstrate that the Ring-Diagram framework remains computationally efficient in this context and exhibits more favorable scaling behavior compared to the Hilbert series approach. This highlights both the practicality and general applicability of our method for analyzing flavor invariants in models with extended symmetry structures.

In Table~\ref{tab:tab4}, we present a summary of the lowest order of CP-odd invariants due to matrix properties (excluding their complex nature), the highest order of basic invariants, and the lowest order of joint invariants for the SM, SMEFT, $\nu$SMEFT and type-I seesaw models. The final column indicates the orders greater than or equal to the sum of the lowest order of CP-odd invariants and the lowest order of joint invariants for each framework, except when the number of the lowest order of joint invariants is smaller than that of the highest order basic invariants. In such scenarios, it lists orders greater than the sum of the lowest order of CP-odd and the highest order of basic invariants. This column clarifies in which order the invariants cease, which aligns with the appearance of negative leading numbers in the PL.

\begin{table}[t]
\begin{center}
\begin{tabular}{l|c c c|c}
\textbf{Model} & \textbf{LO CP-odd} & \textbf{HO Basic} & \textbf{LO Joint} & \textbf{Termination} \\ \hline
SM & 6 & 4 & 6 & $\geq$ 12 \\ \hline
SMEFT dim-6 (dim-$2n$) & 3 & 4 & 4 & $\geq$ 7 \\ \hline
$\nu$-SMEFT dim-5 & 6 & 6 & 7 & $\geq$ 13 \\ \hline
$\nu$-SMEFT dim-5,6 & 4 & 6 & 5 & $>$ 10 \\ \hline
$\nu$-SMEFT dim-5,7 & 5 & 6 & 6 & $\geq$ 11 \\ \hline
Type-I seesaw & 4 & 8 & 5 & $>$ 9 \\ \hline
\end{tabular}
\caption{\textit{Summary of the occurrences and termination of different types of invariants for SM,  SMEFT dim-6, $\nu$-SMEFT dim-5, $\nu$-SMEFT dim-5,6 , $\nu$-SMEFT dim-5,7 and type-I seesaw models. The second column outlines the lowest order of CP-odd invariants, while the third and the fourth columns detail the highest order of basic invariants and the lowest order of joint invariants, respectively. The final column delineates the order of invariant termination.}}
\label{tab:tab4}
\end{center}
\end{table}

This method can be applied in the construction of invariants from high-rank tensors and complex structures as well as the frameworks of multi-Higgs doublet models ($n$HDMs) and $n$HDM-Effective Field Theory ($n$HDM-EFT)~\cite{Darvishi:2019dbh,Birch-Sykes:2020btk,Darvishi:2023fjh,Darvishi:2025gdl} that we plan to report in our upcoming publications. This integrated approach aims to fill existing gaps in the literature and provides a more complete picture of the intricate relationship between CPV and EDMs.

\section*{Acknowledgments} 
The work of ND is supported by STFC under the grant number ST/T006749/1.
This work is supported by the National Science Foundation of China under Grants No. 12347105, No. 12375099 and No. 12047503, and National Key Research and Development Program of China Grant No. 2020YFC2201501, No. 2021YFA0718304. 
\bibliographystyle{unsrt}
\bibliography{biblio.bib}
 
\end{document}